\documentclass[floatfix,showpacs,superscriptaddress,nofootinbib,twocolumn,showkeys]{revtex4-2}

\usepackage[utf8]{inputenc}

\usepackage{graphicx}
\usepackage{xcolor}
\usepackage{amsmath}
\usepackage{hyperref}

\usepackage{amssymb}
\usepackage{multirow,array}
\usepackage{lipsum}
\usepackage{enumitem}
\usepackage{soul}

\newcommand{\trento}{\texttt{T$_\mathtt{R}$ENTo}}

\newcommand{\SMASH}{\texttt{SMASH}}
\newcommand{\music}{\texttt{MUSIC}}
\newcommand{\urqmd}{\texttt{UrQMD}}
\newcommand{\vish}{\texttt{VISH}}
\newcommand{\trajectum}{\texttt{TRAJECTUM}}

\newcommand{\sqrts}{\sqrt{s_\textrm{NN}}}

\begin{document}

\title{Bayesian quantification of strongly-interacting matter with color glass condensate initial conditions}

\author{Matthew Heffernan}
\affiliation{Department of Physics, McGill University, Montr\'{e}al QC H3A\,2T8, Canada}

\author{Charles Gale}
\affiliation{Department of Physics, McGill University, Montr\'{e}al QC H3A\,2T8, Canada}
\author{Sangyong Jeon}
\affiliation{Department of Physics, McGill University, Montr\'{e}al QC H3A\,2T8, Canada}

\author{Jean-Fran\c{c}ois Paquet}
\affiliation{Department of Physics and Astronomy, Vanderbilt University, Nashville TN 37235}
\affiliation{Department of Mathematics, Vanderbilt University, Nashville TN 37235}

\date{\today}

\begin{abstract}
A global Bayesian analysis of relativistic Pb + Pb collisions at $\sqrts$ = 2.76 TeV is performed, using a multistage model consisting of an IP-Glasma initial state, a viscous fluid dynamical evolution, and a hadronic transport final state. The observables considered are from the soft sector hadronic final state. Posterior and Maximum a Posteriori parameter distributions that pertain to the IP-Glasma and hydrodynamic phases are obtained, including the shear and bulk specific viscosity of strong interacting matter. The first use of inference with transfer learning in heavy-ion analyses is presented, together with Bayes Model Averaging. 
\end{abstract}

\maketitle

\section{Introduction}
    Much is known about the behavior of Quantum Chromodynamics (QCD) -- the theory of the nuclear strong interaction -- in situations where ``cold'' strongly interacting systems are investigated by high-energy probes. This success is owed in great part to asymptotic freedom, the running of the strong constant whose value decreases as the energy scale increases, rendering controlled and systematically improvable perturbative calculations possible. In regimes where perturbative approaches converge poorly, lattice QCD has proven to be a powerful tool to investigate QCD both at zero and finite temperatures \cite{[{See, for example, }][{, and references therein.}]Karsch:2022opd}.
    Comparing with the status of ``cold'' QCD, much less is known about the behavior of the theory in conditions of extreme temperatures and energy density, although some features have nevertheless been predicted with certainty. In such environments, lattice calculations have predicted a crossover transition to occur from composite hadronic degrees of freedom to a partonic state (the quark-gluon plasma (QGP)) at a temperature $\approx 150$ MeV, for vanishing net baryonic density  \cite{[{See, for example, }][{, and references therein.}]Ratti:2018ksb}. Furthermore, exploring the QCD phase diagram away from the axis where baryon density vanishes, several theoretical studies support the existence of a first-order phase transition line, terminating at a critical end point (CEP) \cite{[{See, for example, }][{, and references therein.}]An:2021wof}. 
    
    In addition to an active global research effort in the theory of strongly-interacting systems under extreme conditions, a vigorous experimental program exists to further study and characterize the QGP through experiments performed at facilities around the world and also through observations involving dense stellar objects such as neutron stars \cite{Annala:2019puf}. In terrestrial laboratories, this exotic state of QCD has been experimentally observed at the Relativistic Heavy-Ion Collider (RHIC, at Brookhaven National Laboratory) and at the Large Hadron Collider (LHC, at CERN) involving the relativistic collisions of large nuclei (``heavy ions'') \cite{[{See, for example, }][{, and references therein.}]Busza:2018rrf}, and much activity is currently also being devoted to studies involving comparatively smaller hadronic systems \cite{Strickland:2018exs}.

    One of the theoretical breakthroughs in modeling relativistic heavy ion collisions has been the realization of the effectiveness of relativistic viscous fluid dynamics, which features collective hadronic flow that accurately describes experimental observations in heavy-ion collisions \cite{[{See, for example, }][{, and references therein.}]Gale:2013da}.  
    Along with this milestone in theory, the importance of deviation from ideal fluid dynamics is quantified with the evaluation of transport coefficients that represent fundamental features of QCD. The main ones represent the shear and bulk viscosity of the strongly interacting matter \cite{Gale:2013da,Florkowski:2017olj}. 
    Much research has been devoted to the evaluation of those shear and bulk viscosity coefficients using a variety of models and approaches, both perturbative and non-perturbative \cite{[{See, for example, }][{, and references therein.}]Romatschke:2017ejr}. 
    Up to now, direct calculations have been met with limited success. One of the contributing factors is the fact that the conditions created in nuclear collisions and reconstructed by hadronic probes span a parameter space where QCD is inherently non-perturbative and strong non-equilibrium features render the use of fluctuation-dissipation techniques \cite{forster1995} problematic. 
    The difficulty in directly calculating the transport coefficients has been highlighted in several presentations and reviews and is illustrated by a wide spread in theoretical results \cite{Csernai:2006zz,Moore:2020pfu,Astrakhantsev:2017nrs,Astrakhantsev:2018oue,Christiansen:2014ypa,Chakraborty:2016ttq,Heffernan:2020zcf}. 
    Consequently, data-driven techniques -- chiefly Bayesian inference -- have been developed and currently are successful in extracting the  transport coefficients from heavy-ion collision data through systematic model-to-data comparison. 
    
    The efforts to obtain the temperature dependence of the coefficient of shear viscosity over the entropy density, $\eta/s$, and of the bulk viscosity over the entropy density, $\zeta/s$, have relied on multistage models constructed to describe the entire space-time history of the collision process. Prior to this work, the modeling of the different reaction stages has included 
    \begin{itemize}
        \item \trento\, \cite{trento} supplemented by freestreaming \cite{freestream1, freestream2} for the early stage, pre-hydrodynamic era
        \item \music\, \cite{Schenke:2010nt,Schenke:2010rr,Paquet:2015lta}, \vish2+1 \cite{Shen:2014vra}   
        and -- more recently -- \trajectum\, \cite{Nijs:2020roc}, for the relativistic viscous fluid dynamics epoch 
        \item \urqmd\, \cite{Bass:1998ca} and \SMASH\, \cite{smash} for the late time evolution and dynamical freezeout 
    \end{itemize}
    
    Various combinations of those elements have been assembled  as {\it ab initio} simulations to interpret measurements. Traditional modeling and simulation techniques would simply entail simulating the nuclear reaction as faithfully as possible, obtaining a good fit to the final state(s), and extracting physically relevant information from the exercise. 
    In the field of relativistic heavy-ion collisions, this avenue of investigation is not practical at scale for a variety of reasons.  Firstly, the sophistication of the multistage models used to simulate and interpret experiments probing nuclear matter under extreme conditions comes at considerable computational expense. In addition, the many-body environment typical of relativistic nuclear collisions final states generate a wide variety of observables, many of which are correlated with each other and increase the dimensionality of the Bayesian inference.  Those two aspects have led to the development of modern approaches combining principal component analysis (PCA) with surrogate modeling in an effort to minimize possible correlations and accelerate calculations.
    
    Many of these aspects have been used with Bayesian inference (described in the following sections) to determine the temperature dependence of shear and bulk viscosity within some statistically-relevant intervals \cite{Bernhard:2019bmu,SIMSPRL,SIMSPRC,Nijs:2020ors,Nijs:2020roc, Liyanage:2023nds}. This work \cite{Heffernan:2023kpm} shares similar goals but with important differences. It is known that the extraction of QCD transport coefficients from the analysis of heavy-ion collisions is influenced by the physics of very early times \cite{Keegan:2016cpi,NunesdaSilva:2020bfs,Gale:2021emg}. In this context, the calculations made up to now have been made using \trento\ and freestreaming, for the initial very early stages. While this model and scenario are practical and versatile, it is worthwhile to explore analyses based on an approach with some degree of microscopic support. In this vein, this work will use the IP-Glasma model \cite{Schenke:2012wb} which we summarize in a later section, and which has a history of successful phenomenology for heavy-ion collisions \cite{Gale:2012rq}.  Another novel aspect highlighted in this work is the first use of transfer learning in a realistic analysis involving the physics of relativistic heavy-ion collisions. We describe this aspect later in our paper, and also in a companion letter \cite{Heffernan:2023gye}. Finally, this study also comprises a series of technical innovations \cite{Heffernan:2023kpm} ({\it e.g.} the prior distributions, the design space, etc.) which will be presented and discussed in turn, as appropriate.  To keep this survey as simple and transparent as possible, we focus on Pb + Pb collisions at an energy of $\sqrts$ = 2.76 TeV. Even at this degree of resolution, the required numerical work needed to establish the surrogate modeling on a firm statistical basis was considerable. Extending our approach to other systems and energies is left for future work. 
    
    This paper is structured as follows: section \ref{Bayesrev} reviews the basic tenets and usage of Bayesian analysis in the context of this work, and goes over the basics of surrogate modeling and of transfer learning. The following section - Section \ref{models} - then discusses the different components of our physical multistage model. Section \ref{priors} covers the approach to defining priors, together with a description of the physical parameters studied in this work. Our approach to the design phase is also outlined. Section \ref{closure} discusses the important milestone of closure tests and self-consistency of the surrogate modeling: this step is crucial in order to ensure that the model is self-consistent. We then address comparing the model with data: postdictions and predictions. The statistics approach to selecting a particular physical model over competitors is discussed in Section \ref{BMS}. The paper ends with a summary and conclusion. 

\section{Bayesian Inference and Surrogate Modeling}

\label{Bayesrev}
    \subsection{Bayes Theorem}
        
        We begin by defining the statistical notation used throughout this work.
        $p(A)$ denotes the probability density $p(\cdot)$ of a proposition $A$. $p(A|B)$ denotes the probability density of proposition $A$ conditional on proposition $B$, \emph{i.e.} the probability density of $A$ \emph{given} $B$. There may be multiple statements to which the proposition of interest is conditional; these are all contained to the right of the vertical bar, \emph{e.g.} $p(A|B,C,D,\dots)$. 
    
        With this established, we can begin to interpret Bayes' Theorem \cite{bayes1763lii}, a fundamental statement of probability theory:
        \begin{equation}
            p(H|d,I) = \frac{p(d|H,I)p(H,I)}{p(d,I)}. \label{eq:bayes}
        \end{equation}
        $H$ represents a particular \emph{hypothesis}, such as the proposed values of a set of parameters and $d$ represents data to which the hypothesis is compared.
        The Bayes evidence 
        quantifies a balance between quality of fit via the likelihood and predictive power by penalizing increasing dimensionality. It can be used in model selection as the best model is the one that fits the data best with the fewest number of free parameters. 
        Finally, the quantity of interest in Bayesian inference is $p(H|d,I)$, the \emph{posterior}. It quantifies the belief in a given hypothesis $H$ \emph{posterior} to comparison with measured data $d$.
        
        Bayes' theorem formalizes statistical learning by making a prior belief explicit and then comparing it to data, after which the prior belief is determined to be relatively more or less likely. The result \emph{posterior} to comparison with data is the new state of understanding. 
    
    \subsection{Surrogate modeling with a statistical emulator}
        
        Surrogate modeling is a strategy for computation with expensive likelihood functions. Likelihood functions are expensive because they require detailed model evaluation. A cheaper model (the ``surrogate'') is trained to emulate the expensive model using calculations from the more expensive model. This less computationally expensive surrogate can be considered a low-fidelity model, or a model-of-a-model, and compromises a limited degree of accuracy for great reduction of computational expenditure. It does this by mapping inputs to outputs and learning the functional relationship between them rather than attempting to produce a coarse version of the intermediary physics. These methods have had success in heavy ion physics \cite{Bernhard:2015hxa, Bernhard:2016tnd, Bernhard:2017vql, Bernhard:2019bmu, BernhardPhD, Everett:2021ruv, Everett:2021ulz, SIMSPRL, SIMSPRC, Nijs:2020ors, Nijs:2020roc}.
        
        Given a set of training points, there are infinitely many functions that can describe the points. Gaussian processes (GPs) -- the surrogate models used in this study -- assign a probability to each of these functions, meaning that the output is a probability distribution of the characterization of the data. Conveniently, this also allows one to determine the relative confidence in the prediction. 
        The only assumptions by the GP are that it assumes the function is continuous and smoothly varying with respect to the length scales of the observations. 
        
    
    \subsection{Transfer learning}
        \label{sec:transfer-learning}

        A surrogate modeling technique only recently considered in the context of heavy ion collisions is transfer learning \cite{5288526, Liyanage:2022byj}. This learns about a ``task'' of interest (the target task) by using information from related tasks (source tasks). In heavy ion collisions, inductive transfer learning -- where the source and target have the same input domain -- can be readily deployed. This allows for transfer learning between models of viscous corrections at particlization that do not introduce additional parametric flexibility~\cite{Liyanage:2022byj}. Efficient transfer was also found to be possible for collisions of slightly different nuclei at different collision energies~\cite{Liyanage:2022byj}.
        
        Transfer learning is performed by first having a trained surrogate model for a source task. Then, the discrepancy between the source and the target is found and is encapsulated in a discrepancy function. The advantage of transfer learning is to use comparably little new training information about the target to learn the discrepancy function.
        
        More formally, if $f_S(x)$ is the source and $f_T(x)$ the target, one can propose a simple relationship
        \begin{equation}
            f_T(x) = \rho f_S(x) + \delta (x)
        \end{equation} 
        where $\rho$ is a linear correlation between the source and target estimated via the maximum likelihood and $\delta(x)$ is the discrepancy between the source and target models. $f_S(x)$ and $\delta(x)$ are considered to be independent Gaussian processes. This is derived from multifidelity emulation, where the source is a computationally-inexpensive low-fidelity model and the target is a computationally-expensive high-fidelity model \cite{kennedy2000predicting}.
        
        This vastly reduces the computational cost of training new models that are similar to an already-trained source. In the case of (linear) viscous corrections, particularly those between Grad's 14-moment viscous corrections \cite{https://doi.org/10.1002/cpa.3160020403,Denicol:2010xn}  and the Chapman-Enskog viscous corrections \cite{chapman1990mathematical,SIMSPRC}, an order of magnitude fewer training points are required to reach a specified accuracy when using transfer learning as opposed to training a new surrogate model \cite{Liyanage:2022byj}. This work is the first time transfer learning methods will be used for Bayesian inference in heavy ion collisions. We use transfer learning to implement a second viscous correction model with Grad's 14-moment viscous corrections as the source $f_S$ and Chapman-Enskog relaxation time approximation (RTA) viscous corrections as the target $f_T$.

\section{Physical Models}
\label{models}
    As all of the individual elements of our hybrid modeling exist in the literature and have been used extensively in a variety of applications, only a brief summary is provided here and the reader will be referred to the appropriate references. 
    
    \subsection{Pre-equilibrium -- IP-Glasma}
        The very first instants of the heavy-ion collisions considered in this work are modeled by IP-Glasma, an approach which is derived from the Color-Glass Condensate (CGC) \cite{[{See, for example, }][{, and references therein.}]Gelis:2010nm}. More specifically, the CGC action can be written as \cite{Krasnitz:1998ns} 
        \begin{eqnarray}
        S_{CGC} = \int d^4 x \left(- \frac{1}{4} F^a_{\mu \nu} F^{a\, \mu \nu} +J^{a\,\mu}A^a_\mu \right)
        \end{eqnarray}
        where $F^{a\, \mu \nu}$ is the non-Abelian field strength tensor with color index $a$, and $J^{a\, \mu}$ is the current representing the hard partons that source soft gluons. The CGC can be viewed as an effective field theory representation of QCD. Its implementation here follows the IP-Glasma model of initial conditions \cite{Schenke:2012wb,Schenke:2012hg,Gale:2013da}, where the IP-Sat approach \cite{Bartels:2002cj,Kowalski:2003hm} is used to determine the fluctuating initial color configuration in the two highly energetic approaching nuclei. These color charges then act as sources for the small $x$ soft gluon fields, which have a large occupation number and therefore can be treated classically.         
        Their evolution obeys the Yang-Mills equation:
        \begin{eqnarray}
        \left[D_\nu, F^{\mu \nu}\right]^a = J^{a\, \mu}
        \end{eqnarray}
        with $D^a_\mu = \partial_\mu - i g A_\mu t^a$,  and $t^a$ the color SU(3) matrices. The color current $J^{a\,\mu} =  \delta^{\mu \pm} \rho^a_{A (B))} (x^\mp, {\mathbf x_\bot})$ is generated by nucleus A (B) moving along the light-cone direction $x^+ (x^-)$, and $\rho^a$ represents the color charge distribution extracted from IP-Sat. The Glasma distributions resulting from solving the Classical Yang-Mills equations event-by-event then serve as an input to fluid dynamics, at proper time $\tau_0$.  
        For the purpose of this work, an important parameter of IP-Glasma is $\mu_{Q_s}$, the constant of proportionality relating the color charge per unit transverse area $g^2 \mu (x, {\mathbf b}_\bot)$ to the \cite{Lappi:2007ku} squared saturation scale $Q_s^2$. This is one of the parameters considered in this study. 


         To connect IP-Glasma to hydrodynamics, we follow the same procedure as Ref.~\cite{McDonald:2020xrz} and previous literature. The energy density $\varepsilon$ and flow velocity $u^\mu$ are defined by Landau matching with the Classical Yang-Mills energy-momentum tensor 
        \begin{equation}
            T^{\mu}_\nu u^\nu = \varepsilon u^\mu
        \end{equation}
        and finding the timelike eigenvector for $u^\mu$. The shear stress energy-momentum tensor is the traceless, transverse part of the Classical Yang-Mills energy-momentum tensor. The bulk pressure is defined as $\Pi\equiv \varepsilon/3-P(\varepsilon)$ where $P(\varepsilon)$ is the pressure from the QCD equation of state.
        
    \subsection{Viscous hydrodynamics -- MUSIC}

    \label{sec:viscous_hydro}

        Hydrodynamics is an effective theory of long-wavelength modes. Practically, this means that the evolution of a collection of differential elements can be described not by tracking microscopic particles, but considering their long-wavelength (or spatially-coarse) collective motion. Analogously, to model a hurricane or storm front, it is not necessary (or even relevant) to model the behavior of every constituent water droplet or molecule of air. Instead, the collective dynamics at a much larger scale reveal the physics of interest.
    	
        Most of the modern approaches to relativistic fluid dynamics perform a gradient expansion of hydrodynamic equations up to second order, following original work by M\"{u}ller, Israel, and Stewart \cite{Muller1967,ISRAEL1979341} (MIS), who added transport coefficients in the form of shear and bulk relaxation times that 
        characterize the timescale on which shear stress tensor and bulk pressure approach their first-order solutions, ensuring causal solutions under a range of conditions \footnote{Recent works have further investigated the constraints imposed by causality on relativistic fluid dynamics \cite{Plumberg:2021bme, Noronha:2021syv}.}. 
  
        In this study, second-order transient relativistic hydrodynamics is used to describe the plasma, specifically the DNMR formulation \cite{Denicol:2012cn} with both shear and bulk viscosity as implemented numerically in MUSIC (\emph{MUSCl for Ion Collisions}) \cite{Schenke:2010nt,Schenke:2010rr,Paquet:2015lta}. The equation for the conservation of energy and momentum is coupled with relaxation equations for the shear tensor and the bulk pressure, with parametrized shear and bulk viscosities (discussed below) and second-order transport coefficients related to the first-order ones \cite{Denicol:2014vaa}. 

        The transport coefficients of interest here, $\eta/s$ (the shear viscosity over the entropy density) and $\zeta/s$ (the bulk viscosity over the entropy density), characterize the first-order deviation from ideal fluid dynamics. The Equation of State (EoS) is where the specific material properties of QCD matter inform the hydrodynamic stage and as a result, must be constructed to be consistent with the model choices. The EoS at high temperature is matched to lattice calculations \cite{PhysRevD.90.094503}. At low temperature, the EoS matches that of the particle list used in the hadronic transport (to be discussed in a later section), which ensures that the EoS is continuous across the transition between the two stages. The matching between the high and low temperature results must be done in a manner consistent with what is currently known about QCD: it must have a smooth crossover between degrees of freedom rather than a sharp phase transition at vanishing baryochemical potential.\footnote{Discussions of non-zero baryochemical potential are beyond the scope of this work, but are a vibrant field which features a search for a possible QCD critical point \cite{Noronha-Hostler:2019ayj, Noronha-Hostler:2019nve, Du:2021fyr, Busza:2018rrf, Luo:2017faz, Karthein:2021nxe, Dore:2020jye}.} Attempts to constrain the equation of state directly from hadronic observables have shown promise, but as of yet still have significant remaining uncertainty \cite{Pratt:2015zsa, Sorensen:2021zxd}. Active learning techniques are also being applied to the efforts to characterize the equation of state \cite{Mroczek:2022oga}. 
     
        The equation of state used in this work smoothly connects the HotQCD calculation \cite{PhysRevD.90.094503} at high temperatures to a list of stable resonances at low temperatures, and matches that of Ref. \cite{SIMSPRC} and the code that produced it is publicly available with the default parameters \cite{eosmaker}.         
    
    \subsection{Particlization -- iS3D}
        To particlize the hydrodynamic medium, one defines a surface at constant temperature, energy density, or entropy - these choices are equivalent in the case of zero baryochemical potential, which this study strictly respects. This temperature is the switching, or particlization, temperature. 
        Once this surface has been drawn, particles can be sampled stochastically, respecting energy and momentum conservation \emph{on ensemble average}. This means that the sampled distribution converges to the true distribution of particles, momenta, etc. it is useful to \emph{over}sample this surface. This is either done a fixed number of times (typically 100 to 300 times), or until a sufficient number of particles has been sampled. The way the sampling is performed is via the Cooper-Frye prescription \cite{PhysRevD.10.186}, implemented in iS3D \cite{iS3D}. Given an isothermal (or isentropic, etc.) hypersurface $\Sigma$ with normal vector $\sigma_\mu(x)$, the invariant momentum spectra of a particle species $i$ with degeneracy $g_i$ is
    	\begin{equation}
    		E\frac{dN_i}{d^3p} = \frac{g_i}{(2\pi)^3}\int_{\Sigma} f_i(x,p) p_\mu d\sigma^{\mu}(x)
    	\end{equation}
        where $f_i(x,p)$ is the phase-space distribution, and $g_i$ is a degeneracy factor. This distribution function reproduces the energy-momentum tensor of hydrodynamics at the particlization surface,
        \begin{equation}
    		T^{\mu \nu}(x) = \sum_i \frac{g_i}{(2\pi)^3} \int \frac{p^\mu p^\nu f_i(x,p)}{E} d^3p.
        \end{equation}
        Here, $f_i(x,p)$ is species-specific, representing either Bose-Einstein or Fermi-Dirac statistics.
    	
        The out-of-equilibrium nature of the system generates interesting physics, but presents significant challenges. If at the time of particlization the hydrodynamic medium were in equilibrium, the choice of the distribution function would simply be the equilibrium form, and the rest frame velocity and temperature would be fixed by the hydrodynamic velocity and the energy density in the local rest frame. However, the medium is generally not in equilibrium and consistency between the kinetic description of particles and viscous hydrodynamics must be attempted. The existence of shear and bulk stress contributions  -- $\pi^{\mu \nu}$ and  $\Pi$ -- produce deviations of the microscopic distributions and yields from the equilibrium ones. As mentioned earlier, this study will exclusively consider 
        Grad's 14-moment approximation and the linear Chapman-Enskog expansion in the relaxation time approximation.

        The distribution function for a fluid out of local equilibrium may be separated as
	\begin{equation}
		f_i(x,p) = f_{eq,i}(x, p) + \delta f_i(x,p)
	\end{equation}
	where $f_{eq,i}(x,p)$ is the equilibrium distribution function (Bose-Einstein or Fermi-Dirac for different particle species) and $\delta f_i(x,p)$ is the non-equilibrium correction.
    
    Unfortunately, the separation of the distribution function into the equilibrium contribution and a viscous correction, despite the constraints from matching to hydrodynamics, does not fully specify the momentum-dependence of $\delta f_i(x,p)$. This means that the choice of correction remains a modeling choice with inherent ambiguity that can impact calculations of hadronic observables \cite{Dusling:2009df}. To constrain further, the reasonable assumption is made that hydrodynamics and relativistic kinetic theory are simultaneously applicable at the transition between them. 
	
	Linearized viscous corrections linearize the correction $\delta f_i$ in the shear stress tensor, bulk viscous pressure, and baryon diffusion current. In this study, baryon diffusion is not considered. In linearized viscous corrections, the expansion coefficients are adjusted to exactly reproduce $T^{\mu\nu}$.
	
	Grad's 14-moment approximation expands the correction $\delta f_i (x,p)$ in momentum moments of the distribution function \cite{https://doi.org/10.1002/cpa.3160020403}, only truncating at the level with terms involving $p^\mu$ and $p^\mu p^\nu$, \emph{i.e.} at hydrodynamic order.	
	
	The Chapman-Enskog expansion is a gradient expansion around $f_{eq,i}$. The relaxation time approximation (RTA) is used for the collision term of the Boltzmann equation. Expanding $f_i$ into its equilibrium component and correction and assuming hydrodynamic gradients are small in comparison to the relaxation time, a first order gradient correction for the thermal distribution may be derived \cite{PhysRevC.90.044908},
    
    \subsection{Hadron Cascade -- SMASH}
        Once hadrons have been sampled from a hydrodynamic hypersurface, they can be evolved using kinetic theory via the SMASH transport code \cite{smash}. The particles interact with each other, scattering, decaying, and forming resonances. These are computed in SMASH using measured particle properties and channels \cite{ppreview} via a tower of coupled Boltzmann equations.
        \begin{equation}
            p^\mu \partial_\mu f_i(x,p) = C[f_i]
        \end{equation}
        where $i$ is an index over species. Once again, $f_i(x,p)$ is species-specific, representing Bose-Einstein statistics for particles with integer spin and Fermi-Dirac statistics for particles with half-integer spin. This list of species is given in the SMASH documentation.\footnote{Available at \url{smash-transport.github.io}} This work uses SMASH Version 1.8. 

\section{Prior Specification and Experimental Design}
\label{priors}

    \subsection{Priors}
    
        The prior distribution, or state of knowledge, for each parameter must be justified for each study. However, the general form of the prior deserves some attention. 
        Most Bayesian inference studies in heavy ion collisions to date have used uniform priors. 
        Existing guidance from Bayesian practitioners in the statistics community suggests using uniform priors with sharp cutoffs only if that is an accurate reflection of the underlying constraint and not as a general non-informative choice. Additionally, priors may be chosen with features such as boundary-avoidance or invariance under reparametrization \cite{stanprior}. Another important consideration is to interrogate what ``weakly'' or ``non-informative'' means in the absence of explicit reference to the likelihood. If the dominant constraint comes from the prior, then the prior is informative. Conversely, if the likelihood is the dominant source of constraint, then the prior is less informative.
        
        In order to bias the priors as little as possible and to smoothly move beyond the uniform distribution, this study uses the symmetric Generalized Normal distribution with varying mean $\mu$, location $\alpha$, and shape parameter $\beta$. The shape parameter $\beta$ controls the tails of the distribution. When $\beta=\infty$, the distribution becomes the uniform distribution. When $\beta=2$, the distribution is Gaussian, while when $\beta=1$, the distribution is Laplacian. This provides a flat plateau with power law tails, smoothly interpolating between the current practice (effectively $\beta=\infty$) and priors more reflective of the underlying physics. This distribution has support on the whole real line and can be shifted. Additionally, the Half Generalized Normal distribution exists for instances where a sharp cutoff is reasonable, e.g. positive specific viscosity for non-decrease of entropy. Quantities such as the probability density function (PDF) and cumulative distribution function (CDF) are well defined, as is the entropy of the distribution.
        
        The probability density function of the Generalized Normal distribution is
        \begin{equation}
            p(x,\mu, \alpha,\beta) = \frac{\beta}{2 \alpha \Gamma (1/\beta)} e^{(-|x-\mu|/\alpha)^\beta}
        \end{equation}
        where $\Gamma$ is the Gamma function.
    
    \subsection{Parameters}
    \label{sec:parameters}
    
        In this section, the physical meaning of the free parameters investigated is described. The specific choices for individual priors are explicitly specified but the general form of the prior remains the same: a Generalized Normal distribution with a specified shape parameter $\beta$ and a central 99\% interval. Rather than specify values of the location and scale, the central 99\% interval is chosen and the parameters that produce this interval are found through numerical optimization. This is more interpretable as it specifies a 99\% degree of belief that the parameters are within a certain range and is directly comparable to the 100\% central interval used to characterize the uniform distribution.
        
        In this study, only parameters in IP-Glasma and MUSIC are varied.  The choice of viscous correction is in effect a parameter in the Copper-Frye particlization sampling (as implemented in iS3D), but is fixed for each calculation. 
    
        The parameters in IP-Glasma are mostly fixed via the IP-SAT model's comparison to deep inelastic scattering experiments. Two parameters in IP-Glasma can be considered  poorly constrained and are thus included in the Bayesian study: (i) the proportionality between the saturation scale and color charge densities, and (ii) the onset of hydrodynamics. In this study the strong coupling has been fixed to $g = 2$, a value compatible with the bulk of heavy-ion phenomenology at the energies of the LHC \cite{Schenke:2020mbo}.
        
        Each parameter in the initial stage model is now described in more detail and is given a shorthand notation.
        
        \begin{enumerate}
            \item $\mu_{Q_s}$: Multiplier from the saturation scale to the color charge density profile ($Q_s = \mu_{Q_s} g^2 \mu$). In the CGC, these quantities are proportional but an {\it a priori} constraint on this proportionality is not known from theory. 
            \item $\tau_0$: Proper time of the transition between IP-Glasma and hydrodynamics. 
            In IP-Glasma, the Glasma phase stabilizes within approximately 0.2 fm while flow continues to build as shown in Fig.~\ref{fig:2dipg-pressures} reproduced from \cite{McDonald:2017MSc}. The onset time of hydrodynamics is not known with certainty, but estimates have been guided by the fact that, parametrically, gluon saturation should be attained for momentum scale smaller than $Q_s$ \cite{Mueller:1999fp}, which corresponds to a time scale $\approx 1/Q_s$. Practically, IP-Glasma initial states with a proper time span 0.2~$<~\tau~<$~0.4~fm have been used \cite{Gale:2012rq,McDonald:2016vlt,Schenke:2020mbo}. Recent studies with freestreaming \cite{BernhardPhD, SIMSPRC, Nijs:2020roc} extract a longer time to the onset of hydrodynamics than those typically used with IP-Glasma. This work will allow for switching times up to $\approx 1.2$ fm, informed by longer hydrodynamic onset time and the approach to hydrodynamics \cite{Schlichting:2019abc}, to determine if these long onset times are favored when using a pre-equilibrium model with microscopic dynamics. 
            \begin{figure*}
                \centering
                \includegraphics[width=0.75\textwidth]{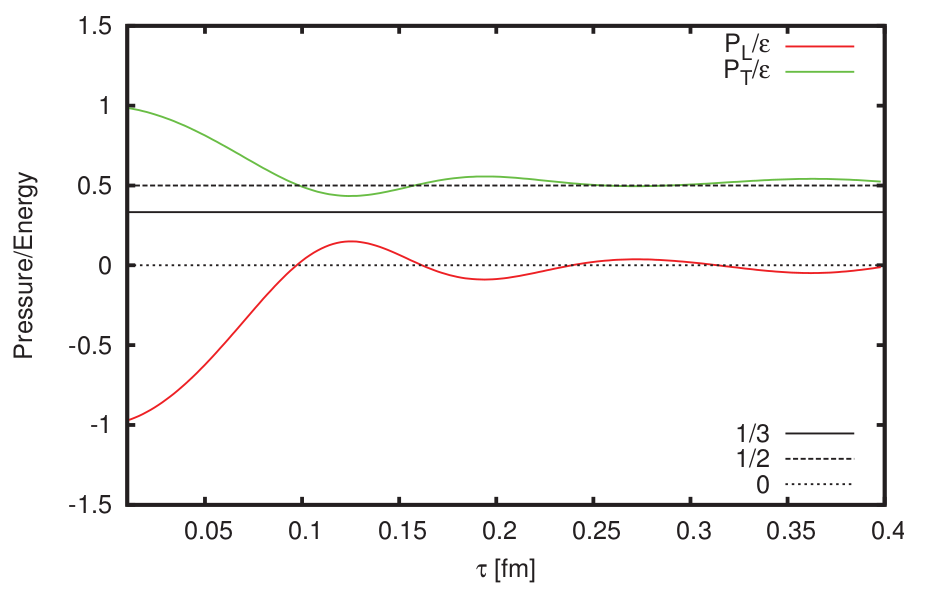}
                \caption{Longitudinal and transverse pressure, scaled by the energy density, as a function of proper time in (2+1)D IP-Glasma. Adapted from \cite{McDonald:2017MSc}. 
                }
                \label{fig:2dipg-pressures}
            \end{figure*}
        \end{enumerate}

        In the relativistic hydrodynamic phase, the temperature dependence of both shear and bulk viscosity is varied.
        Because of the parametric flexibility in the viscosity, these parameters dominate the analysis. One more parameter is the particlization temperature between hydrodynamics and hadronic transport.
        
        It has been proposed to avoid using a parametrization, which correlates values of the viscosity at different temperatures \cite{Everett:2021ruv}, an idea that has recently been explored in other contexts in heavy ion collisions~\cite{Xie:2022ght}. This is beyond the scope of this work.
        This work uses the viscosities as parametrized in \cite{SIMSPRL, SIMSPRC} but widens the prior ranges.
        
        The specific shear viscosity is parametrized as
        \begin{eqnarray}
        	\eta/s(T) &=& (\eta/s)_{kink} + a_{\eta,low} (T-T_{\eta,kink})\Theta(T_{\eta,kink}-T) \nonumber \\
        	&&+ a_{\eta,high} (T-T_{\eta,kink})\Theta(T - T_{\eta,kink}) \label{eq:shearparam}
        \end{eqnarray} 
        where the function has four parameters: $(\eta/s)_{kink}$, $a_{\eta,low}$, $a_{\eta,high}$ and $T_{\eta,kink}$. These control the value of $\eta/s$ at some kink, the slope below and above the kink, and the temperature of the kink. In practice, this can be less than 0, so the value used is ${\tt MAX}(0,\eta/s)$. Generally, strong coupling implies low viscosities, and the strongest coupling should be in the deconfinement region (below, quarks are confined, and above asymptotic freedom reduces the strength of the interaction). This minimum in shear viscosity has been observed for a large number of systems. \cite{Bernhard:2019bmu,Csernai:2006zz,Adams:2012th}  For QCD, direct calculations and experimental extractions currently produce a variety of results \cite{Gonzalez:2020bqm,Rose_2020} with variable temperature dependence.
        
        The specific bulk viscosity is parametrized as the probability density function of a skewed Cauchy distribution,
        \begin{align}
        	\zeta/s(T) &=\frac{(\zeta / s)_{\max } \Lambda^{2}}{\Lambda^{2}+\left(T-T_{\zeta,c}\right)^{2}} \label{eq:bulkparam} \\
        	\Lambda &=w_{\zeta}\left[1+\lambda_{\zeta} \operatorname{sign}\left(T-T_{\zeta,c}\right)\right] \nonumber
        \end{align}
        where the function again has four parameters: the maximum of the bulk viscosity $(\zeta/s)_{max}$, the temperature at which the bulk viscosity is maximum $T_{\zeta,c}$, the width $w_\zeta$ of the bulk viscosity peak  and the skewness $\lambda_{\zeta}$. This parametrization, as highlighted in \cite{SIMSPRC}, is based on the expectation that the specific bulk viscosity for QCD matter reaches a peak near the deconfinement transition; this is related to the trace anomaly of QCD or a corresponding dip in the speed of sound in-medium~\cite{Heffernan:2020zcf, Kharzeev_2008, Rose_2020, Karsch:2007jc, PhysRevLett.103.172302}. At high temperature, QCD becomes increasingly conformal and the specific bulk viscosity is expected to smoothly approach zero~\cite{PhysRevD.74.085021}.  
        
        \begin{figure*}[!htbp]
            \centering
            \includegraphics[width=0.75\textwidth]{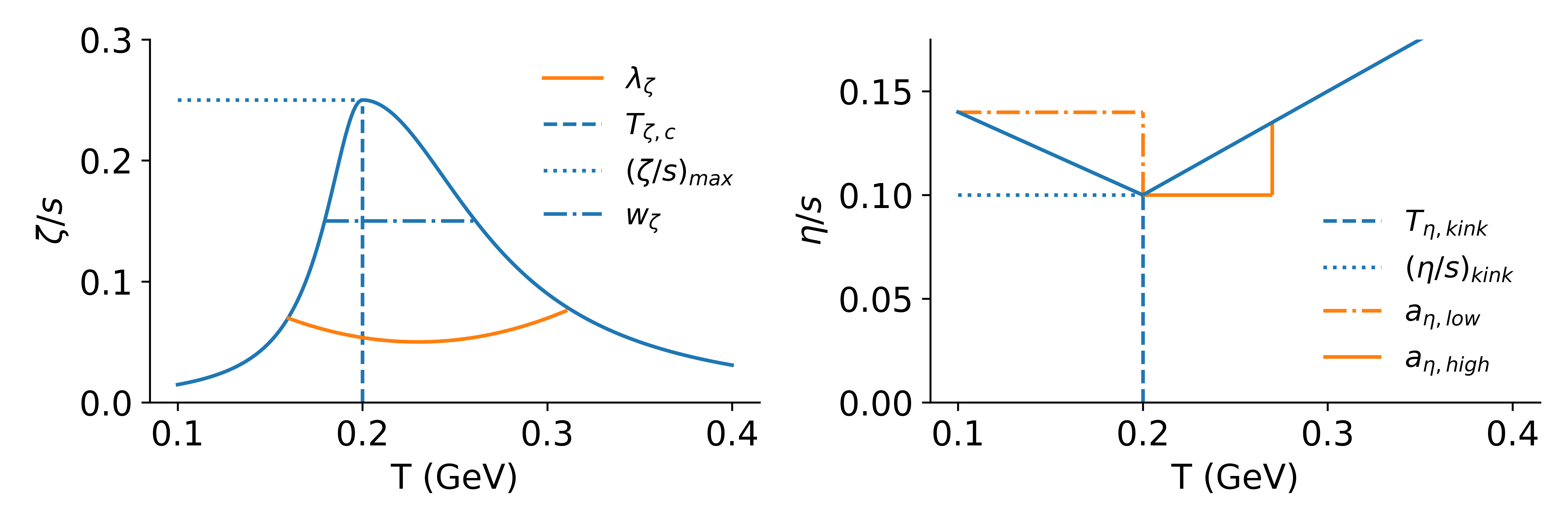}
            \caption{The parametrization of the viscosities.}
            \label{fig:viscosities}
        \end{figure*}
        
        The full list of the parameters varied in the relativistic hydrodynamic phase is as follows:
        \begin{enumerate}
            \item $(\eta/s)_{kink}$: The value of $\eta/s$ at the kink temperature.
            \item $T_{\eta,kink}$: The temperature at which $\eta/s$ changes slope.
            \item $a_{\eta,low}$: The slope of the $\eta/s$ below the kink temperature. This is broadly expected to be negative or $0$, but has not yet been constrained conclusively by model-to-data comparison.
            \item $a_{\eta,high}$: The slope of $\eta/s$ above the kink temperature. This is anticipated to be positive definite, but has not yet been constrained conclusively by model-to-data comparison. A theoretical exception to this expectation can be found in the NJL model for SU(3) \cite{PhysRevC.88.045204}.
            \item $(\zeta/s)_{max}$: The maximum of $\zeta/s$.
            \item $T_{\zeta,c}$: The temperature of the maximum of $\zeta/s$.
            \item $w_\zeta$: The width of the peak in $\zeta/s$.
            \item $\lambda_{\zeta}$: The asymmetry of the peak in $\zeta/s$. %
            \item $T_{sw}$, also known as the particlization or switching temperature. 
            A surface at constant temperature is drawn (assuming no baryochemical potential) from which hadrons are sampled with the Cooper-Frye formula, implemented in the iS3D code. The individual hadrons are then described with hadronic transport (SMASH).
        \end{enumerate}

    As stated earlier, the parameters related to IP-Glasma are $\mu_{Q_s}$ and $\tau_0$, the latter defining the boundary with the hydrodynamics phase. With other parameters held fixed, $\mu_{Q_s}$ was varied and the final multiplicity dependence was observed and used to determine a broad range for this parameter. Allowing for an approximate factor of two in the prior yields a 99\% prior range, as seen in Table \ref{tab:fullpriors}. The prior for $\tau_0$ is extended to times considered late by most applications, to ensure those values are properly explored.
    The remaining parameters whose priors must be motivated are those of the hydrodynamic stage: the 8 parameters of the specific shear and bulk viscosity as well as the particlization temperature. For the parameters of the specific shear viscosity, the parametrizations presented earlier and shown in Fig. \ref{fig:viscosities} are used. The priors for the parametrization of the specific shear viscosity were widened, when compared to those used in a previous study \cite{SIMSPRC}, and both signs of the slope below and above the kink were explored. The shape of the specific bulk viscosity allowed for a peaked distribution, with both variable width, asymmetry and normalization. The form of the priors for each parameter is again the symmetric Generalized Normal Distribution or the half symmetric Generalized Normal Distribution if the quantity is commensurate with a sharp cutoff (e.g. is required to be positive definite). The full set of parameter priors, with the central 99\% range and the Generalized Normal distribution shape parameter $\beta$ is collected in Table \ref{tab:fullpriors}. 
        
        \begin{table*}
            \begin{center}
                \begin{tabular}{lcccc}
                    Parameter & 0.5$^{th}$ percentile & 99.5$^{th}$ percentile &  $\beta$ & Distribution\\
                    \hline
                    $\mu_{Q_s}$ & 0.55 & 0.90 & 10 & Generalized Normal \\
                    \hline
                    $\tau_0$ [fm] & 0.20  & 1.20 & 20 & Generalized Normal\\
                    \hline
                    $T_{\eta,kink}$ [GeV] & 0.120 & 0.320 & 20 & Generalized Normal\\
                    \hline
                    $a_{\eta, low}$ [GeV$^{-1}$] & -2.10  & 1.20 & 20 & Generalized Normal\\
                    \hline
                    $a_{\eta, high}$ [GeV$^{-1}$] & -1.20  & 2.10 & 20 & Generalized Normal\\
                    \hline
                    $(\eta/s)_{kink}$ & 0.00 & 0.30 & 10 & Half Generalized Normal\\
                    \hline
                    $(\zeta/s)_{max}$ & 0.00 & 0.30 & 10 & Half Generalized Normal\\
                    \hline
                    $T_{\zeta,c}$ [GeV] & 0.100 & 0.350  & 10 & Generalized Normal\\
                    \hline
                    $w_\zeta$ [GeV] & 0.02 & 0.18 & 30 & Generalized Normal\\
                    \hline
                    $\lambda_{\zeta}$ & -1.0 & 1.0 & 20 & Generalized Normal\\
                    \hline
                    $T_{sw}$ [GeV] & 0.135 & 0.180 & 10 & Generalized Normal\\
                    \hline
                \end{tabular}
            \caption{Prior hyperparameters and distributions for each parameter varied.} \label{tab:fullpriors}
            \end{center}
        \end{table*}
    
    \subsection{Maximum Projection Designs}
        
        Previous studies using uniform priors have sampled the allowed parameter space using maximin Latin hypercube sampling (LHS) techniques, which maximize the minimum Euclidean distance between points. Latin hypercubes are designed to provide uniform coverage when projected into 1 dimension while the maximin algorithm helps select points that give a fairly reasonable coverage of the volume.
    
        An issue that may arise in surrogate modeling is that not all parameters are equally impactful; some may even have little impact on the final result. As a result, there is a projection of the full design space that impacts the outputs, called the ``active subspace''. It is not possible to know the active subspace ahead of time, but it is possible to construct a space filling design that maximizes \emph{all} arbitrary projections of the space to lower dimensions. This is the idea behind the Maximum Projection (MaxPro) design strategy \cite{MaxPro}. 
        Specifically, this study will utilize a MaxPro Latin Hypercube Design.
        
\begin{figure}[htb!]
        	\centering
        	\includegraphics[width=0.9\columnwidth]{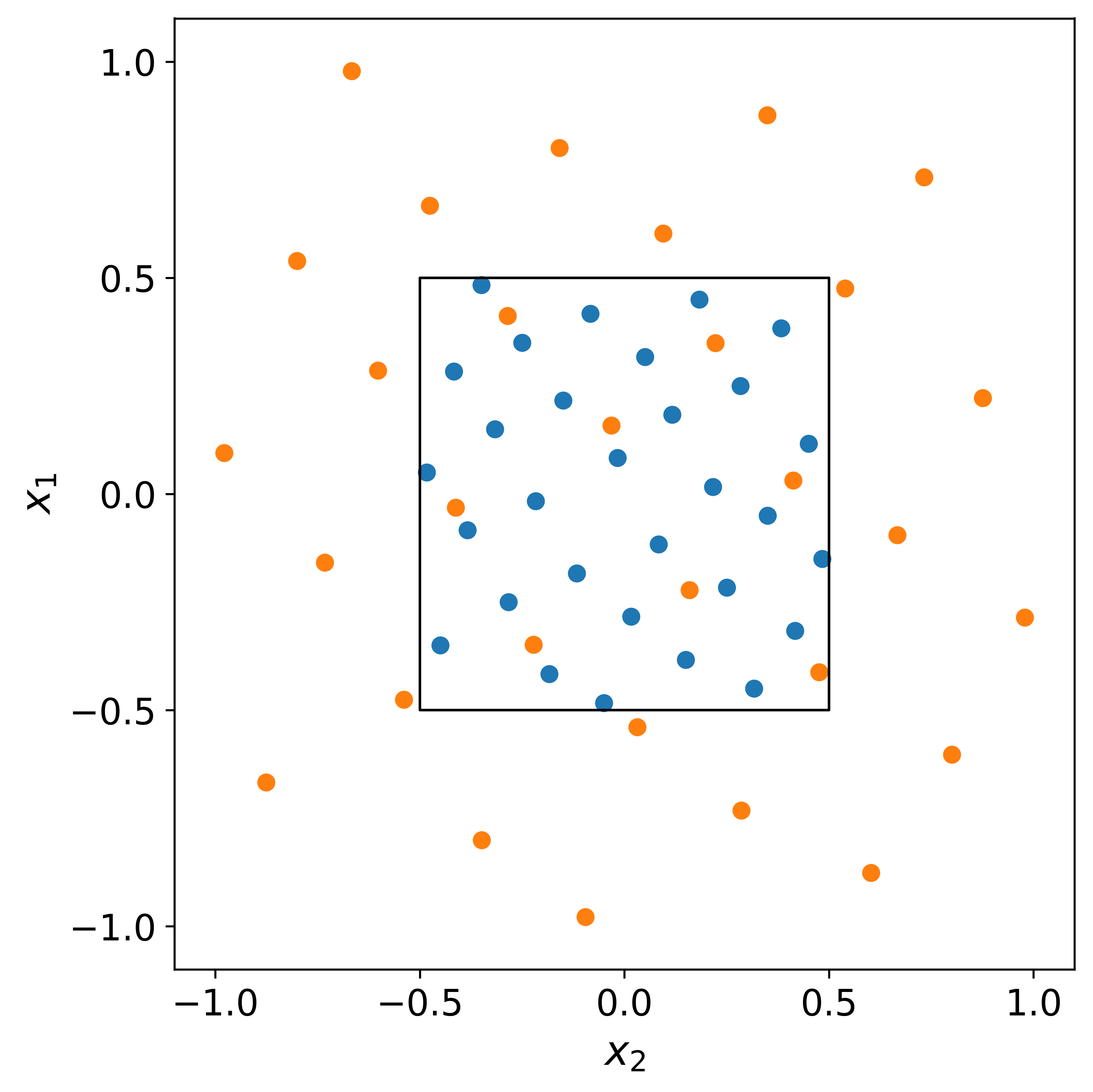}
        	\caption{Deformation of a 2 dimensional Maximum Projection design on the unit hypercube centred at $0$ according to a standard symmetric Generalized Normal distribution with $\beta=10$. The points of the centred unit hypercube are highlighted with a square box and are shown in blue, while points shown in orange have been deformed as described.}
        	\label{fig:sample-deformation}
        \end{figure}

        The sampling must also be made commensurate with the parameter ranges and priors used. This is accomplished by sampling designs on a unit hypercube with the relevant number of dimensions. The priors are chosen and the percent point function (or quantile) can be straightforwardly calculated. The sample location on each dimension of the unit hypercube corresponds to a percentile of the prior range in each dimension. This ensures uniform coverage of the probability volume by weighting by the prior density. 
        This deformation technique is shown for a simple 2D example in Fig.~\ref{fig:sample-deformation} and its success has already been demonstrated \cite{Heffernan:2022swr}. In this study, 350 design points were used for the primary choice of model, which uses Grad's viscous corrections at particlization. An additional 50 design points, which also maximize the MaxPro metric, were generated for model calculations with Chapman-Enskog viscous corrections.

\section{Model Validation}
\label{closure}
    
    It is important to investigate which parameters are both reliably constrained using the underlying hybrid model and are reliably emulated by the Gaussian process surrogate model. This step, revisited at the outset of each study, must be performed to ensure that predictions made by the Gaussian process emulators are sensible and will provide physical -- rather than spurious -- constraint. 
    
    \subsection{Forward model validation} 

    The physical observables we shall consider are divided into two classes that we label ``first generation observables'' and ``next generation observables''. This distinction is somewhat arbitrary but receives some support from chronology. The first generation observables broadly describe large-scale features of the fireball and add four-particle azimuthal Fourier coefficients to the set of observables used in a previous Bayes study \cite{SIMSPRL,SIMSPRC} with the exception of correlated momentum fluctuations. More specifically, the quantities in this class are
    \begin{itemize}
        \item $d N_{\rm ch}/d\eta$: The number of charged hadrons per unit pseudorapidity. Measurements are from the ALICE Collaboration \cite{ALICEPbPb2760}. 
        \item $d N_i/dy$, $i\in$ \{ $\pi$, p, K, etc.\}: Identified charged hadrons per unit rapidity. Measurements are from the ALICE Collaboration \cite{ALICE:2013mez}. 
        \item $d E_{\rm T} /d\eta$: Transverse energy, defined as $E_{\rm T} = \sqrt{m^2 + p_{\rm T}^2}$, per unit rapidity. Measurements are from the ALICE Collaboration \cite{ALICE:2016igk}.
        \item $\langle p_{\rm T} \rangle_i$, $i \in \{\pi$, p, K\}: Mean transverse momenta of identified hadrons. Measurements are from the ALICE Collaboration \cite{ALICE:2013mez}.
        \item $v_n\{2\}$: Two-particle azimuthal Fourier coefficients. Measurements are from the ALICE Collaboration \cite{ALICE:2011ab}.
        \item $v_n \{4\}$: Four-particle azimuthal Fourier coefficients. Measurements are from the ALICE Collaboration \cite{ALICE:2016ccg}.
    \end{itemize}

    \begin{figure*}[tbp]
        \begin{center}
            \includegraphics[width=\textwidth]{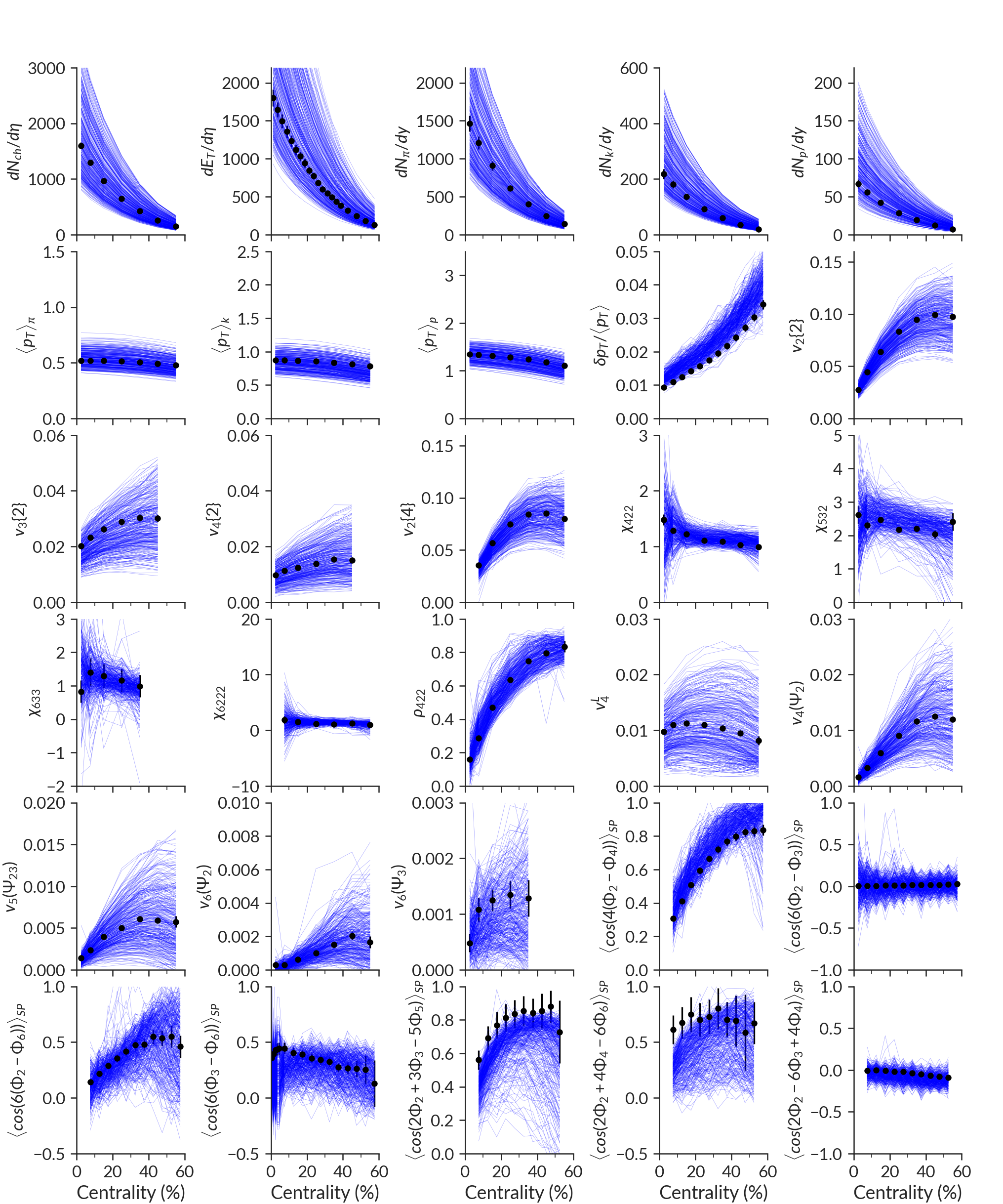}
            \caption{Calculations at each design point forming the prior predictive distribution for each observable. Points are experimental data.}
            \label{fig:full-allobs-prior-predictive}
        \end{center}
    \end{figure*}
    
    \begin{figure*}[tbp]
        \begin{center}
            \includegraphics[width=\textwidth]{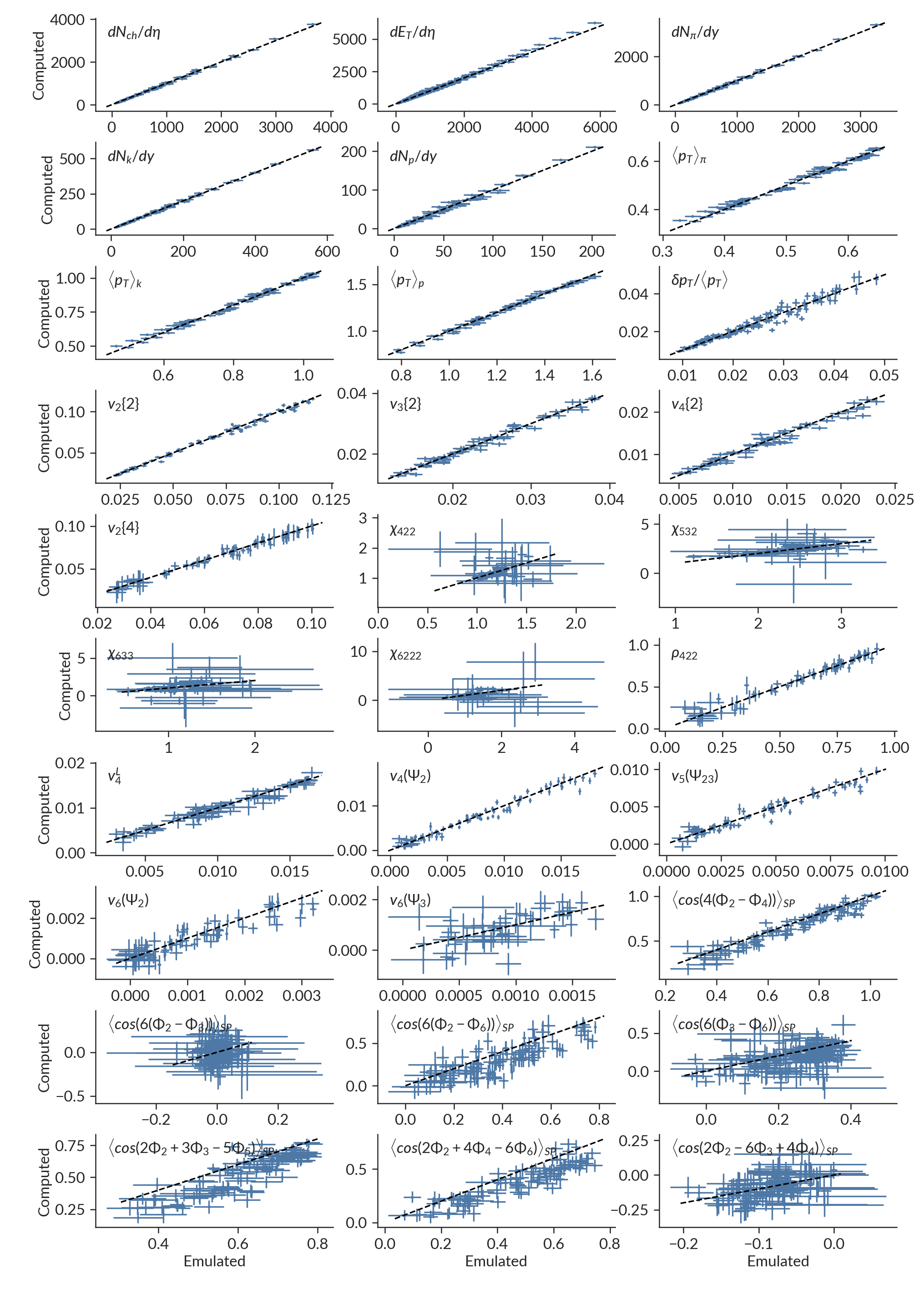}
            \caption{Emulated vs. computed for all observables considered. Successful emulation is clustered around $y=x$.
            }
            \label{fig:full-study-emu-vs-computed-obs-selection}
        \end{center}
    \end{figure*}

    The ``next generation observables'' explore  correlations between geometric features or momentum fluctuations and decompositions of observables into a linear and non-linear response of the medium. Again, more specifically:
    \begin{itemize}
        \item Two- and three-plane Scalar Product Event Plane Correlators: Correlations between expansion coefficients $v_n$ reveal patterns of fluctuations in the initial state and non-linear effects in hydrodynamics. Measurements, as well as detailed definitions, are from the ATLAS Collaboration \cite{ATLAS:2014ndd}. These patterns are coupled and reproduction of them in parametric models has been shown to be highly model-dependent \cite{Qiu:2012uy}. The ALICE Collaboration measures similar quantities, which are also used, statistics allowing \cite{Acharya:2017zfg}. 
        \item  $\chi_{n, m k}$: Nonlinear response coefficients that quantify mixing between higher- and lower-order modes. These decompose higher order $v_n$ into a linear component from the corresponding position space energy density Fourier coefficients ($\epsilon_n$) and a non-linear component from lower modes. For example, $v_5 = v^{\rm L}_5 + \chi_{5, 3 2}\, v_3 v_2 $. Measurements and more details may be found in \cite{Acharya:2017zfg}.
        \item Linear and non-linear flow modes: these quantify the linear and non-linear response of the flow to collision geometry, similar to the event plane correlators and $\chi_{n, m k}$ above \cite{Acharya:2017zfg}.
        \item $\delta p_{\rm T}/ \langle p_{\rm T}\rangle$: Correlated transverse momentum fluctuations. This quantifies the correlations between deviations from the mean transverse momentum. If the deviations are uncorrelated over all events, this quantity is 0 \cite{ALICE:2014gvd}. 
    \end{itemize}
    The purpose of using these carefully chosen observables is to efficiently constrain the properties of strongly-interacting matter. For example, the multiplicities constrain the overall energy of the system, the azimuthal Fourier coefficients constrain the momentum- space geometry of the hydrodynamic stage, and next-generation observables couple various aspects of the medium evolution. 

    The set of observables that are reliably calculated and distinguishable from statistical fluctuations are again all of what we will call ``first generation observables''; the nonlinear response coefficients $\chi_{4,22}$, $\chi_{5,23}$, $\chi_{6,222}$, and $\chi_{6,33}$; the linear and nonlinear flow modes $v_4^L$, $v_4(\Psi_2)$, $v_5(\Psi_{23})$, $v_6(\Psi_2)$, $v_6(\Psi_3)$; and the event plane correlations $\rho_{422}$, $\langle \cos(4(\Phi_2-\Phi_4)) \rangle$, $\langle \cos(6(\Phi_2-\Phi_3)) \rangle$, $\langle \cos(6(\Phi_2-\Phi_6)) \rangle$, $\langle \cos(4(\Phi_3-\Phi_6)) \rangle$, $\langle \cos(2\Phi_2+3\Phi_3-5\Phi_5) \rangle$, $\langle \cos(2\Phi_2+4\Phi_4-6\Phi_6) \rangle$, and $\langle \cos(2\Phi_2-6\Phi_3+4\Phi_4) \rangle$.  The calculation of these observables at each design point is shown in Fig.~\ref{fig:full-allobs-prior-predictive}. Principal component analysis (PCA) is now performed.

    In a space defined by the observables, where each dimension corresponds to a particular observed quantity, it is possible to identify correlations. Principal component analysis is a simple technique to ``rotate'' in observable space into a linear combination of the original axes such that every dimension of the data is linearly independent.
    This rotation is also invertible, meaning that predictions can be made for the transformed space and inverted back to the observable space. This is useful as it is no longer necessary to interpolate between hundreds of dimensions in the observable space, but rather only interpolate in a $\mathcal{O}(10)$ dimensional space, which is much more feasible.
    Another way to think of this rotation is by a decomposition of the data in question to its eigenvalues and eigenvectors. The eigenvalues are the fraction of the total variance in the data described by each eigenvector. 

    A truncation of the eigenvectors is commonly used as a dimensionality reduction technique, separating signal from noise by eliminating principal components that only describe a tiny fraction of the total variation. This is a widely-used practice in the analysis of large data sets, such as is common in modern machine learning applications \cite{reddy2020analysis}. 
    In this work, 20 principal components explain 90.6\% of the variance in the observables across the design points. Surrogate models are trained on the principal components vs. the parameters that are varied in this study and predictions are made in this reduced space before being transformed back to the observable space. This process can be validated by producing model calculations at validation points not used in training the surrogate model emulators and the predictions can be compared to calculations. These emulator predictions at validation points vs. the computed results are shown in Fig.~\ref{fig:full-study-emu-vs-computed-obs-selection}. 
        
    Observables that are only loosely clustered along the $y=x$ lines in Fig.~\ref{fig:full-study-emu-vs-computed-obs-selection} are not kept for the final analysis and observables that are extremely uncertain are also not included. 
    This constitutes ``forward model validation.'' Given a known set of inputs, the predictions are compared to model calculations and observables the surrogate model predicts poorly are inappropriate for inclusion in a physics study. Finally, the $\langle \cos(4(\Phi_2-\Phi_4)) \rangle$ event plane correlator is not included as it quantifies the same correlation as the $\rho_{422}$ correlator and has an overall bias. The finalized set of observables for testing self-consistency and comparison to data is comprised of the first generation observables; the flow modes $v_4^L$, $v_4(\Psi_2)$, $v_5(\Psi_{23})$, $v_6(\Psi_2)$; and the plane correlations $\rho_{422}$, $\langle \cos(2\Phi_2+3\Phi_3-5\Phi_5) \rangle$, and $\langle \cos(2\Phi_2+4\Phi_4-6\Phi_6) \rangle$. While not included in the Bayesian calibration, the excluded observables remain  excellent candidates for predictions with higher-statistics calculations to test the posterior state of knowledge. 
    
\begin{figure*}[!htbp]
        \begin{center}
            \includegraphics[width=0.75\textwidth]{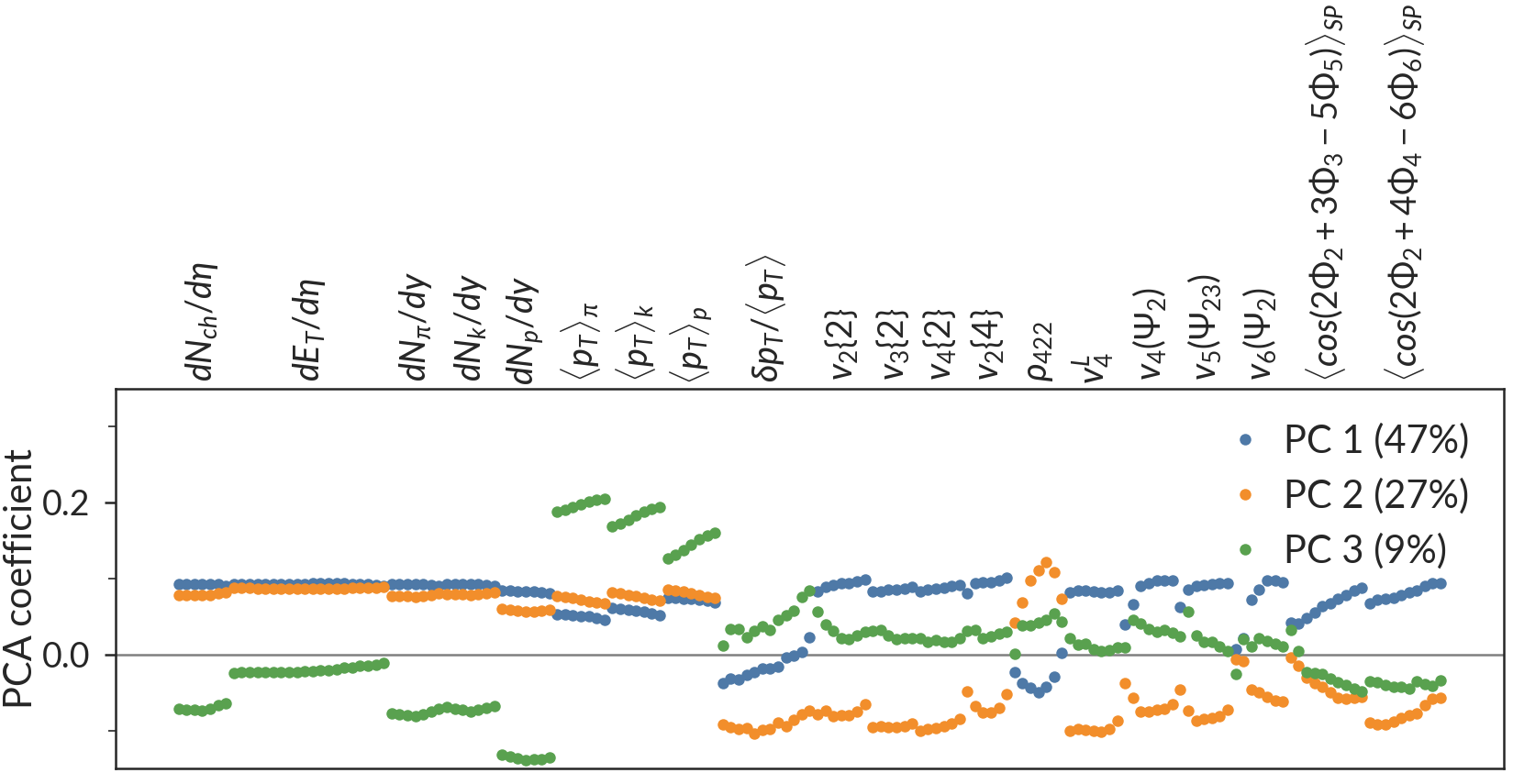}
            \caption{Observable relation to the first three principal components.}
            \label{fig:full-study-pca-vectors-variance}
        \end{center}
    \end{figure*}

    Once these observables have been selected, the principal component analysis and Gaussian process emulation is repeated and are found to be sufficiently reliable for performing self-consistency tests and comparisons to data. Further details of the principal component analysis for the final observable set are shown in Fig.~\ref{fig:full-study-pca-vectors-variance}, where the relationship between the first three principal components (PCs) is shown.  
    The first few principal components contain the majority of the variance of the data and it can be clearly seen that the first three PCs relate clearly to the observables, further supporting the idea that they are successfully reducing the dimensionality of the data with minimal loss of underlying signal. With the final observable set, 30 principal components explain 97.94\% of the variance in the data. The full set of principal components to explain the total variance in the data consists of 161 PCs, meaning that the remaining 131 principal components represent 2.06\% of the variance in the data, which is almost certainly dominated by noise in the underlying calculations. 
    Note that the presence of exclusively linear correlations between observables must be (and has been) investigated for the final set of chosen observables, but is sufficiently large (a 334 x 334 matrix of plots to show pairwise combinations of every observable in every centrality) as to not fit in this work.

    \subsubsection{Transfer learning for Chapman-Enskog $\delta f$}

        Viscous corrections at particlization are an important source of uncontrolled theoretical uncertainty to quantify. An extremely computationally-efficient way to control the uncertainty is using transfer learning. This uses information learned from a source system -- in this study, the already validated Grad viscous correction -- to learn about a similar target system, the Chapman-Enskog RTA $\delta f$. By construction, these are both linearized viscous corrections and are designed to be small corrections to the equilibrium distribution function. This is a prime opportunity to use transfer learning to enable Bayesian inference for the first time in heavy ion collisions. In this study, the Grad design points are used to construct the ``source'' emulator described in Sec.~\ref{sec:transfer-learning}. Then, 50 calculations using Grad viscous corrections are performed at 50 design points chosen according to the Maximum Projection metric, ensuring coverage of the design space as well as its lower-dimensional projections. The difference between the Grad and Chapman-Enskog calculations is used to train a Gaussian Process emulator of the difference between these two models. 
        
        Transfer learning is implemented using \emph{emukit} and \emph{GPy}'s \cite{emukit2019, gpy2014} multifidelity emulation framework and follows the proof-of-concept in \cite{Liyanage:2022byj}.  
        We build on this proof of concept by additionally incorporating principal component analysis and evaluating the covariance matrix necessary for evaluating the likelihood function, thereby enabling the use of transfer learning in full-scale Bayesian inference studies. 
        
        The information contained in the principal component analysis for the Grad viscous corrections (see Fig.~\ref{fig:full-study-pca-vectors-variance}) is exploited so that the transfer learning can take place on the principal components. The Grad PCA, trained on a large number of design points, can be understood to perform a critical covariance-revealing and noise-filtering function. By acting as a rotation in the observable space, the true underlying signal is contained in the first $N$ PCs and noise fluctuations are reduced. This reveals mutual information between observables, \emph{e.g.} that one can be fairly confident of $dN_{ch}/d\eta$ in the 30-40\% bin given its value in the 0-5\% bin. It also means that observables that require higher statistics to calculate reliably, such as $\delta p_T/\langle p_T \rangle$, become correlated with observables that do not, resulting in noise reduction and more successful surrogate modeling. Additionally, by training the transfer learning emulator on the same principal components as the source emulator, the comparison between the two is put on an even footing. 
        
        A second improvement to the transfer learning is using ``transformed parameters'', introduced and used in \cite{designtransformation,SIMSPRL, SIMSPRC}. Although the parametrization of the specific shear and bulk viscosity may appear intuitive and concise, it can present challenges to nonparametric models such as Gaussian processes, since the relationship between the observables and these parameters can be highly non-linear and non-uniform. However, observables are often more straightforwardly-dependent on the value of the specific shear and bulk at a given temperature. For any one set of parameters in Eqs.~\ref{eq:shearparam} and \ref{eq:bulkparam} there exists one and only one set of values of $\eta/s$ and $\zeta/s$ at a set of temperatures, and a one-to-one mapping takes place. Thus, no information is gained or lost by performing this transformation. By using the transformed observables, the transfer learning emulator's mean squared error was reduced by a factor between $2$ and $20$ for every observable considered as well as corresponding improvement in the distance between the coefficient of determination $R^2$ and its maximum value of one. Finally, software changes were made to make it indistinguishable from the original Emulator object and therefore compatible with existing MCMC software and ready for use. 
        
        \begin{figure*}[!htb]
        	\begin{centering}
        		\includegraphics[width=0.9\textwidth]{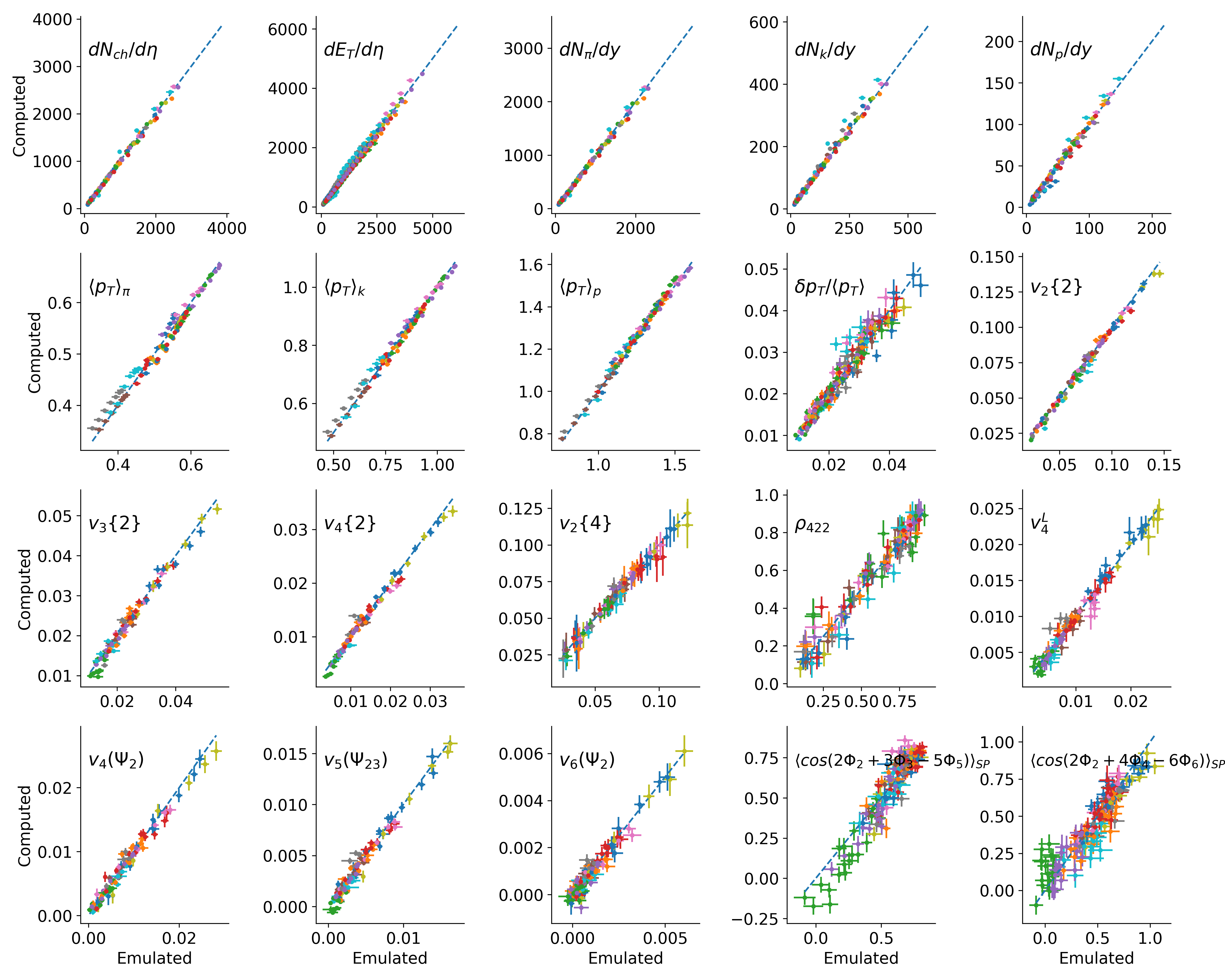}
        		\caption{Transfer learning emulated vs. computed for all observables considered. Validation points are shown with a consistent color to identify correlations between points. The diagonal dashed line is located at $y=x$, and denotes perfect prediction.}
        		\label{fig:full-study-TL-emulated-vs-computed}
        	\end{centering}
        \end{figure*}
        
        The transfer learning emulator validation begins with comparing emulated predictions to computed values at validation points not used in training, shown in Fig.~\ref{fig:full-study-TL-emulated-vs-computed}. In this figure, the colors denote that the points all come from the same validation design point but are merely different centralities of the same observable. This helps to show the correlations between these points. All the observables considered for the study with Grad viscous corrections are well predicted by the transfer learning model, in some cases even better than the source emulator trained on the full design. Uncertainties are often larger in the transfer learning model than in the Grad emulator, but this does not interfere significantly with the quality of predictions and is consistent with having two Gaussian Processes, each with their own variance, rather than just one. Predictions by the transfer learning emulator are broadly consistent with the true values and the emulator uncertainty is well balanced with the computed uncertainty in the most statistics-hungry observables. Were one source of uncertainty systematically larger than the other, this would suggest imbalance between the number of design points and the number of model runs at each design point~\cite{Weiss:2023yoj}, which must be judged by the most statistics-hungry calculations. In this case, there are the correlated momentum fluctuations and event plane correlators: $\delta p_T/\langle p_T \rangle$, $\rho_{422}$, and $\langle \cos(2\Phi_2 +4\Phi_4 + 6 \Phi_6)\rangle_{SP}$. Further worth highlighting is what appears to be a slight emulator bias in the three-plane correlators in Fig.~\ref{fig:full-study-emu-vs-computed-obs-selection} is resolved in the transfer learning emulation, suggesting yet-more accurate predictions.
        
    \subsection{Inverse model validation}

        Once again, the model is tested for self-consistency with pseudodata generated by the underlying multistage model at known points in the parameter space that were not used in training the surrogate model. The surrogate model is then used for inference with pseudodata and the resulting posterior is investigated to determine how well it recovers underlying truth. Due to the fact that a particular parametrization has been chosen for the specific shear and bulk viscosity, the test for self-consistency is best compared as, for example, $\eta/s$ vs. temperature. After all, despite the motivation for the parametrization, the physics is contained in the temperature dependence of the viscosity, not a particular representation. 
    
\begin{figure}[!htb]
        	\begin{centering}
        		\includegraphics[width=0.8\linewidth]{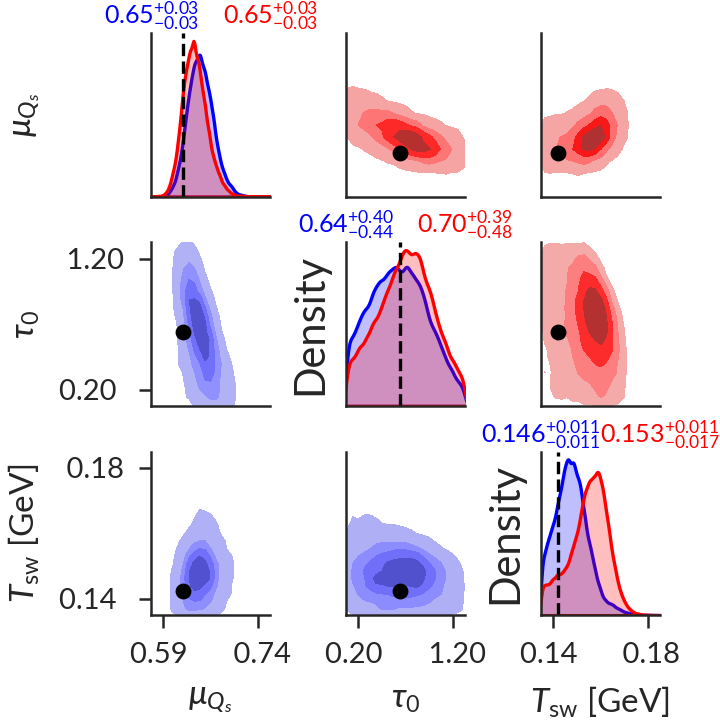}
        		\includegraphics[width=\linewidth]{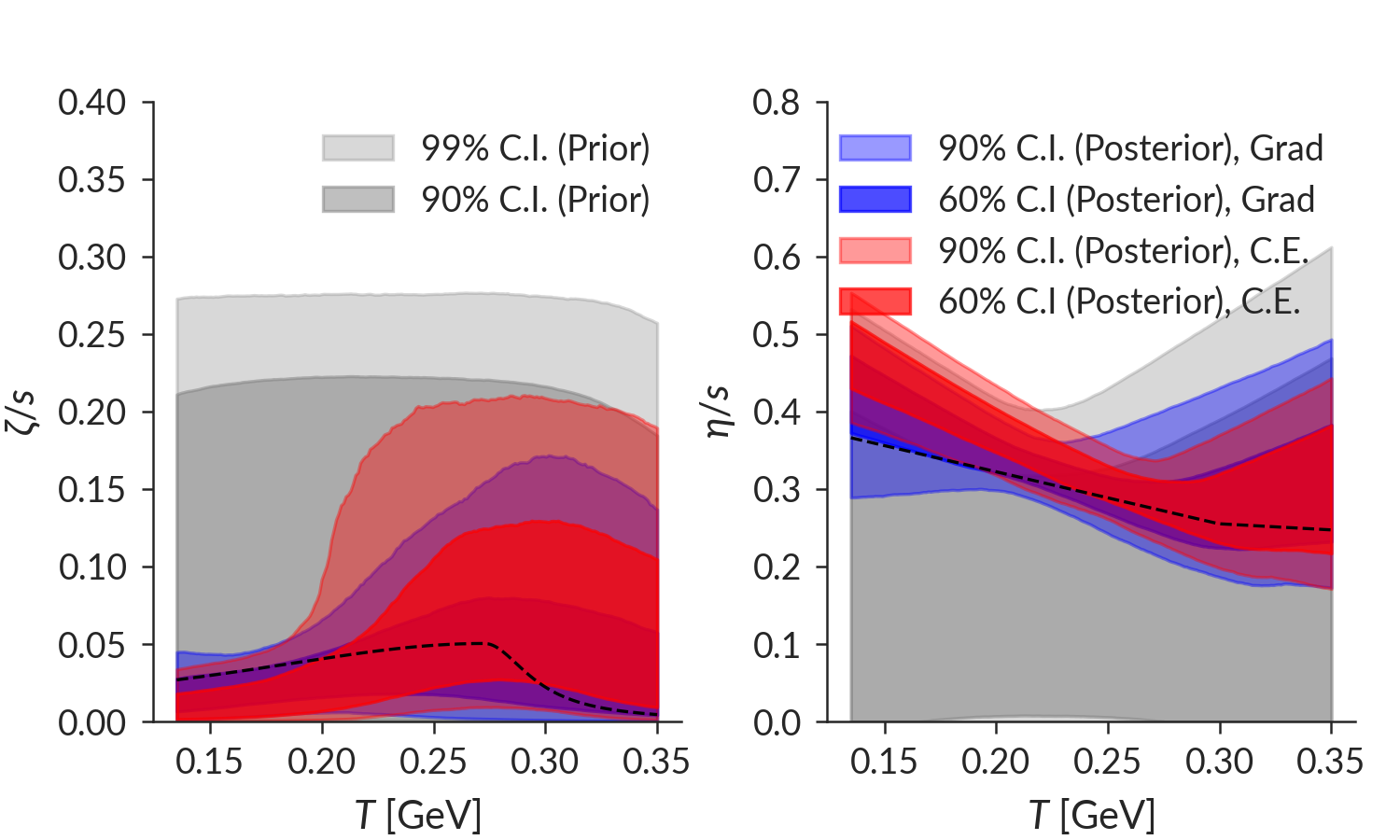}
        		\caption{Posterior distributions of non-viscous (top) and viscous (bottom) parameters for a sample validation point. The true values are highlighted in black (top). The quoted values are the median and 95\% C.I.}
        		\label{fig:full-closure-grad-val-3-posteriors}
        	\end{centering}
        \end{figure}

\begin{figure}[!htb]
        	\begin{centering}
        		\includegraphics[width=0.8\linewidth]{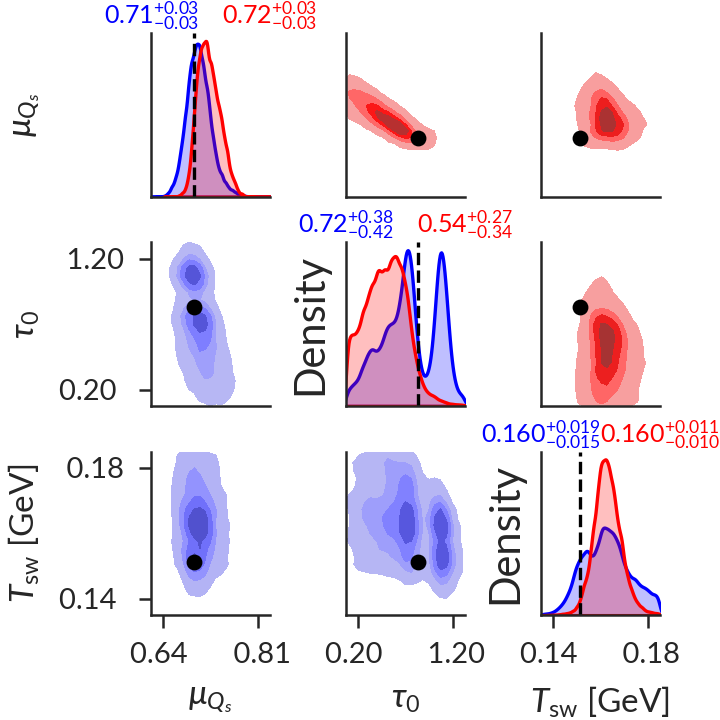}
        		\includegraphics[width=\linewidth]{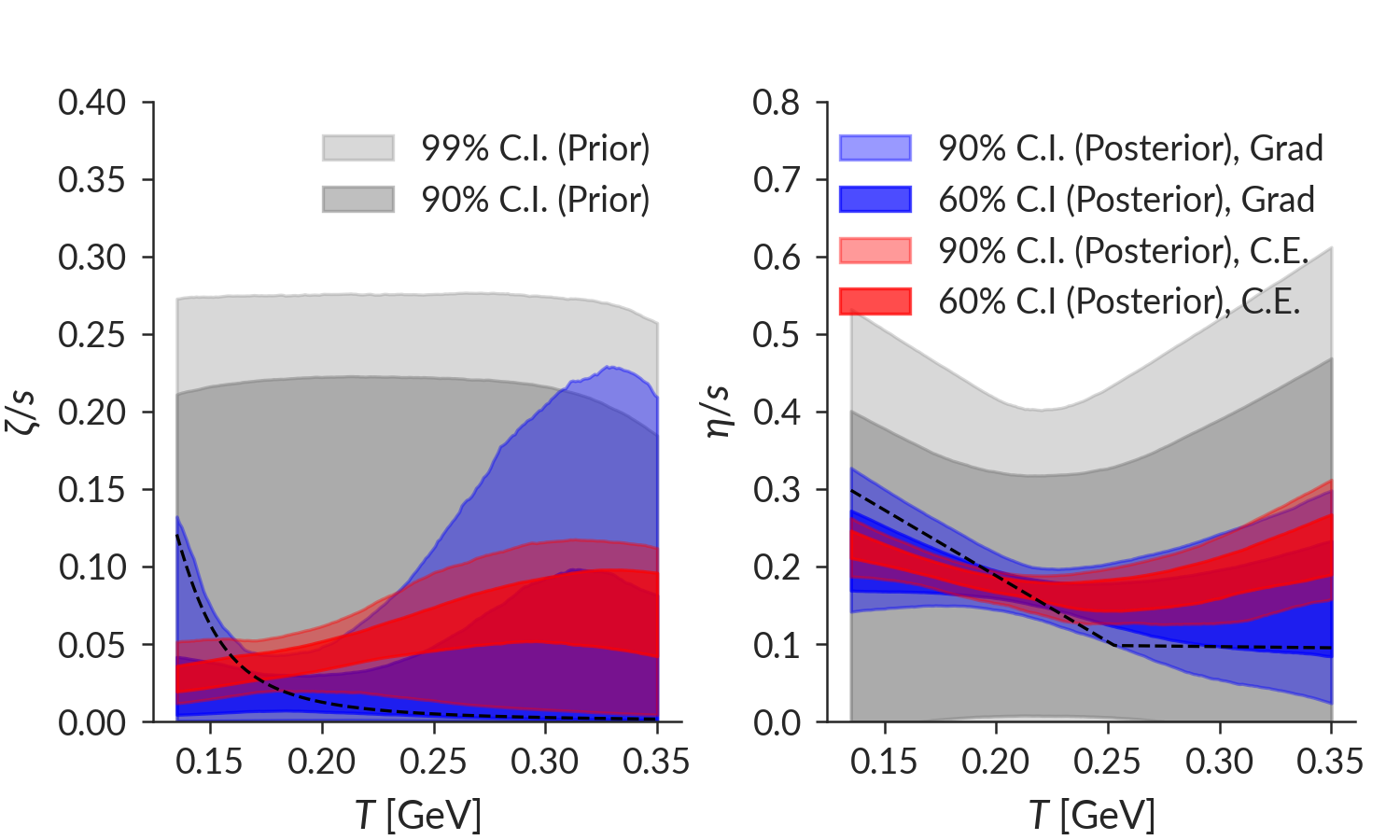}
        		\caption{Posterior distributions of non-viscous (top) and viscous (bottom) parameters for a second sample validation point. The true values are highlighted in black (top). This is an important example of interpretable failure. The quoted values are the median and 95\% C.I.}
        		\label{fig:full-closure-grad-val-2-posteriors}
        	\end{centering}
        \end{figure}

\begin{figure}[!htb]
        	\begin{centering}
        		\includegraphics[width=0.8\linewidth]{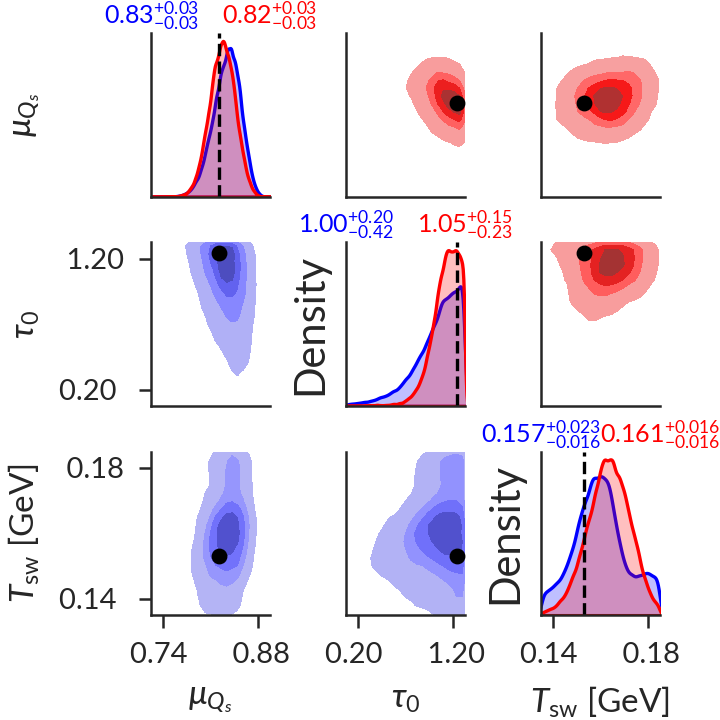}
        		\includegraphics[width=\linewidth]{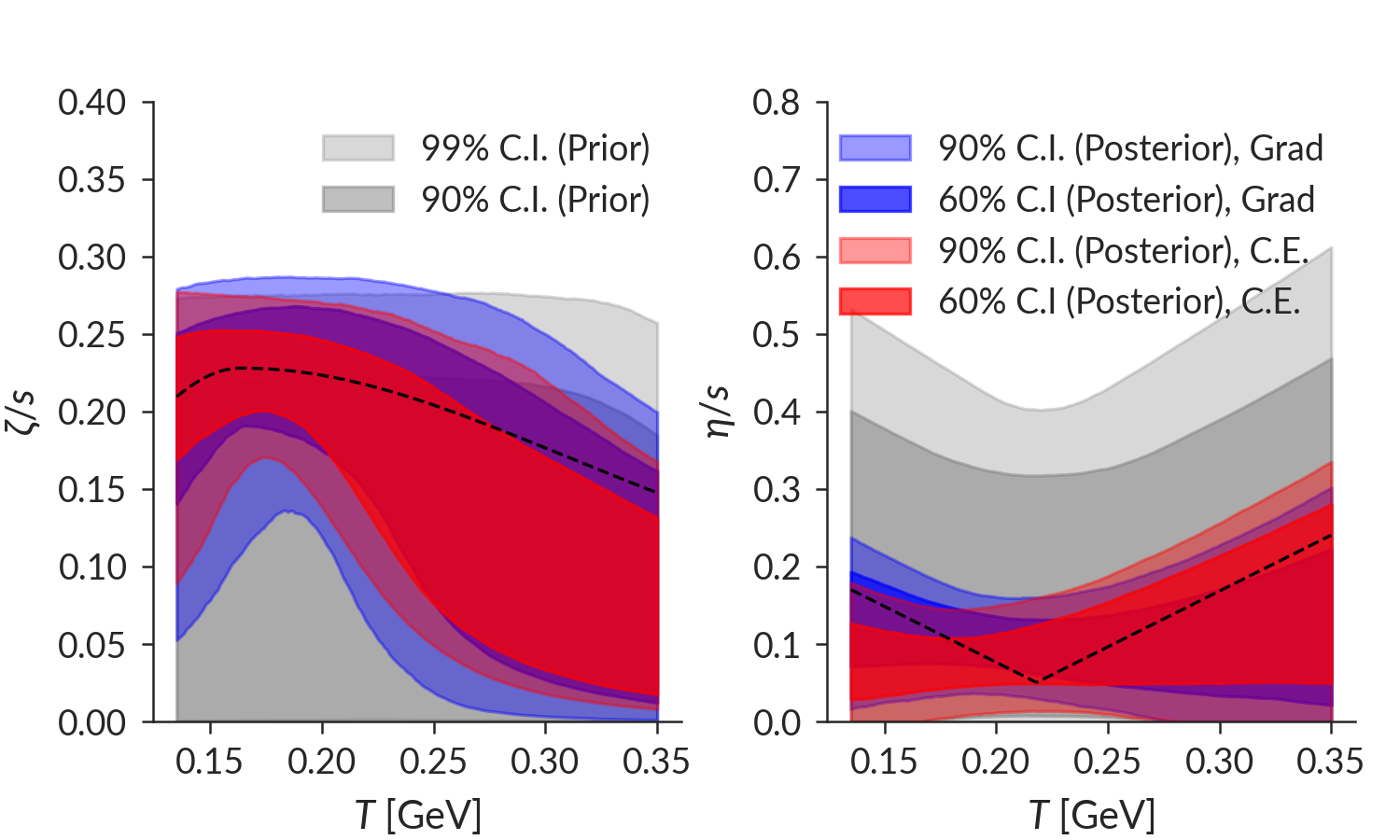}
        		\caption{Posterior distributions of non-viscous (top) and viscous (bottom) parameters for a third sample validation point. The true values are highlighted in black. The quoted values are the median and 95\% C.I.}
        		\label{fig:full-closure-grad-val-6-posteriors}
        	\end{centering}
        \end{figure}

        It is cumbersome to show this result for all validation points, but care is taken to show a representative sample of validation points in this section. 
        The parameters related to the hydrodynamic viscosities
        are shown separately from those not related to viscosity, for former shown as $\eta/s$ or $\zeta/s$ vs. temperature. No discernible covariance is seen between the two groups of parameters. 
        While no covariances are seen, when the model is pushed to the edges of the prior region, the distribution can become bimodal. 
        Examples are shown in Figs.~\ref{fig:full-closure-grad-val-3-posteriors}-\ref{fig:full-closure-grad-val-6-posteriors}.

        What is important to inspect is if the posterior consistently contains the known truth. For example, does the true value fall within the 90\% credible interval (C.I.) approximately 90\% of the time? If so, then it is plausible that, provided with a 90\% credible interval, a gambler would break exactly even assuming they were presented with fair odds by the bookmaker. This is clearly the case for results shown in Figs.~\ref{fig:full-closure-grad-val-3-posteriors} and \ref{fig:full-closure-grad-val-6-posteriors}. The sample validation points chosen for these figures additionally demonstrate the resolution of a large, relatively flat bulk viscosity (Fig.~\ref{fig:full-closure-grad-val-6-posteriors}) and a bulk viscosity with a comparatively sudden peak at high temperature (Fig.~\ref{fig:full-closure-grad-val-3-posteriors}) in addition to a variety of $\eta/s$. All are recovered well and within the 90\% credible interval, although one needs to consider this in tandem with Figs.~\ref{fig:full-study-emu-vs-computed-obs-selection} and \ref{fig:full-study-TL-emulated-vs-computed} to be confident in closure performance.
        
        The posteriors in Fig.~\ref{fig:full-closure-grad-val-2-posteriors} demonstrate a strong bimodality for Grad viscous corrections and bias in Chapman-Enskog $\delta f$ and, while the truths are partially recovered, the posteriors seem at odds with physical intuition and are not in particularly good agreement with each other, such as in $\tau_0$. This occurs because the true value of the bulk viscosity peak is \emph{below} the particlization temperature and a bimodality develops in $\zeta/s$ for Grad $\delta f$, while the Chapman-Enskog $\delta f$ attempts to compensate and does not resolve the second $\zeta/s$ mode and poorly resolves $\eta/s$. For each peak of $\zeta/s$, a different value of the switching time between IP-Glasma and MUSIC is preferred as the model is pushed into a corner, causing bimodality in the posterior of the initial condition and particlization parameters. The observable that couples these quantities is $\delta p_T / \langle p_T \rangle$, whose pseudodata is noisier than the experimental data, further exacerbating the issue. This example of an interpretable failure is an edge case in the parameter space. 
 
        It is intuitive that the model struggles to reproduce true values of hydrodynamic quantities that are located outside the hydrodynamic evolution. 
        Joint priors (\emph{i.e.} requiring the bulk peak temperature to be greater than particlization) have not yet been developed for heavy ion collision studies and doing so is beyond the scope of this work. Note as well that this is a particular feature of the multi-modal bulk viscosity as the true value of $T_{sw}$ in Fig.~\ref{fig:full-closure-grad-val-3-posteriors} is close to the edge but can still be well-constrained. Nonetheless, the ability to interpret these failures of the modeling workflow further strengthens the results derived from this study.
        
\begin{figure*}[!ht]
        	\begin{centering}
        		\includegraphics[width=\textwidth]{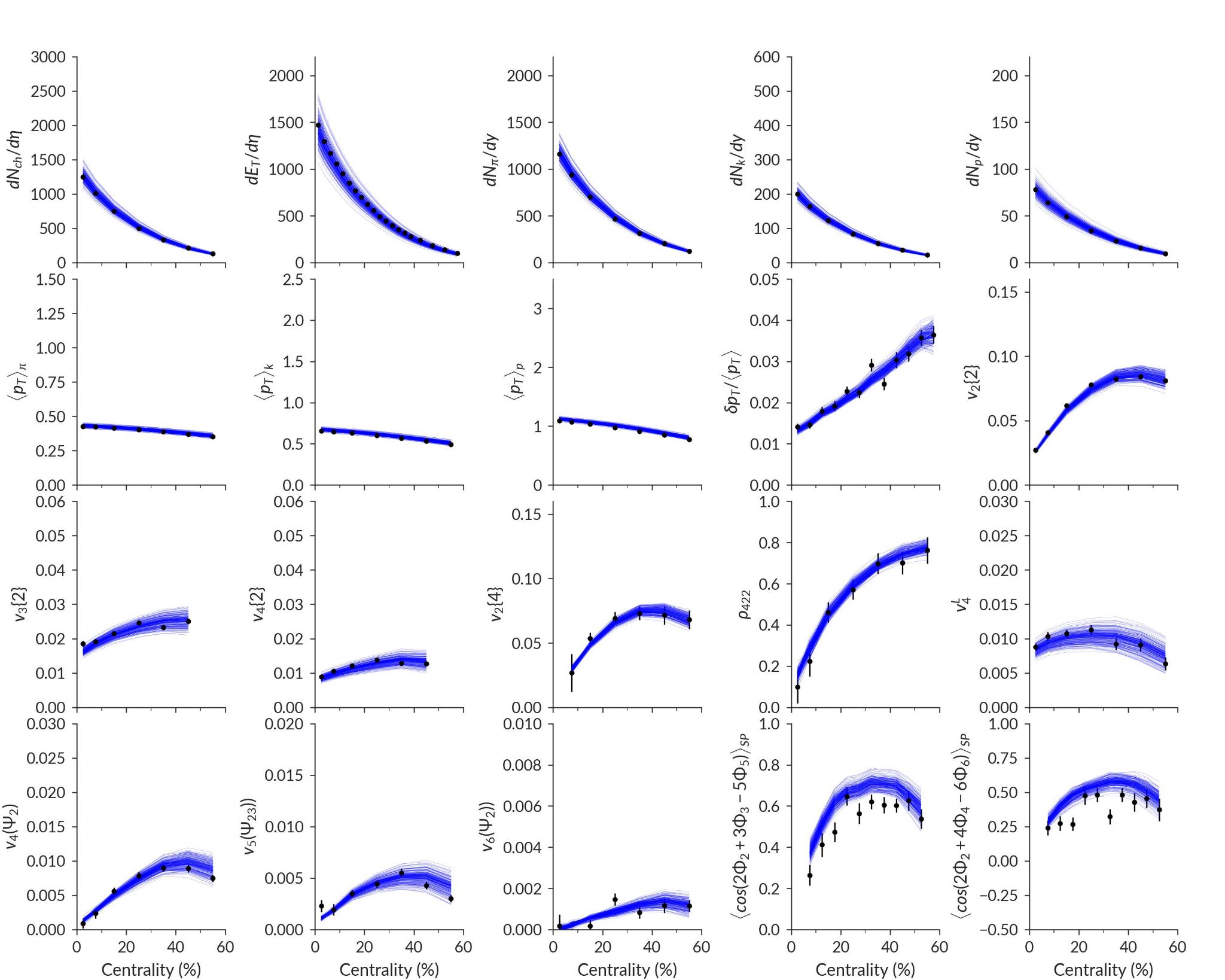}
        		\caption{Posterior predictive distributions with Grad viscous corrections for the posterior shown in Fig.~\ref{fig:full-closure-grad-val-6-posteriors} with pseudodata used for comparison shown as data points.}
        		\label{fig:full-closure-grad-val-6-obs-validation}
        	\end{centering}
        \end{figure*}

        A reassuring feature of the inferential framework is that all of the closure points reproduce the pseudodata well, as exemplified in Fig.~\ref{fig:full-closure-grad-val-6-obs-validation}. As can be seen, the emulator is not overfitting by going through every potentially noisy data point, but is instead robust to statistical fluctuations in the underlying data. This further suggests that the model is behaving well and is well-conditioned for the problem at hand while also not exhibiting strong bias. An example of low-bias can also be seen in the marginal distributions for $\mu_{Q_s}$ in Figs.~\ref{fig:full-closure-grad-val-3-posteriors}-\ref{fig:full-closure-grad-val-6-posteriors} -- the truth is not always exactly located at the peak of the marginal distribution, but instead the peaks are distributed around the true value. Additionally, the two $\delta f$ models are differentiable and provide further evidence that the transfer learning model is not simply reproducing the source model's results.
        
        An exciting feature in these closure tests in comparison to previous studies is the constraint on $\eta/s$ and $\zeta/s$ at higher temperatures. In previous studies, constraint was limited to the low temperature regions and the model was insensitive to the high-temperature (or early-time) behavior of the fireball evolution unless the temperature dependence was explicitly specified by the parametrization \cite{SIMSPRL, SIMSPRC, BernhardPhD, Nijs:2020ors, Nijs:2020roc}.  In these closure test, for the first time, constraint on the viscosity can be achieved even at high temperature. 
        This raises the exciting prospect that the viscosity of strongly-interacting matter in heavy ion collisions may be constrained to an unprecedented precision without sacrificing accuracy.

\section{Inference with LHC Data} 
    
    \subsection{Grad and Chapman-Enskog posteriors}
        
        Now that the surrogate model is known to behave in accordance with expectations for test points and failures are interpretable, the validation pseudodata is exchanged for real experimental data. The previous section has confidently established that the Bayesian parameter estimation produces reasonable results for known inputs, leading to the belief that this should plausibly reveal the underlying properties of experimentally-produced quark-gluon plasma in heavy ion collisions. The repeated validation, observable selection, closure testing, and sanity checks of the surrogate modeling and inference have established that the models are reliable and well-conditioned for the problem at hand. 
    
\begin{figure*}[!htbp]
        	\begin{centering}
        		\includegraphics[width=0.9\linewidth]{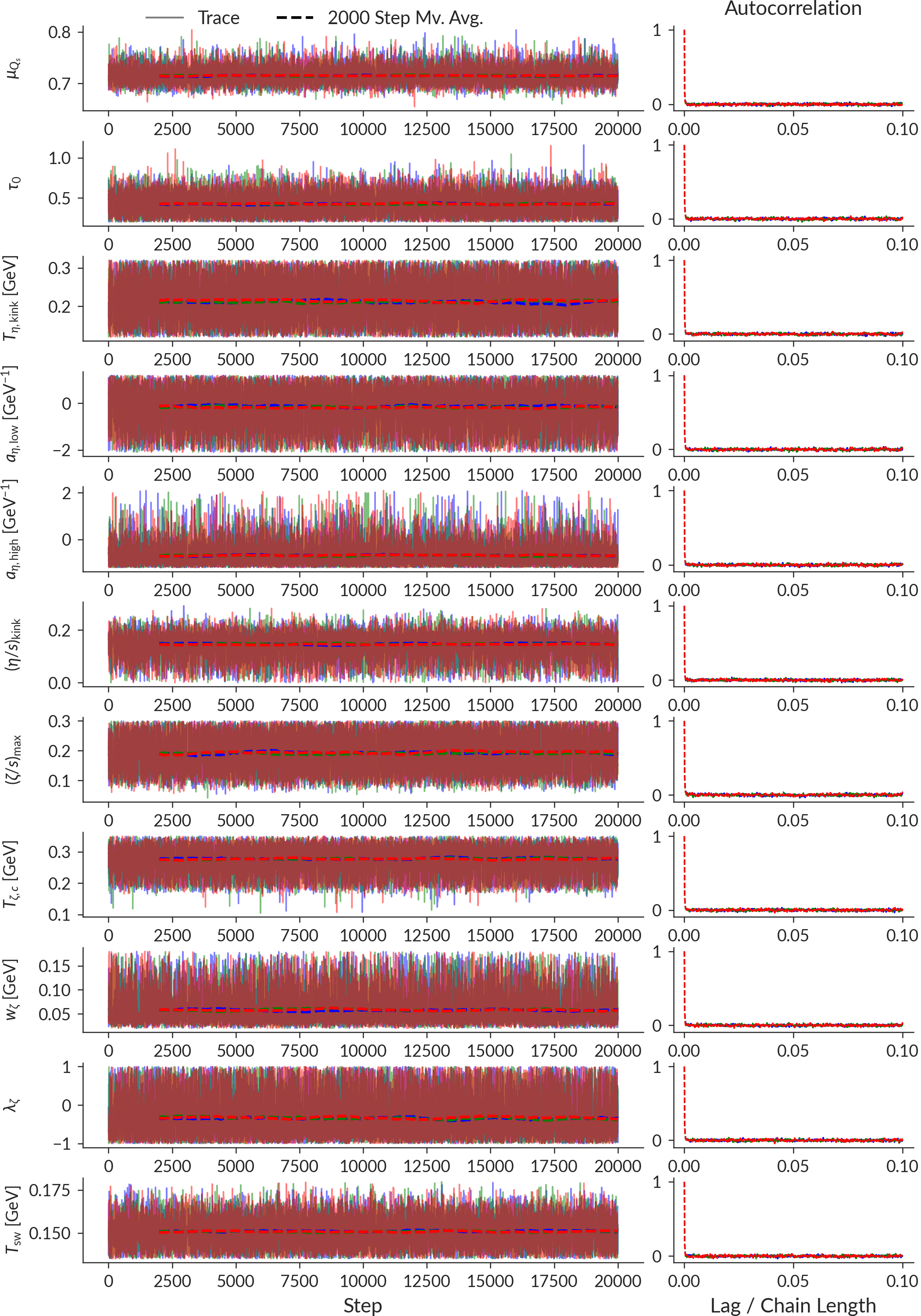}
       			\caption{MCMC trace, moving average, and autocorrelation from comparison to experimental data with Grad viscous corrections. The C.-E. MCMC behavior is comparable.}
        		\label{fig:full-study-grad-data-mcmc}
        		\end{centering}
        \end{figure*}

        The calculations at the design points form the prior predictive distribution and were shown in Fig.~\ref{fig:full-allobs-prior-predictive} for a superset of observables. These calculations cover the experimental results well, although correlations between calculations are difficult to discern and likely introduce some tension. The MCMC is again performed using a parallel tempering algorithm. The above closure test and the below comparison to data are performed using Grad's 14-moment viscous corrections and while the above closure tests were performed with 10,000 MCMC steps with 10 walkers per dimension and 10 rungs in the parallel tempering temperature ladder; the below comparison to data is performed with 20,000 MCMC steps with 50 walkers per dimension and 20 rungs in the parallel tempering ladder for improved sampling resolution. The trace, moving average, and autocorrelation of the final MCMC chain is shown in Fig.~\ref{fig:full-study-grad-data-mcmc} for three sample walkers. It is important to note that these walkers have clearly thermalized, i.e. there is no directed walk, as the trace exhibits no discernible autocorrelation and are thus sampling from the target distribution. This is not trivial, especially as the number of parameters increases and insufficient verification of this can result in incorrect sampling of the posterior.
        
\begin{figure}[!htb]
        	\begin{centering}
        		\includegraphics[width=0.9\linewidth]{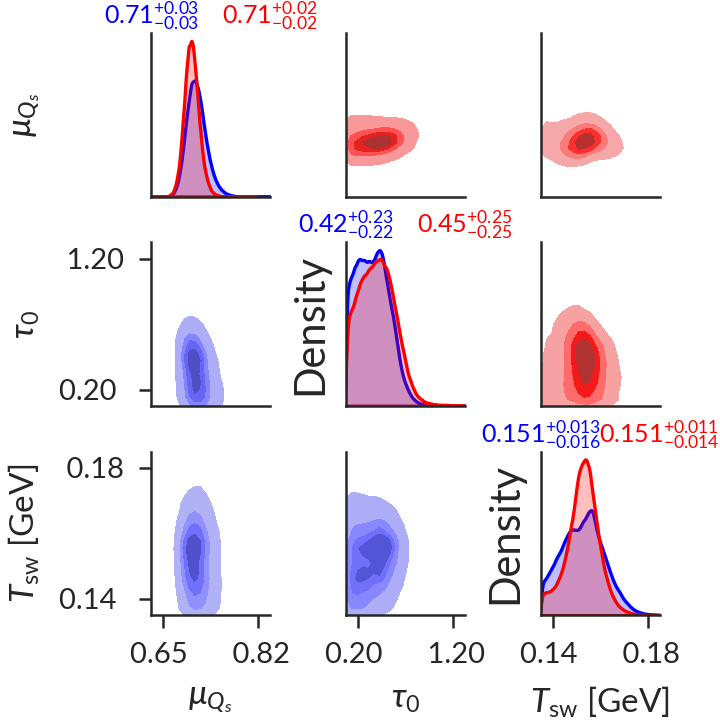}
        		\caption{Posterior distributions of non-viscous parameters from comparison to experimental data with Grad $\delta f$ (blue, lower triangle of the sub-figure matrix) and Chapman-Enskog $\delta f$ (red, upper triangle). The quoted values along the diagonal are the median and 95\% C.I. of the 1-dimensional marginal distribution.}
        		\label{fig:full-study-grad-and-CE-data-posterior}
        	\end{centering}
        \end{figure}

\begin{figure*}[!htb]
        	\begin{centering}
        		\includegraphics[width=0.8\linewidth]{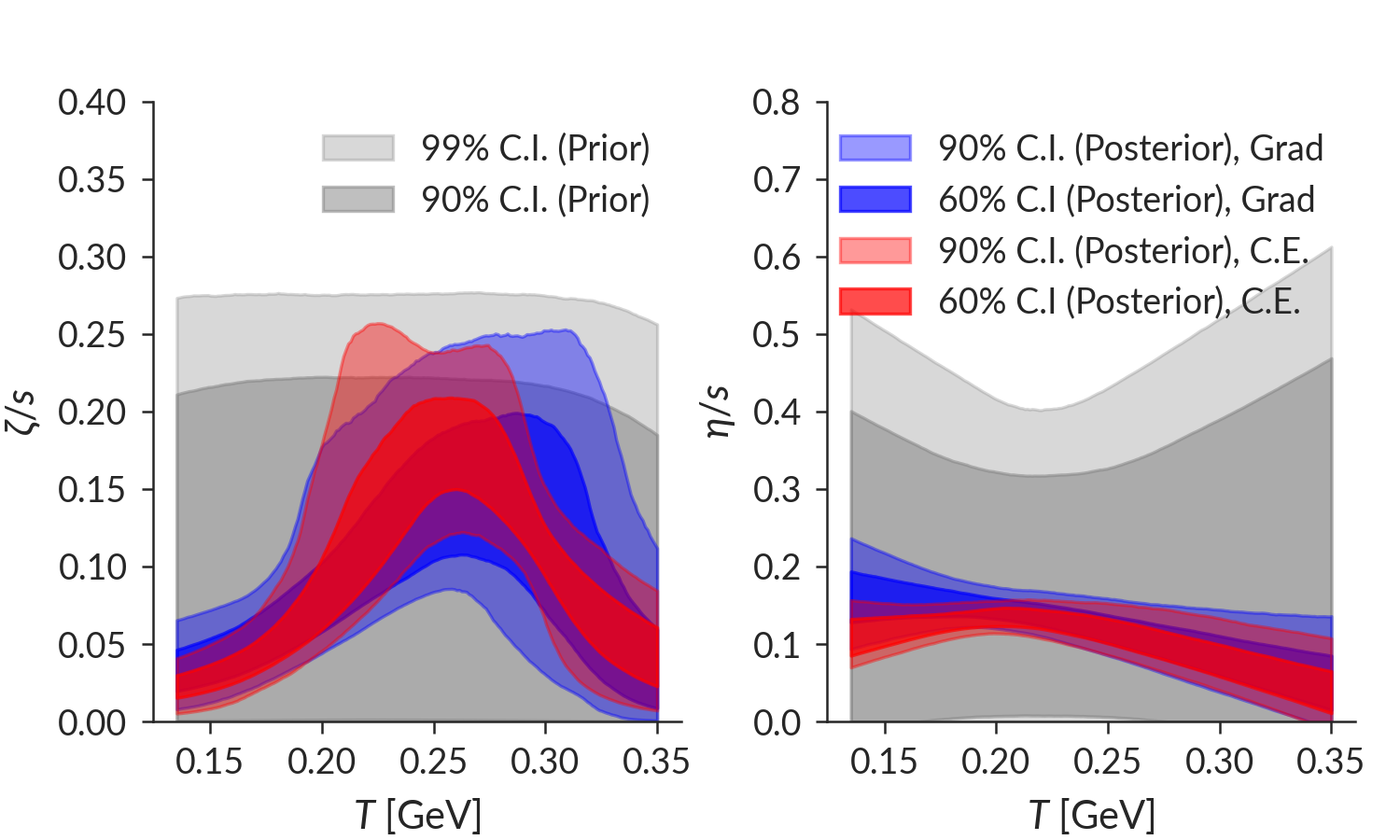}
        		\caption{Viscous posterior with Grad viscous corrections (blue) and Chapman-Enskog viscous corrections (red) from comparison to experimental data.}
        		\label{fig:full-study-grad-and-CE-data-viscous-posterior}
        	\end{centering}
        \end{figure*}

\begin{figure*}[!htb]
        	\begin{centering}
        		\includegraphics[width=\textwidth]{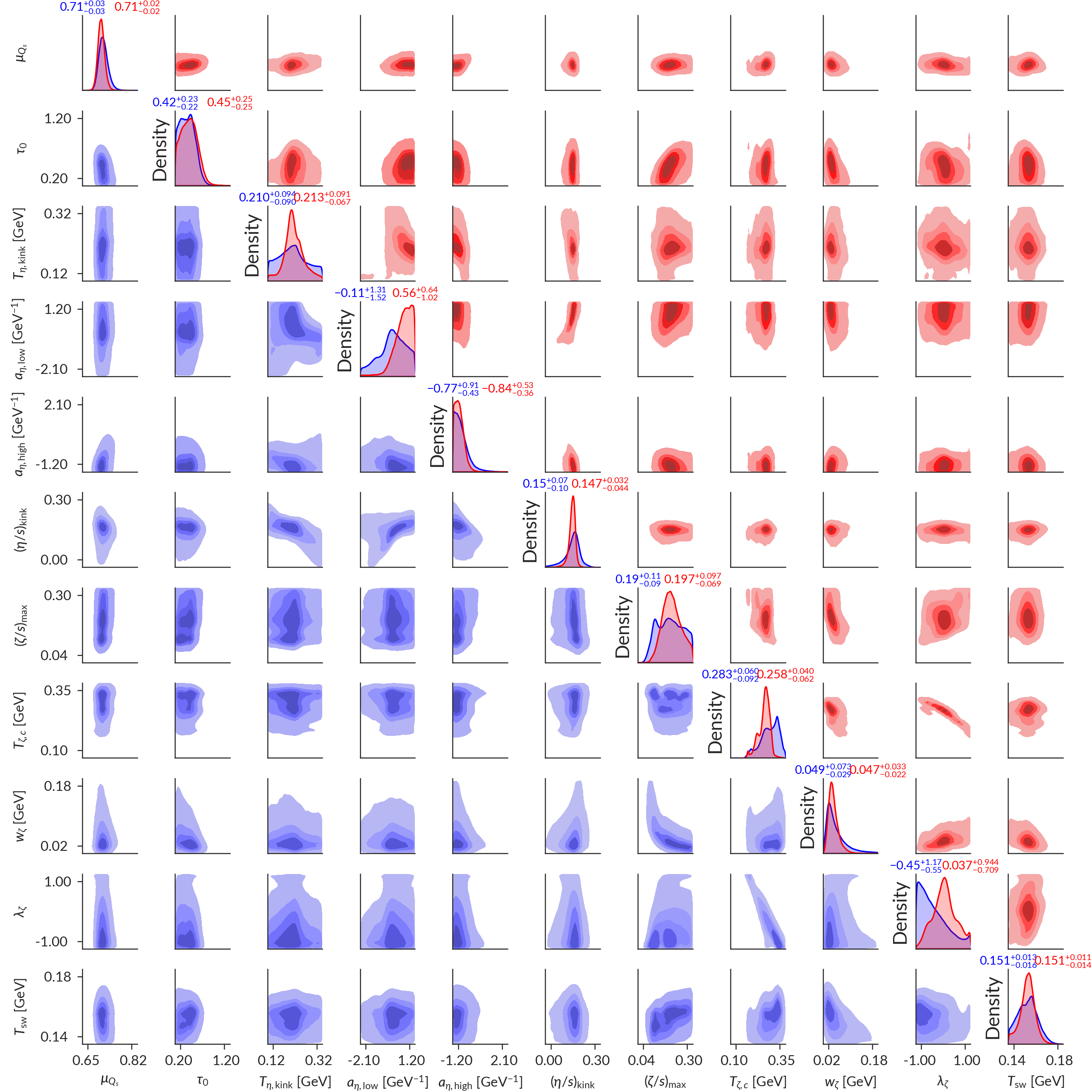}
        		\caption{11-dimensional posterior showing marginal and joint marginal distributions  with Grad viscous corrections (blue, lower triangle) and Chapman-Enskog viscous corrections (red, upper triangle) from comparison to experimental data. Values along the diagonal are the median and 95\% C.I. of the 1-dimensional marginal distribution.}
        		\label{fig:full-study-grad-and-CE-data-full-posterior}
        	\end{centering}
        \end{figure*}

        With confidence in the MCMC, it is finally time to look at the posterior distribution after comparison with data. The non-viscous parameter posterior for both viscous corrections is shown in Fig.~\ref{fig:full-study-grad-and-CE-data-posterior}, the viscous posterior for both viscous corrections is shown in Fig.~\ref{fig:full-study-grad-and-CE-data-viscous-posterior}, and the marginal and joint marginal distributions of the 11-dimensional posterior are shown in Fig.~\ref{fig:full-study-grad-and-CE-data-full-posterior}. The non-viscous parameters demonstrate clear constraint, particularly in the case of the normalization $\mu_{Q_s}$. The switching time between IP-Glasma and MUSIC is well-localized to early times $\tau_0 \lesssim 0.7$ fm, which is in accordance with previous experience and appears not to favor very late hydrodynamic onset times. 
        
        The particlization temperature $T_{sw}$ is also well-constrained within the prior region. A recent estimate of the crossover temperature from lattice QCD places it at $T_c = 156 \pm 1.5 $ MeV \cite{Steinbrecher:2018phh,Guenther:2020jwe}, precisely in the region of highest posterior density for the particlization temperature. The constraint of the particlization temperature is particularly interesting as the chemistry (equation of state) of the hydrodynamic medium and the hadron resonance gas are identical to those of \cite{SIMSPRC}, which required a much lower particlization temperature with the same viscous correction.\footnote{Note that the equation of state used in this work is fully described by a hadron resonance gas up to a temperature of 165 MeV, and subsequently matched to lattice calculations from Ref.~\cite{PhysRevD.90.094503} (see Section~\ref{sec:viscous_hydro}). We thus highlight that a proper transition from hydrodynamics to hadronic degrees of freedom can be achieved somewhat above the pseudocritical temperature $156$~MeV.} This also provides a limit on the lifetime over which the viscosity can act by reducing the lifetime of the hydrodynamic phase during which the viscosity acts. 
        We do not constrain the viscosity to be small at particlization, which means there is potential for large viscous corrections to influence final observables. As will be seen shortly, in the temperature region probed by particlization -- approximately bounded by 0.14 and 0.18 GeV -- the data itself prefers the specific bulk viscosity to be small. In testing for self-consistency, it was found that the model can recover large viscosity at particlization (Fig.~\ref{fig:full-closure-grad-val-6-posteriors}), meaning that the demand for small viscous corrections is an authentic feature of the data.

        \begin{figure}[!htb]
        	\begin{centering}
        		\includegraphics[width=\columnwidth]{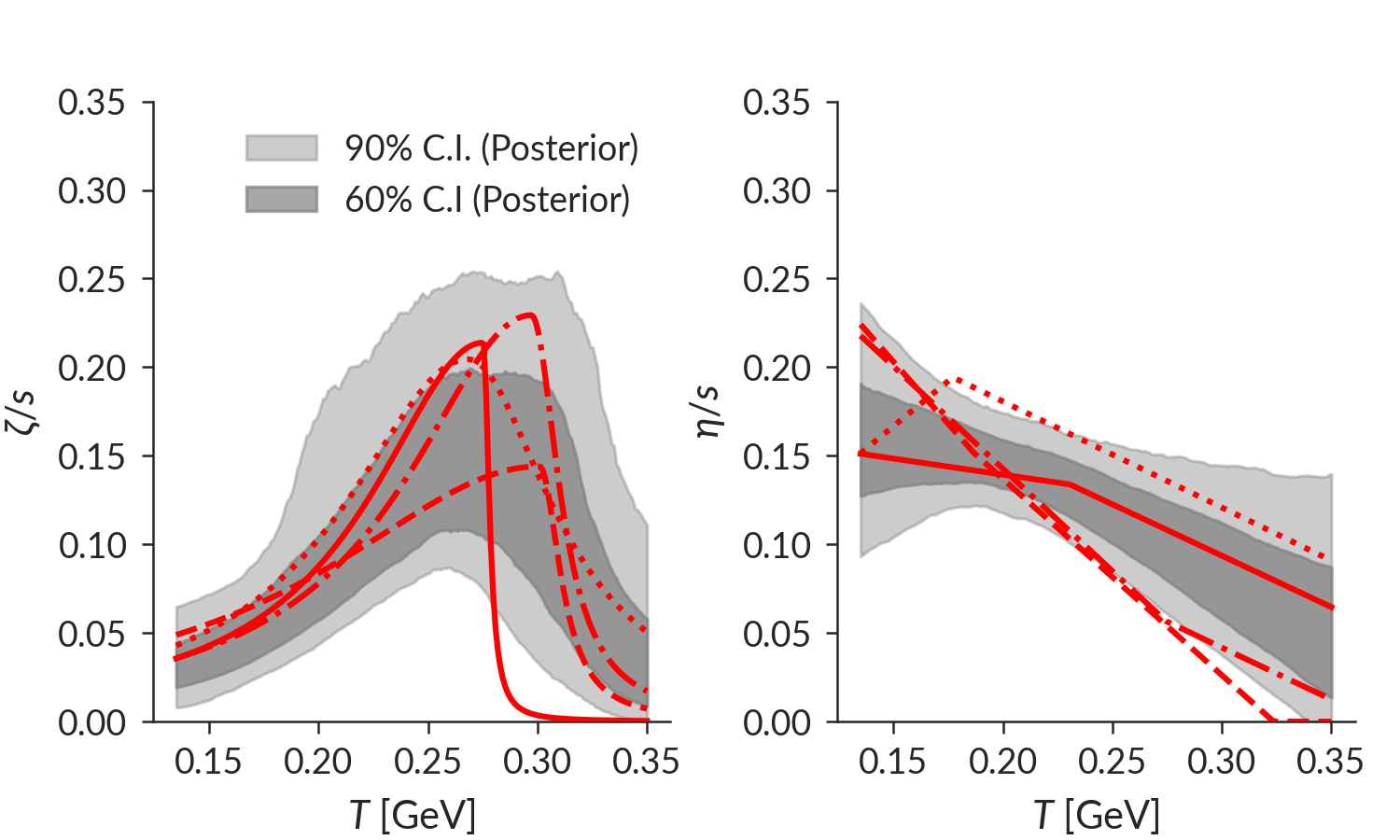}
        		\caption{Samples from the viscous posterior for Grad viscous corrections after comparison to experimental data.}
        		\label{fig:full-study-grad-data-viscous-posterior-w-samples}
        	\end{centering}
        \end{figure}

        In Fig.~\ref{fig:full-study-grad-and-CE-data-viscous-posterior}, the temperature-dependent specific bulk viscosity $\zeta/s$ demonstrates a clear peak and the 99\% C.I. is inconsistent with zero 
        for temperatures between 160 and 300 MeV
        for Grad viscous corrections while for Chapman-Enskog, it is inconsistent with zero over the entire range shown. Randomly drawn example samples from the Grad posterior are shown in Fig.~\ref{fig:full-study-grad-data-viscous-posterior-w-samples}, demonstrating the diversity of choices that are compatible with data. The constraint certainly weakens at high temperature, but the peaked specific bulk viscosity is well-constrained at low and intermediate temperatures. The significant values of bulk viscosity favored by the posterior contrasts with previous Bayesian studies, which either favored small bulk viscosity or could not provide precise constraints on its value. There is theoretical support for a non-negligible value of bulk viscosity in the deconfinement region\cite{Karsch:2007jc, PhysRevLett.103.172302, Bazavov:2019lgz}, decreasing to zero at higher temperatures~\cite{Arnold:2006fz}. The peak of the specific bulk viscosity shifts slightly between the two viscous correction models, but the posteriors are broadly consistent with each other, particularly the 60\% credible intervals. 
        
        An unexpected feature of the viscous posterior is a negatively-sloped specific shear viscosity at higher temperatures. This is driven in part by peripheral $v_3\{2\}$, a fluctuation-driven quantity, and central $v_4\{2\}$. As higher temperatures correspond to earlier times in the fireball evolution, this decreasing high-temperature $\eta/s$ dissipates initial-state fluctuations more slowly. However -- and importantly -- the high-temperature $\eta/s$ posterior is still statistically compatible with a flat line through the 99\% credible interval as will be discussed in more detail throughout the remainder of this work.
        This is a consideration worth investigating in more depth\footnote{Note that a decreasing specific shear viscosity is also the result of a Bayesian analysis with parametric initial conditions which allows for a varying nucleon size \cite{Nijs:2022rme}.}.

\begin{figure*}[!htb]
        	\begin{centering}
        		\includegraphics[width=\textwidth]{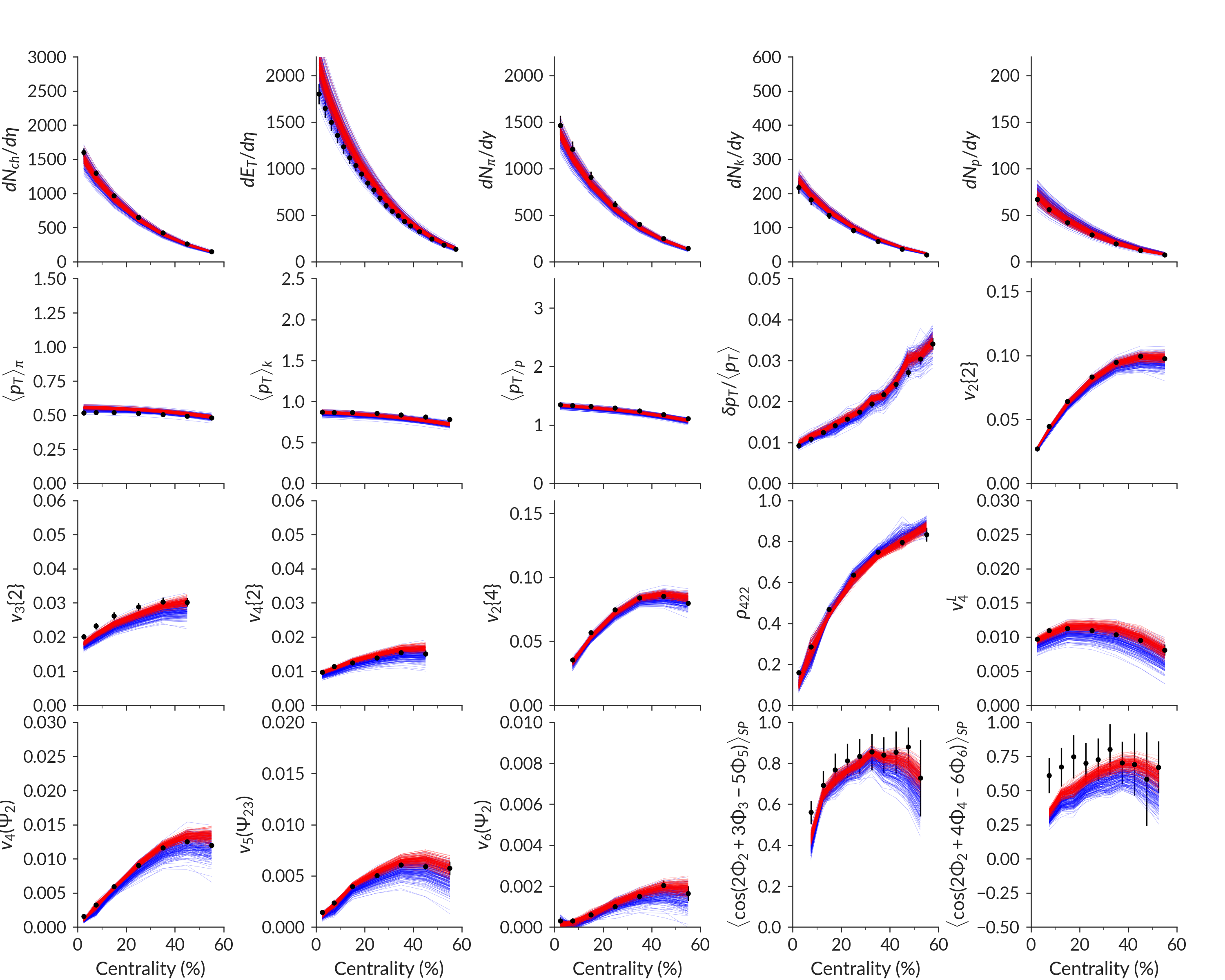}
        		\caption{Posterior predictive distribution with Grad viscous corrections (blue) and Chapman-Enskog (C.-E.) viscous corrections (red) after comparison to data.}
        		\label{fig:full-study-grad-and-CE-data-posterior-predictive}
        	\end{centering}
        \end{figure*}

\begin{figure*}[!htb]
        	\begin{centering}
        		\includegraphics[width=\textwidth]{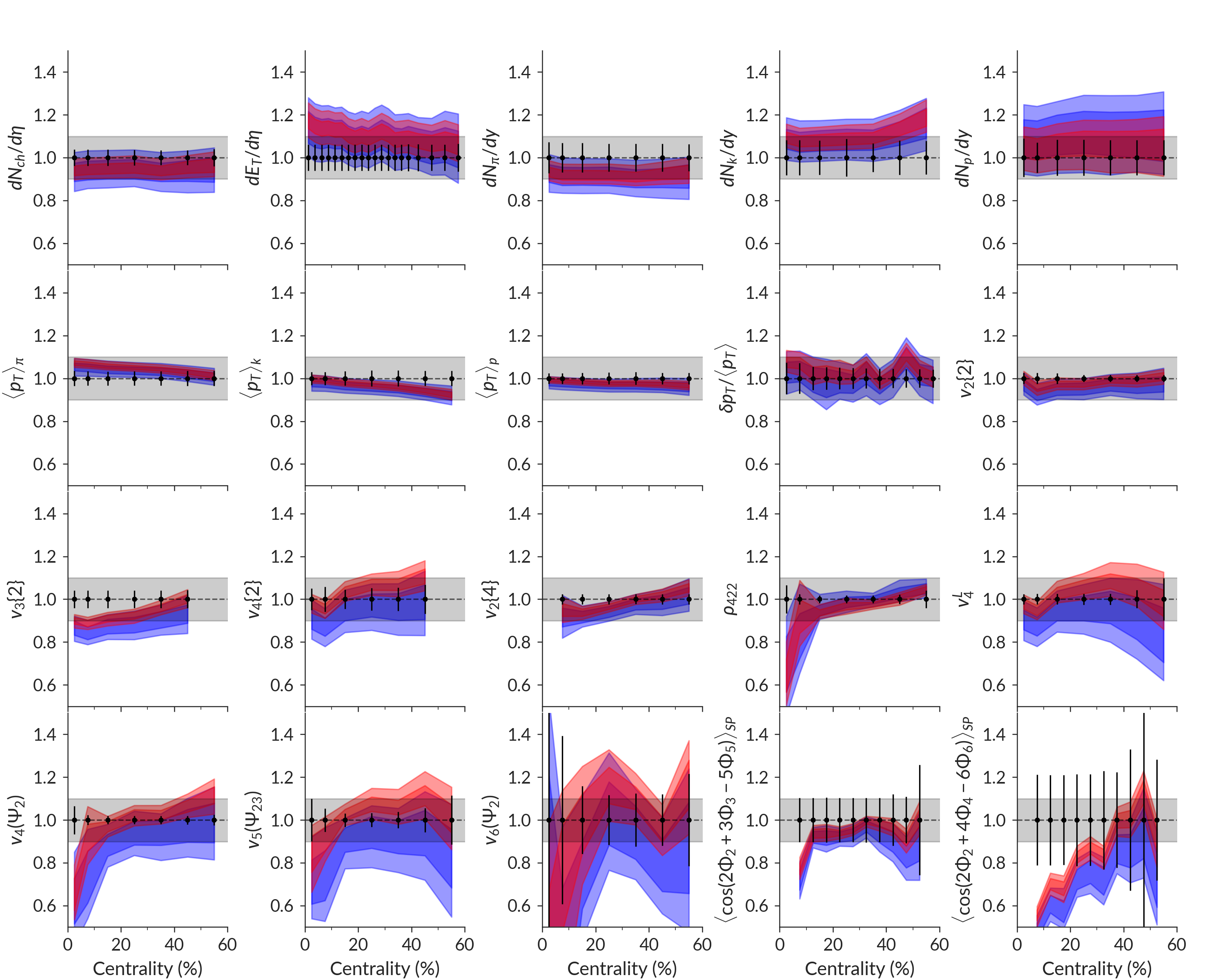}
        		\caption{Posterior predictive ratio (theory/data) with Grad (blue) and Chapman-Enskog (red) $\delta f$ after comparison to data.}
        		\label{fig:full-study-grad-and-CE-data-posterior-predictive-ratio}
        	\end{centering}
        \end{figure*}

        To ensure the quality of the fit and to identify tension in the model, one can inspect the posterior predictive distribution (Fig.~\ref{fig:full-study-grad-and-CE-data-posterior-predictive}) and the ratio of the posterior predictive distribution to experimental data (Fig.~\ref{fig:full-study-grad-and-CE-data-posterior-predictive-ratio}). It is clear from the posterior predictive distributions that the model fits the data well but exhibits tension, seen in the transverse energy and the three-plane correlators. 
        
\begin{figure*}[!htb]
        	\begin{centering}
        		\includegraphics[width=\textwidth]{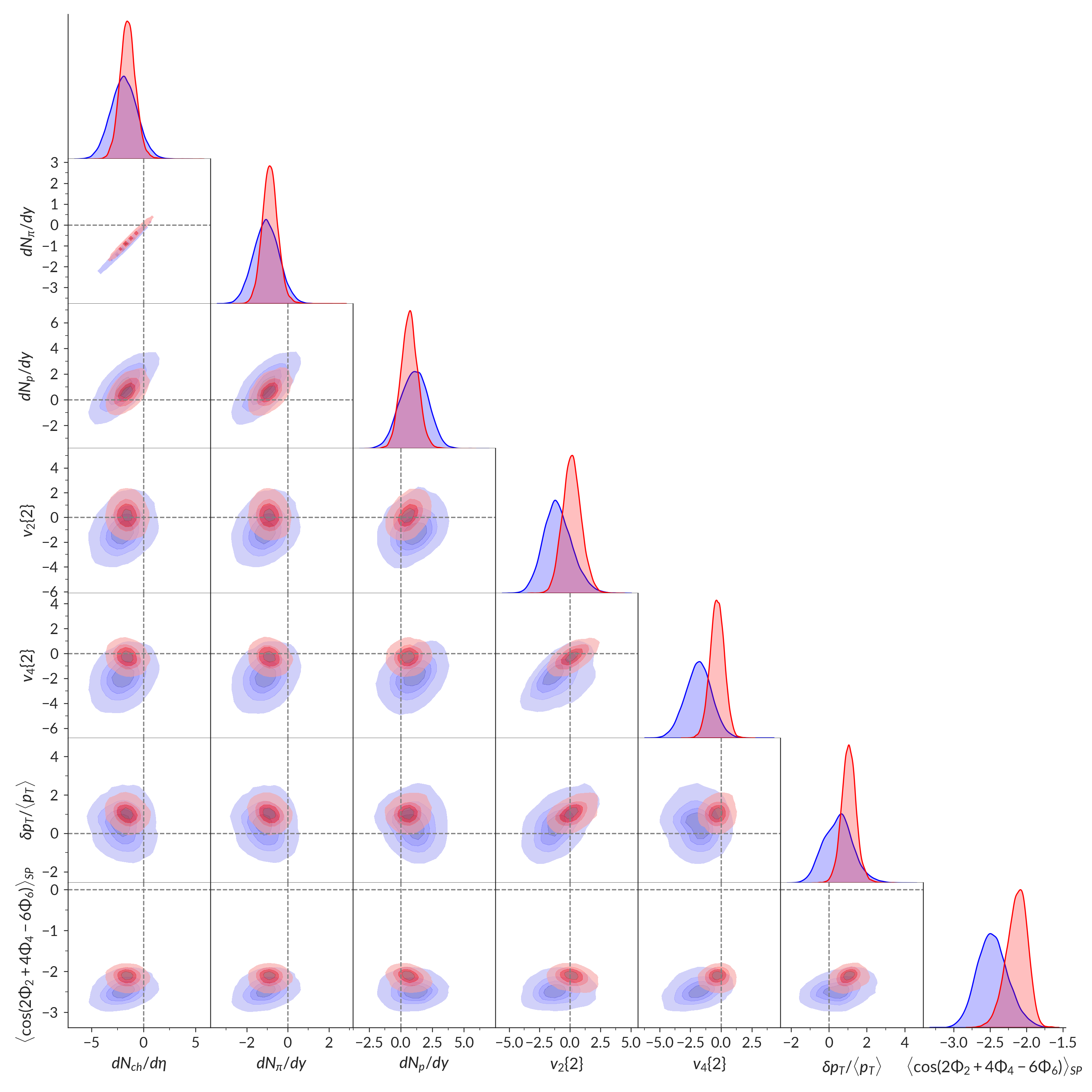}
        		\caption{Correlations between posterior predictive distributions for selected observables for central collisions. Dashed lines denote the central experimental result and x- and y-axis units are the experimental uncertainty for the respective observables. Grad viscous corrections are in blue while Chapman-Enskog viscous corrections are shown in red.}
        		\label{fig:full-study-grad-data-posterior-predictive-correlations}
        	\end{centering}
        \end{figure*}

        The difficulty reproducing the three-plane correlators is also not new, but the postdictions shown in Fig.~\ref{fig:full-study-grad-and-CE-data-posterior-predictive} are consistent with previous studies \cite{Heffernan:2023kpm}.  
        Of note is that the Chapman-Enskog $\delta f$ is closer to reproducing these correlations than the Grad viscous corrections. Insight can be gained by investigating the sensitivity of these observables to various parameters. These comparisons are shown in Appendix~\ref{appendix:A}.
        The dominant sensitivities are to normalization and the shear viscosity kink temperature, similar to the anisotropic flow that naturally influence the correlations. The other potential underlying cause of difficulty in matching these observables is geometric - the observables match as well as they can, but the prior predictive distributions do not cover the data. With the geometry in IP-Glasma fixed by nuclear configurations and deep inelastic scattering, insufficient freedom remains. 
        Before leaving this to future analysis, it must be noted that $\delta p_T/\langle p_T \rangle$ is also at the edge of the prior predictive region. If the three-plane correlators and the $p_T$ fluctuations are correlated, this has potential to reveal further insight. The correlation between these observables at mid-centrality is shown in Fig.~\ref{fig:full-study-grad-data-posterior-predictive-correlations} and reveals that these observables are uncorrelated, suggesting that their tension is independent. A future analysis should attempt to address this by revisiting the constraint from deep inelastic scattering simultaneously with observables from heavy ion collisions.
        
        The posterior predictive distribution for the correlated $p_T$ fluctuations produces the most accurate postdiction of any IP-Glasma calculation and yields the correct centrality dependence, a feature not seen in other models. Investigating this sensitivity, the overall magnitude is reduced by a larger $(\zeta/s)_{max}$ and constraints $\tau_0$ to early times. This suggests yet further that the bulk viscosity must be further investigated for a narrower, taller peak to better reproduce experimental results. This is beyond the scope of this work.\footnote{This narrower, taller peak is difficult to resolve without reparametrization of the width of the bulk viscosity or carefully constructing a scale-invariant prior. This too is beyond the scope of this work, but should be strongly considered in future studies.}
        
        The success of the model with respect to every other observable must be highlighted: nearly every experimental measurement in nearly every observable is consistent with the posterior predictive distribution shown in Fig.~\ref{fig:full-study-grad-and-CE-data-posterior-predictive}. This was by no means guaranteed. Bayesian studies in heavy ion physics have broadly exhibited success with parametric models and fewer observables. 
        This success in describing measurements provides strong support for the IP-Glasma model of initial conditions.
        This represents a step forward in rigorously constructing a hybrid model with each stage containing microscopic physics and testing it via comparison to data. 
    
    \subsection{Post-dictions and predictions with maximum a posteriori parameters}
    
        Scientific models can be evaluated by how well they can describe experimental measurements in systematic model-to-data comparison, as performed up to this point, but also by how well they predict quantities to which they were not explicitly tuned. A model that can only describe quantities to which it is systematically compared is less useful than a model that, once compared to a carefully-selected set, makes accurate predictions. The Bayesian inference performed in this section was performed using a surrogate model trained at a large number of design points, not the underlying model itself. As a result, before moving on to predictions, it is important to explore the veracity of postdictions. The most likely value in the 11-dimensional parameter space is the Maximum a Posteriori estimate, determined by numerical optimization. The full hybrid model is run at the MAP points in Table~\ref{tab:MAP-Grad-and-CE}, but with $6000$ collisions 
        rather than $2500$, with centrality selection performed in the same manner as in \cite{McDonald:2016vlt, Heffernan:2023kpm}. This increase in statistics allows for higher precision results.

         \begin{table*}[!ht]
    	\centering
    	\begin{tabular}{lcccc}
    		Parameter & Grad $\delta f$, $\eta/s$ & Grad $\delta f$, $\eta/s(T)$ & C.-E. $\delta f$, $\eta/s$ & C.-E. $\delta f$, $\eta/s(T)$ \\ \hline
    		$\mu_{Q_s}$ & 0.72341 & 0.70808 & 0.72654 & 0.70858\\ \hline
    		$\tau_0$ [fm] & 0.52127 & 0.51291 & 0.40142 & 0.55159\\ \hline
    		$T_{\eta,\mathrm{kink}}$ [GeV] & 0.150 & 0.22333 & 0.150 & 0.21123\\ \hline
    		$a_{\eta,\mathrm{low}}$ [GeV$^{-1}$] & 0.000 & -0.16259 & 0.000 & 0.65272\\ \hline
    		$a_{\eta,\mathrm{high}}$ [GeV$^{-1}$] & 0.000 & -0.80217 & 0.000 & -0.89472\\ \hline
    		$(\eta/s)_{\mathrm{kink}}$ & 0.13577 & 0.13944 & 0.12504 & 0.14888\\ \hline
    		$(\zeta/s)_{max}$ & 0.28158 & 0.22085 & 0.17391 & 0.20117\\ \hline
    		$T_{\zeta,c}$ [GeV] & 0.31111 & 0.29198 & 0.2706 & 0.25455\\ \hline
    		$w_{\zeta}$ [GeV] & 0.02878 & 0.03625 & 0.05255 & 0.04506\\ \hline
    		$\lambda_{\zeta}$ & -0.96971 & -0.56235 & -0.14178 & 0.06408\\ \hline
    		$T_{\mathrm{sw}}$ [GeV] & 0.15552 & 0.15429 & 0.15069 & 0.1513\\ \hline
    	\end{tabular}
    	\caption{Maximum a Posteriori estimates with Grad's 14-moment and Chapman-Enskog RTA viscous corrections. Estimates with (denoted $\eta/s(T)$) and without (denoted $\eta/s$) temperature-dependent specific shear viscosity are reported.} \label{tab:MAP-Grad-and-CE}
        \end{table*}
        
        A variety of interesting features arise in Table~\ref{tab:MAP-Grad-and-CE}. First, the lattice QCD estimate of crossover temperature -- $T_c = 156 \pm 1.5$ MeV -- is consistent with both Grad MAP estimates of the particlization temperature using Grad viscous corrections, with the MAP estimate from constant $\eta/s$ nearly identical to the central lattice estimate. Using Chapman-Enskog viscous corrections results in a slightly lower estimate of the particlization temperature, but still close to the estimated crossover temperature, suggesting that the hadrons may behave hydrodynamically for a brief period after recombination. Next, the switching time $\tau_0$ is consistent with IP-Glasma's pressures having come to a steady state (see Fig.~\ref{fig:2dipg-pressures}) and with sufficient time for the build-up of pre-equilibrium dynamics that was hypothesized to be of critical importance in describing the strongly-interacting medium. The parameter $\mu_{Q_s}$ relating the saturation scale $Q_s$ to the color charge density profile has a posterior distribution shown in Fig. \ref{fig:full-study-grad-and-CE-data-full-posterior} corresponding to a MAP estimate reported in Table \ref{tab:MAP-Grad-and-CE}. Note the posterior distribution obtained here for $\mu_{Q_s}$ overlaps that reported in \cite{Mantysaari:2022ffw} for a fixed number of hot spots. The value of the specific shear viscosity is broadly consistent with other Bayesian results and the constant $\eta/s$ is very close to past ``chi-by-eye'' fits of $0.13$ \cite{McDonald:2016vlt}. The bulk viscosity maximum and width are consistent with a large, peaked bulk viscosity, further supporting a consistent picture between theoretical expectations and prior modeling success. The asymmetry of the bulk viscosity is of interest as it suggests a bulk viscosity peaked at high temperature and slowly decreasing as it approaches the particlization temperature, where it is well-constrained by the data to be small. While the MAP estimates for the bulk viscosity differ in their parameters between $\eta/s$ and $\eta/s(T)$, the actual value at any temperature differs by a maximum of $\approx $10\% below the region where it nears the lower peak location at $T \approx 0.28$ GeV.
        
        The MAP estimates are used to make predictions of observables not used in the model-to-data comparison. Strictly speaking, this is due to computational limitation: the most appropriate comparison is a full posterior predictive distribution with perhaps a surrogate model trained on a reasonable quantity of high-statistics calculations. At the same time, the MAP estimates are the recommended parameters for use in other studies, such as hard sector studies of jet-medium interactions or photon/dilepton calculations, and therefore represent a faithful picture of how the model will be used in practice.
    
        First, the veracity of the MAP points is determined via postdiction, in which the underlying computationally-expensive multistage model is compared to quantities used in the inference above. In the following figures, the Grad and Chapman-Enskog MAP are shown in blue and red, respectively. The MAP with temperature-dependent $\eta/s$ is shown as a solid line while constant $\eta/s$ is shown as a dashed line. Shaded regions denote aleatoric uncertainty. 

\begin{figure}[!htb]
        	\begin{centering}
        		\includegraphics[width=\columnwidth]{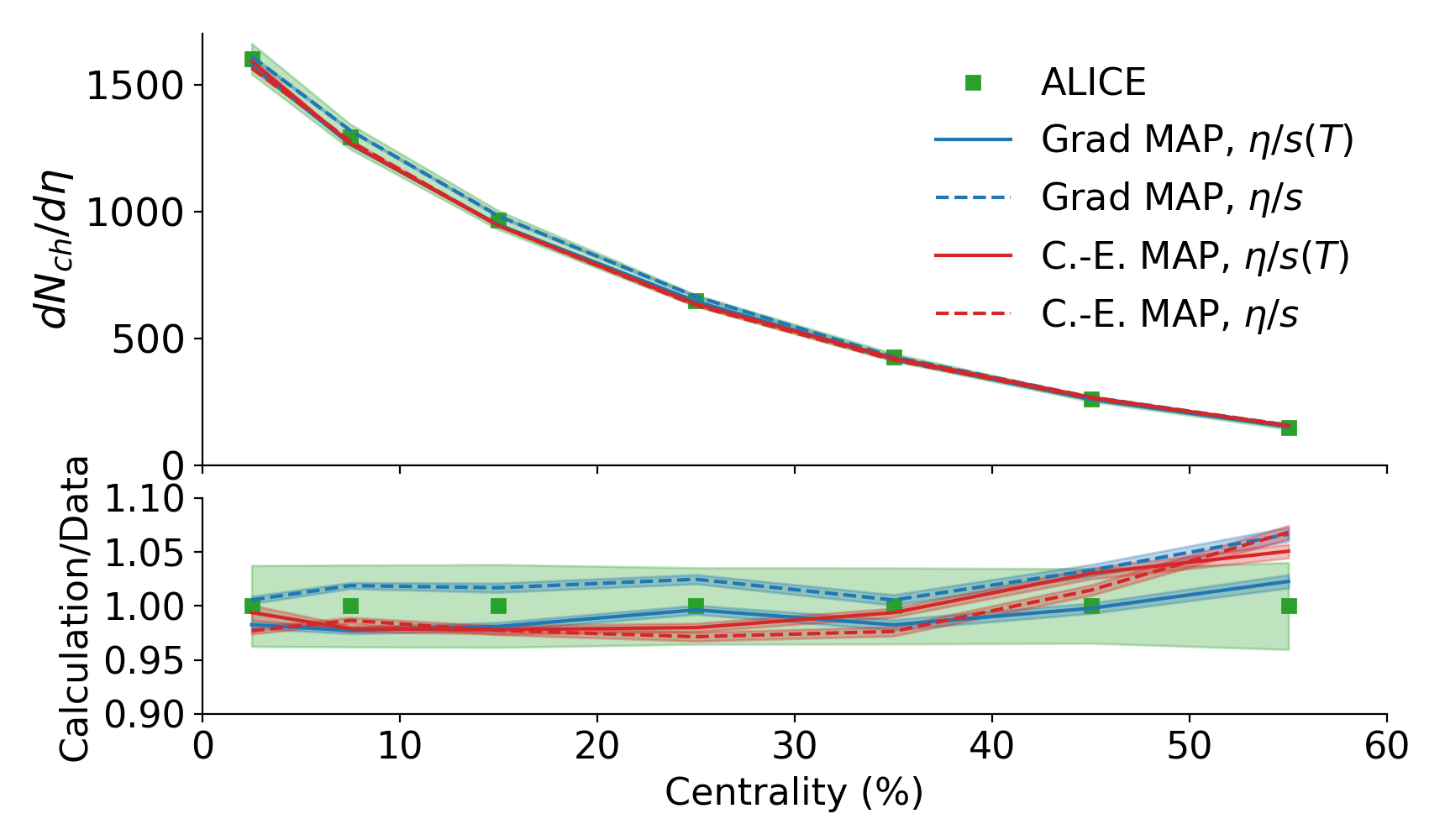}
        		\caption{Postdictions of charged hadron multiplicity at Maximum a Posteriori.}
        		\label{fig:full-study-MAP-dNdeta}
        	\end{centering}
        \end{figure}

\begin{figure}[!htb]
        	\begin{centering}
        		\includegraphics[width=\columnwidth]{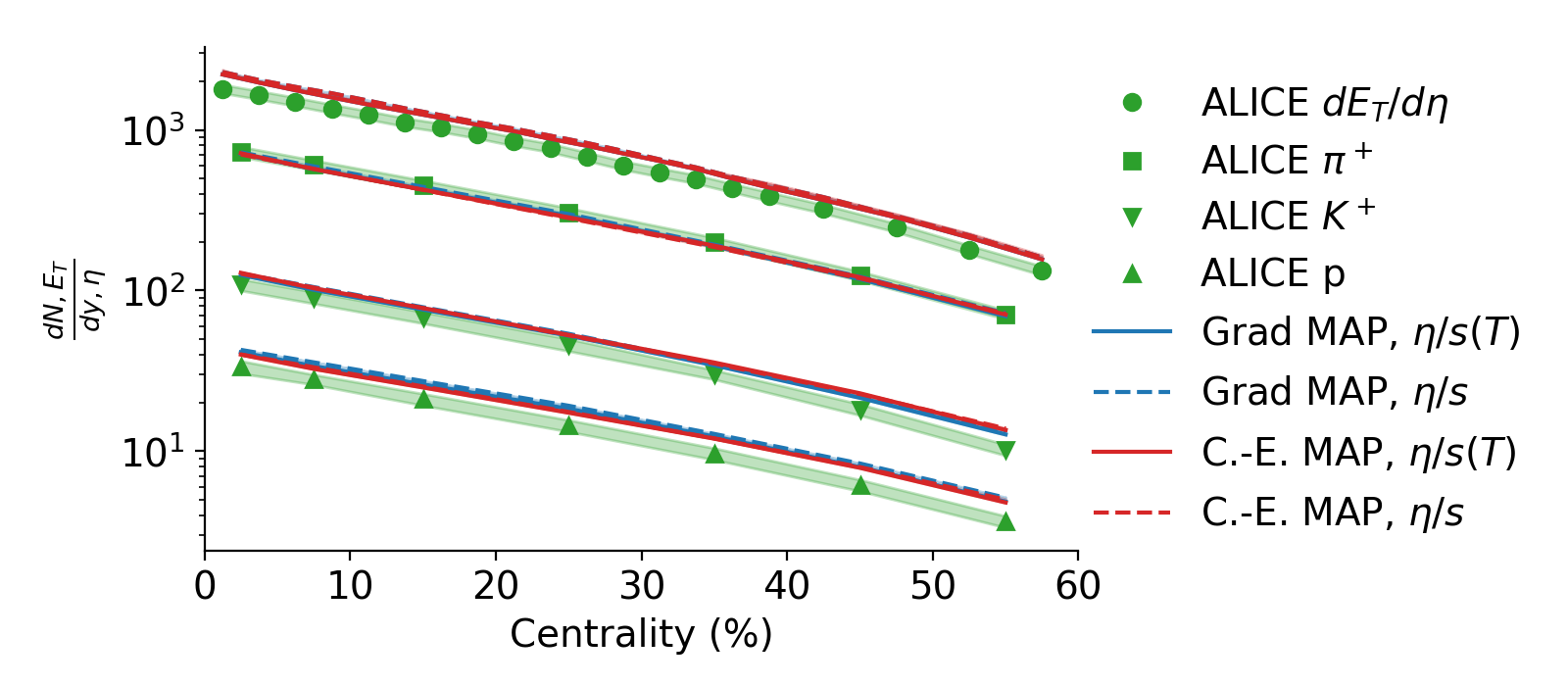}
        		\caption{Postdictions of identified hadron multiplicity at Maximum a Posteriori.}
        		\label{fig:full-study-MAP-dNdy}
        	\end{centering}
        \end{figure}

        The charged hadron multiplicity, Fig.~\ref{fig:full-study-MAP-dNdeta}, compares very favorably with the MAP calculations within the experimental uncertainty for all viscous correction models. A variety of identified particle multiplicities and transverse energy per rapidity slice, Fig.~\ref{fig:full-study-MAP-dNdy}, also compare very well, albeit the proton and kaons are overestimated while the pions are underestimated. This balancing act combined with the overall charged hadron multiplicity shows that aspects of the hadron chemistry are imbalanced. 
        The overestimation of the number of higher-mass particles in turn results in an overestimation of transverse energy. Nonetheless, the differences between the MAP calculations imply an influence of viscous corrections on the hadronic chemistry.
        
\begin{figure}[!htb]
        	\begin{centering}
        		\includegraphics[width=\columnwidth]{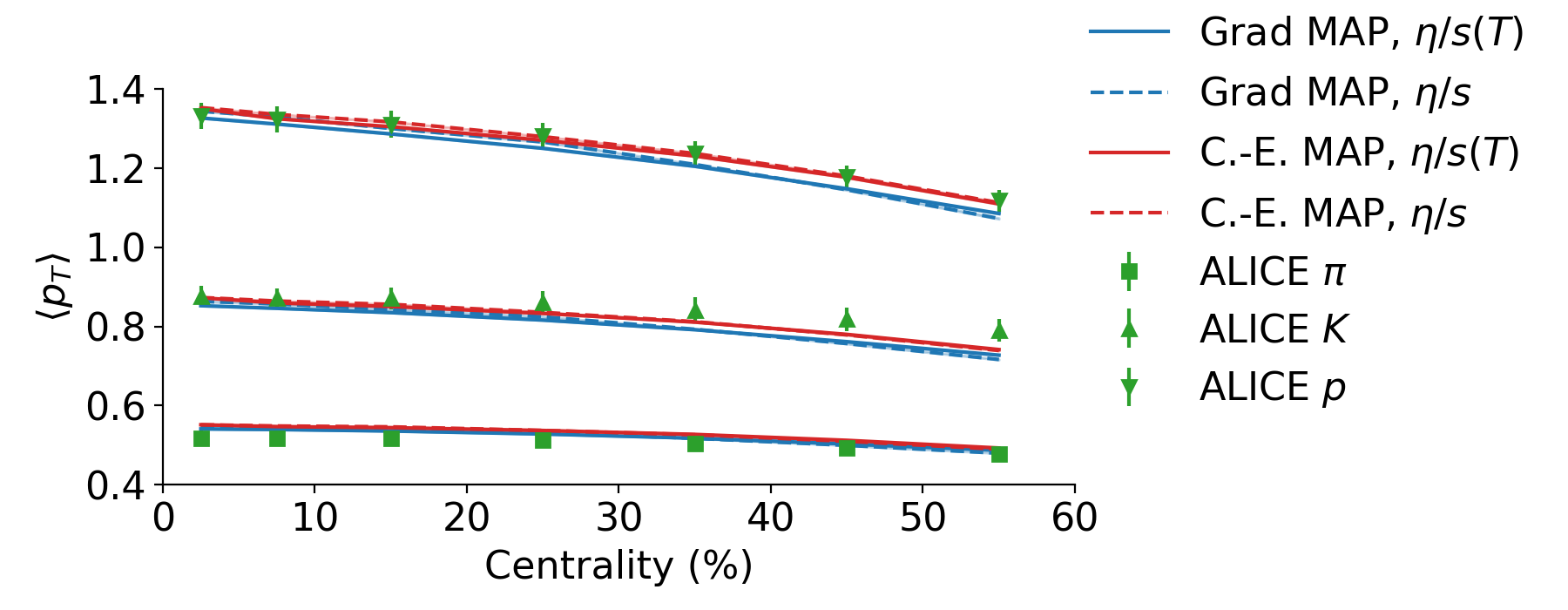}
        		\caption{Postdiction of identified particle $\langle p_T \rangle$ at Maximum a Posteriori.}
        		\label{fig:full-study-MAP-pT}
        	\end{centering}
        \end{figure}

        The mean transverse momentum of identified particles, Fig.~\ref{fig:full-study-MAP-pT}, further reveals the success of the model-to-data comparison while demonstrating how overestimation of multiplicity combined with good estimation of the transverse momentum results in overestimation of transverse energy. The $\langle p_T \rangle$ shows less tension in the chemical makeup than previous results with the same hydrodynamic equation of state, revealing the role of bulk viscosity and a physically-motivated pre-equilibrium model with microscopic dynamics. The primary difference between Grad and Chapman-Enskog MAP calculations is in enhanced proton $\langle p_T \rangle$, in which the Chapman-Enskog MAP better reproduces the experimental results.
        
\begin{figure}[!htb]
        	\begin{centering}
        		\includegraphics[width=\columnwidth]{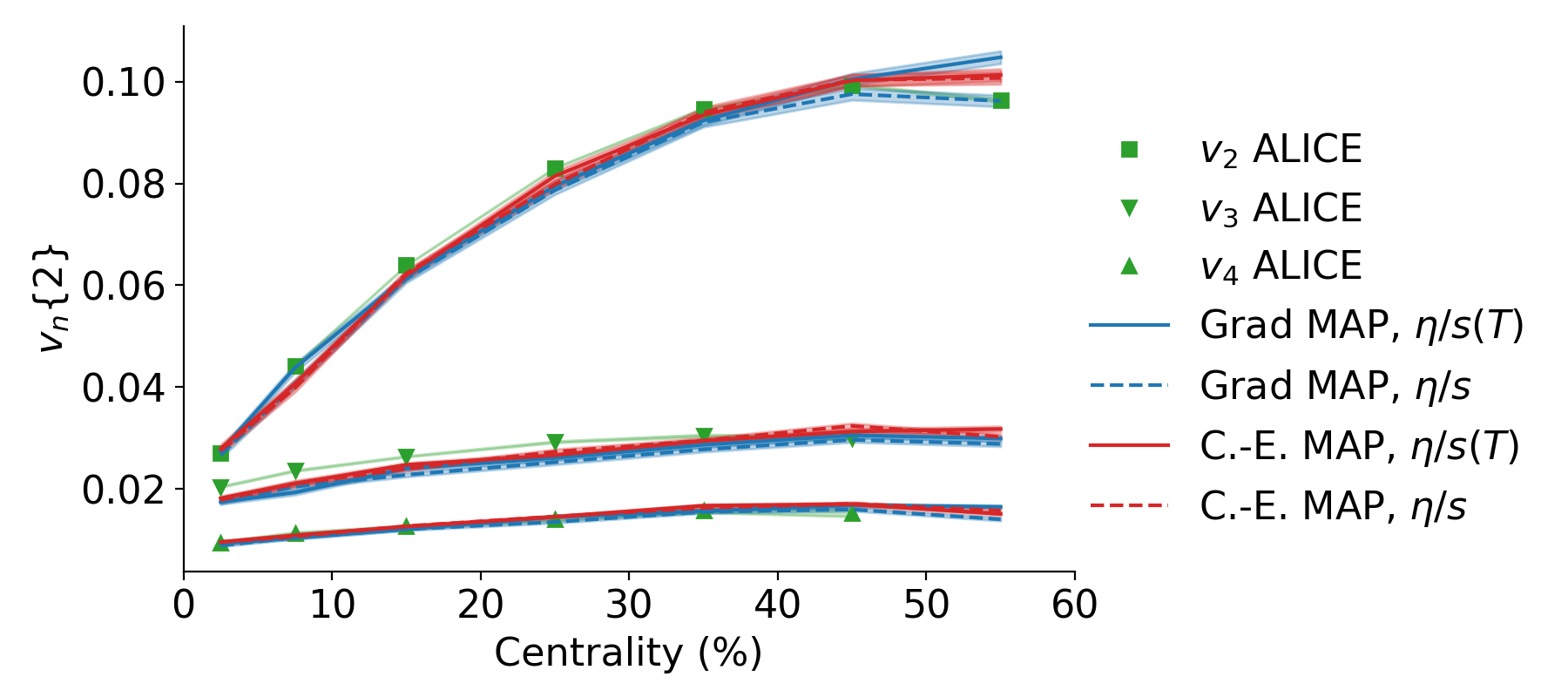}
        		\caption{Postdiction of $v_n\{2\}$ at Maximum a Posteriori.}
        		\label{fig:full-study-MAP-vn2}
        	\end{centering}
        \end{figure}

        The two-particle integrated $v_n$ further reveal good, albeit not perfect, reproduction of experimental results in Fig.~\ref{fig:full-study-MAP-vn2}. Notably, $v_2\{2\}$ and $v_4\{2\}$ are well described, particularly in central collisions, while $v_3\{2\}$ is underestimated. The underprediction of $v_3\{2\}$ is a feature of nearly every study and remains an object of continuing study. 
        
\begin{figure}[!htb]
        	\begin{centering}
        		\includegraphics[width=0.9\columnwidth]{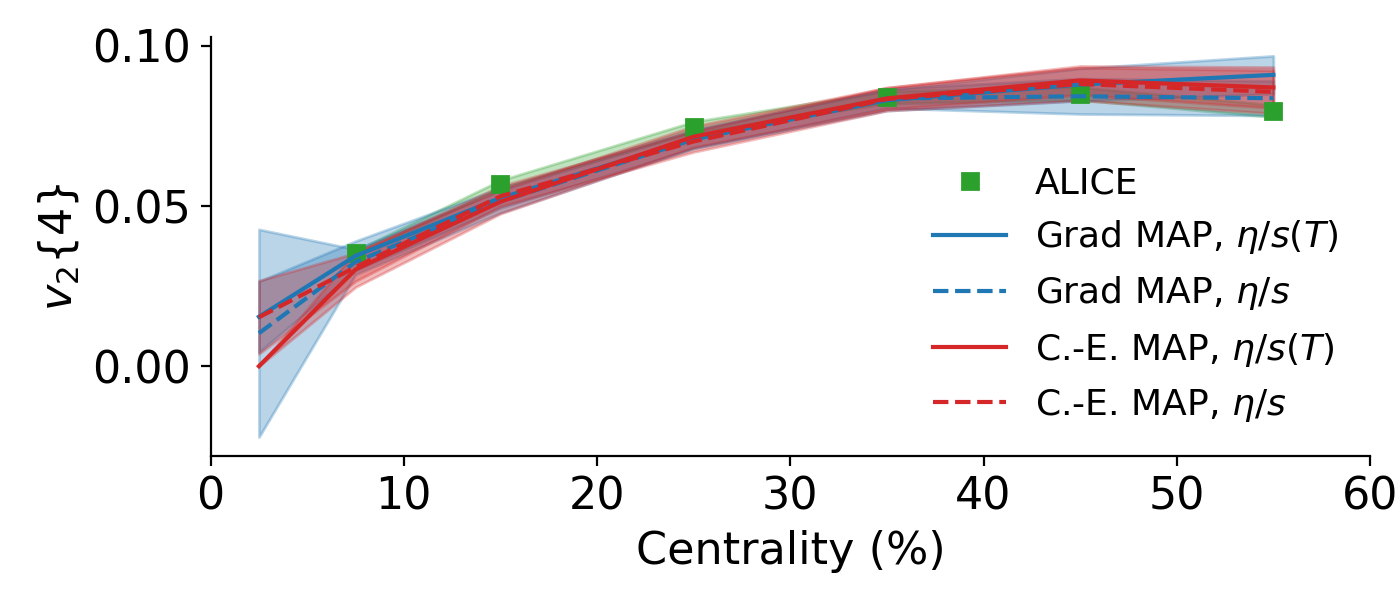}
        		\caption{Postdiction of $v_2\{4\}$ at Maximum a Posteriori.}
        		\label{fig:full-study-MAP-v24}
        	\end{centering}
        \end{figure}

        Peripheral $v_2\{2\}$ reveal that the MAP temperature dependence of the Grad shear viscosity results in an overestimate, while the constant shear more closely reproduces the experimental centrality dependence as do the Chapman-Enskog MAP calculations. For all $v_n\{2\}$, the MAP prediction of this study performs better than the previous state-of-the-art and the tension revealed here produces useful insight both into temperature-dependent $\eta/s$ and remaining progress required in describing the geometric fluctuations that drive $v_3\{2\}$. The four-particle integrated $v_2$ is shown in Fig.~\ref{fig:full-study-MAP-v24}, showing  agreement with data until the most peripheral bin where it is overestimated, consistent with the two-particle $v_2$ in Fig.~\ref{fig:full-study-MAP-vn2}, suggesting that these observables capture broadly similar physics and are similarly-well described by the model, although less tension is observed in $v_2\{4\}$ compared to $v_2\{2\}$.
        
\begin{figure}[!htb]
        	\begin{centering}
        		\includegraphics[width=\columnwidth]{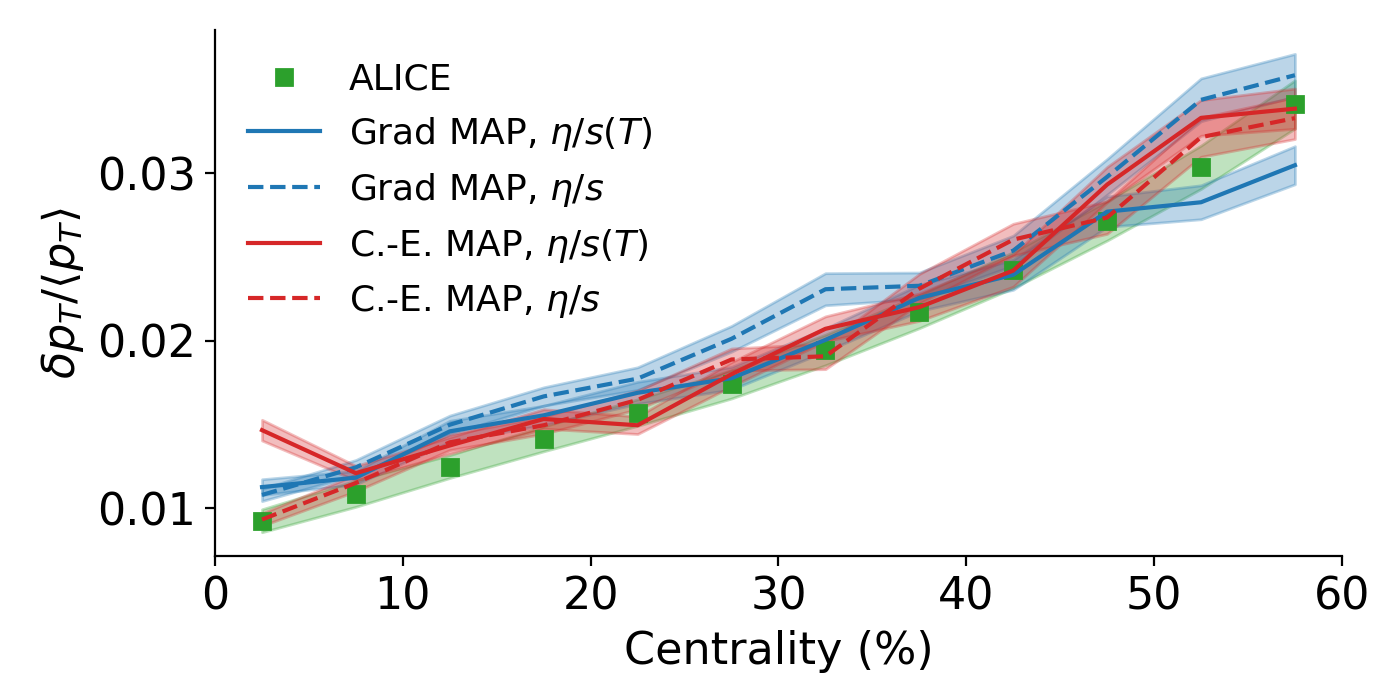}
        		\caption{Postdiction of $\delta p_T / \langle p_T \rangle$ at Maximum a Posteriori.}
        		\label{fig:full-study-MAP-pTfluct}
        	\end{centering}
        \end{figure}

        The correlated momentum fluctuations $\delta p_T / \langle p_T \rangle$, also denoted $\sqrt{C_m}/M$ or $\sqrt{C_m}/\langle p_T \rangle$, in Fig.~\ref{fig:full-study-MAP-pTfluct} are the first calculations to successfully describe this observable from a model with an IP-Glasma pre-equilibrium state and this description is consistent. This study is able to simultaneously describe both quantities with a variety of different viscosities and viscous corrections, resolving tension previously seen with the charged hadron multiplicity \cite{Schenke:2020uqq}. These fluctuations are also sensitive to the temperature dependence of the specific shear viscosity, where the constant $\eta/s$ systematically overestimates the data while correctly reproducing the centrality dependence (itself not seen in either other calculations with IP-Glasma or in the previous state-of-the-art), while the temperature-dependent $\eta/s$ better reproduces the data beginning in mid-central collisions. The Chapman-Enskog MAP reproduces the fluctuations more closely, save for the $\eta/s(T)$ calculation in the most central bin, which is likely the impact of statistical fluctuations.
        
\begin{figure}[!htb]
        	\begin{centering}
        		\includegraphics[width=\columnwidth]{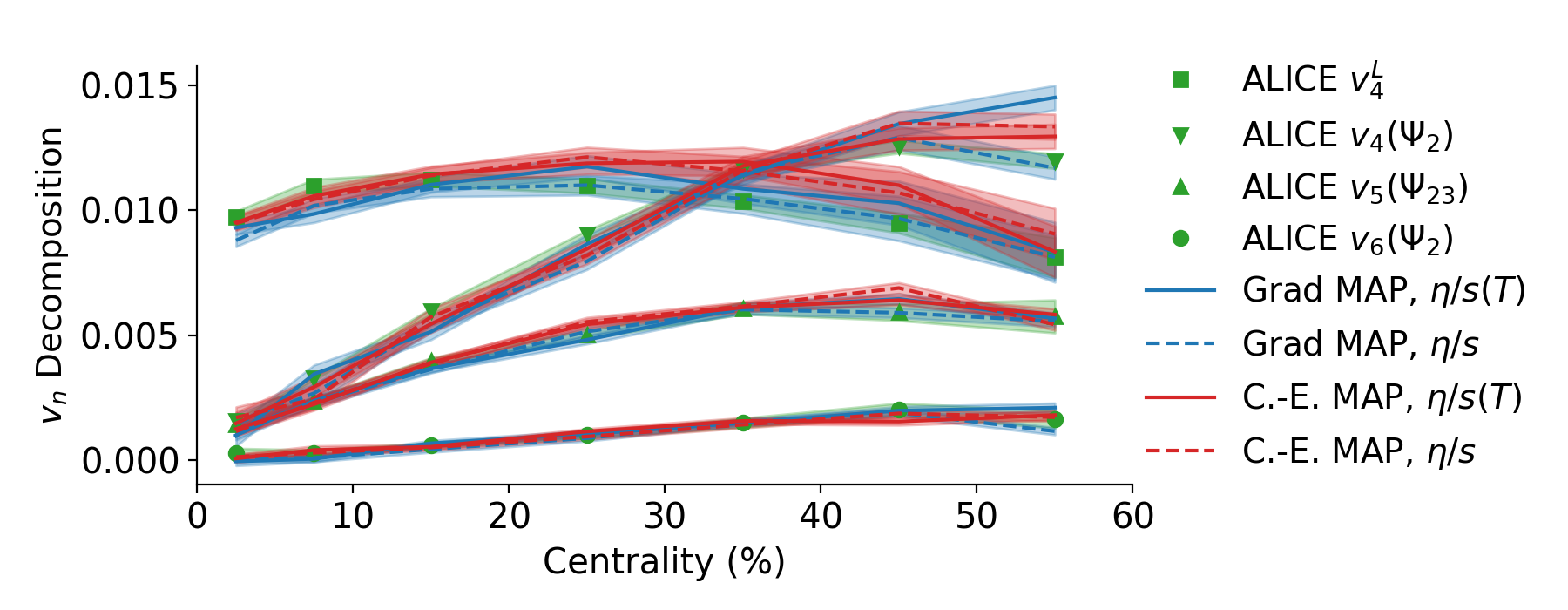}
        		\caption{Postdiction of the decomposition of $v_n$ at Maximum a Posteriori.}
        		\label{fig:full-study-MAP-vn-decomposition}
        	\end{centering}
        \end{figure}

        The decomposition of higher order $v_n$ further reveals the ability to simultaneously describe flow observables in Fig.~\ref{fig:full-study-MAP-vn-decomposition}. For every quantity other than central $v_4^L$, both models produce successful predictions of the experimental data, with the temperature-dependent $\eta/s$ again overpredicting peripheral flow as seen in  $v_4(\Psi_2)$. 
        Although is often consistent with the data within uncertainty, $v_4^L$ is overpredicted by the Chapman-Enskog MAP calculations. Nonetheless, this broad reproduction of the experimental flow decomposition suggests that the momentum-space geometry of the hybrid model successfully reproduces the physical picture in heavy ion collisions.
        
\begin{figure}[!htb]
        	\begin{centering}
        		\includegraphics[width=\columnwidth]{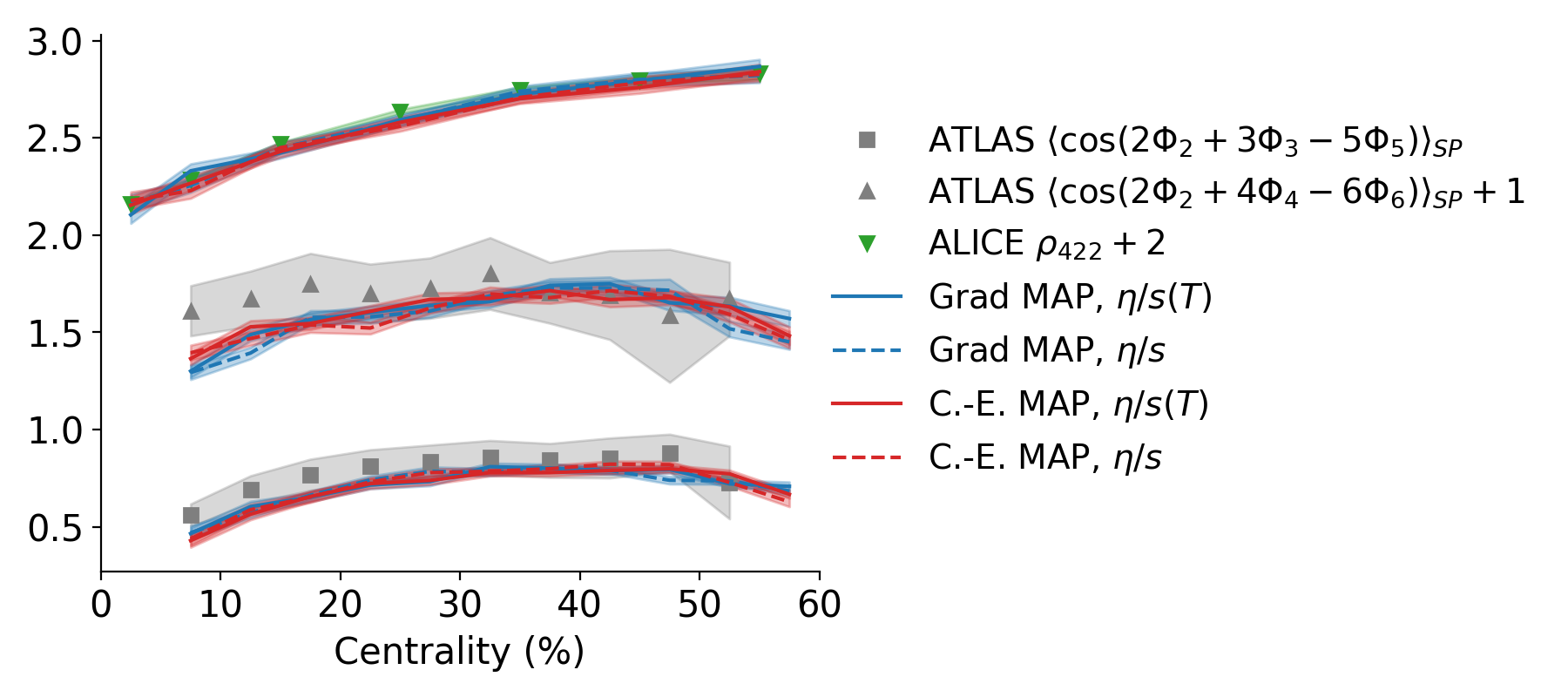}
        		\caption{Postdiction of event plane correlations at Maximum a Posteriori. Data and calculations are shifted for clarity.}
        		\label{fig:full-study-MAP-correlators}
        	\end{centering}
        \end{figure}

        The simultaneous reproduction of flow decomposition and event plane correlation constrains both the initial state geometry and the hydrodynamic evolution. In Fig.~\ref{fig:full-study-MAP-correlators}, the correlators are also well-described by the postdictions and are consistent with experimental uncertainty, save for central $\langle \cos(2 \Phi_2 +4 \Phi_4 -6 \Phi_6) \rangle$. The purely-even correlations are particularly well described and primarily relate the conversion of event planes of initial state geometry to momentum space via hydrodynamics. The mixed even-odd plane correlations reveal that the fluctuation structure is well described and correlates properly with even planes. 
        This postdiction is also well in line with the posterior predictive distributions, further supporting the accuracy of the surrogate modeling.
        
        The postdictions show that MAP parameter values are able to successfully describe the observables used in inference with never-before-seen accuracy for a multistage model with an IP-Glasma pre-equilibrium stage. This alone is a resounding success of the Bayesian inference in this study and conclusively demonstrates the performance of the Gaussian Process emulators as well as the study design. Tension is seen in the hadron chemistry, impacting the transverse energy, as well as in some of the description of the flow harmonics, notably $v_2\{2\}$ and $v_3\{2\}$. However, the decomposition of higher order flow is successful and the overwhelming majority of observables are well-described while the same tension is seen in $v_2\{4\}$, ensuring this effect is not a result of two-particle correlations. In the case of $\delta p_T / \langle p_T \rangle$, successful description is shown for the first time. The impact of viscous corrections is minimal, showing that the different posteriors are accurately accounting for differences in the underlying model calculations.
        
    \subsubsection{Predictions}
    
        Having established the power of the surrogate modeling and demonstrated successful description of a wide range of observables, it is time to turn to predictions of quantities not included in the calibration. Here, ``predictions'' is used to highlight that these observables were not used in systematic comparisons. As a result, the model is blind to these observables beyond information contained in other quantities. If models are differentiable at this stage, perhaps it can shed light on model quality not revealed in the more limited model-to-data comparison. In the following comparisons, centrality bins are chosen to match experimental results and predictions for bins not shown are simply due to dominance by theoretical uncertainty from a small number of events per bin.
        
\begin{figure}[!htb]
        	\begin{centering}
        		\includegraphics[width=\columnwidth]{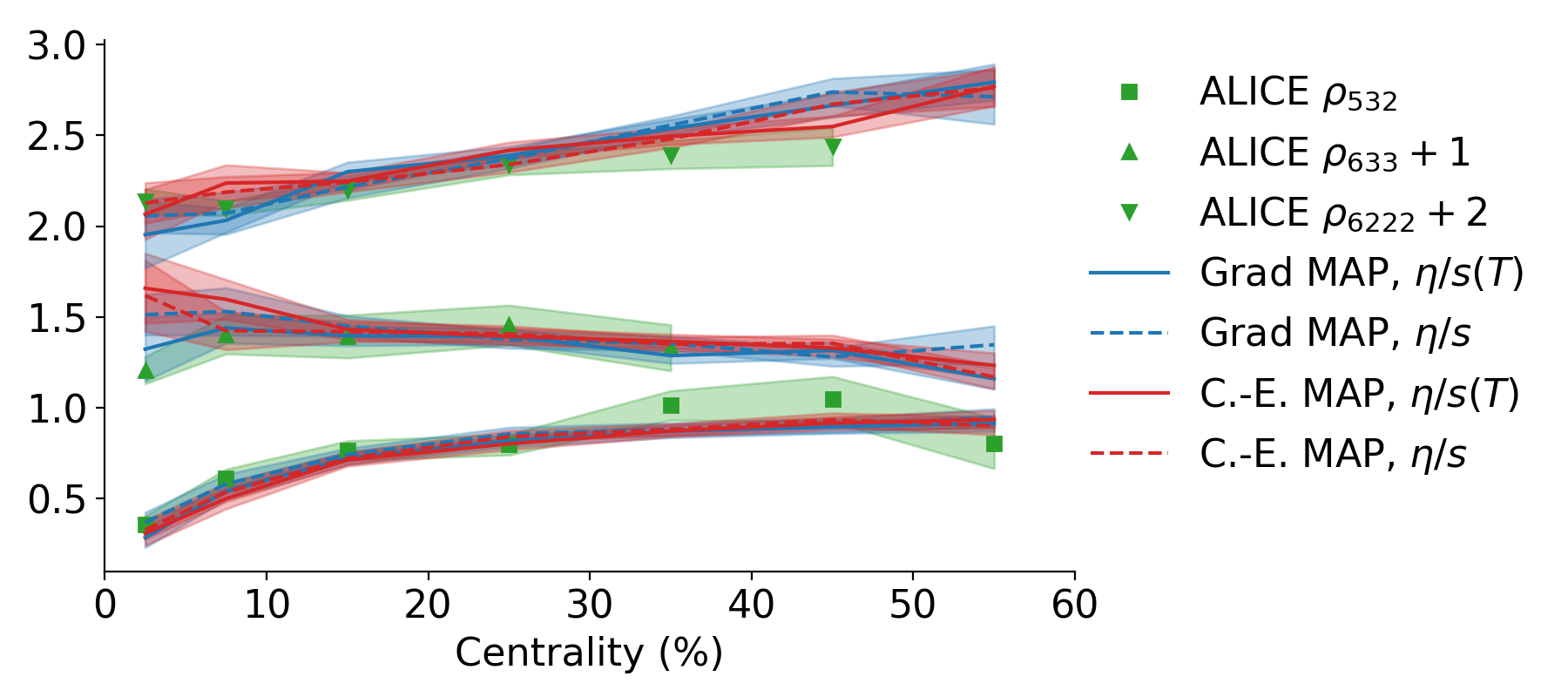}
        		\caption{Prediction of ALICE event plane correlations at Maximum a Posteriori \cite{Acharya:2017zfg}. Data and calculations are shifted for clarity.}
        		\label{fig:full-study-MAP-predictions-correlators-ALICE}
        	\end{centering}
        \end{figure}

\begin{figure}[!htb]
        	\begin{centering}
        		\includegraphics[width=\columnwidth]{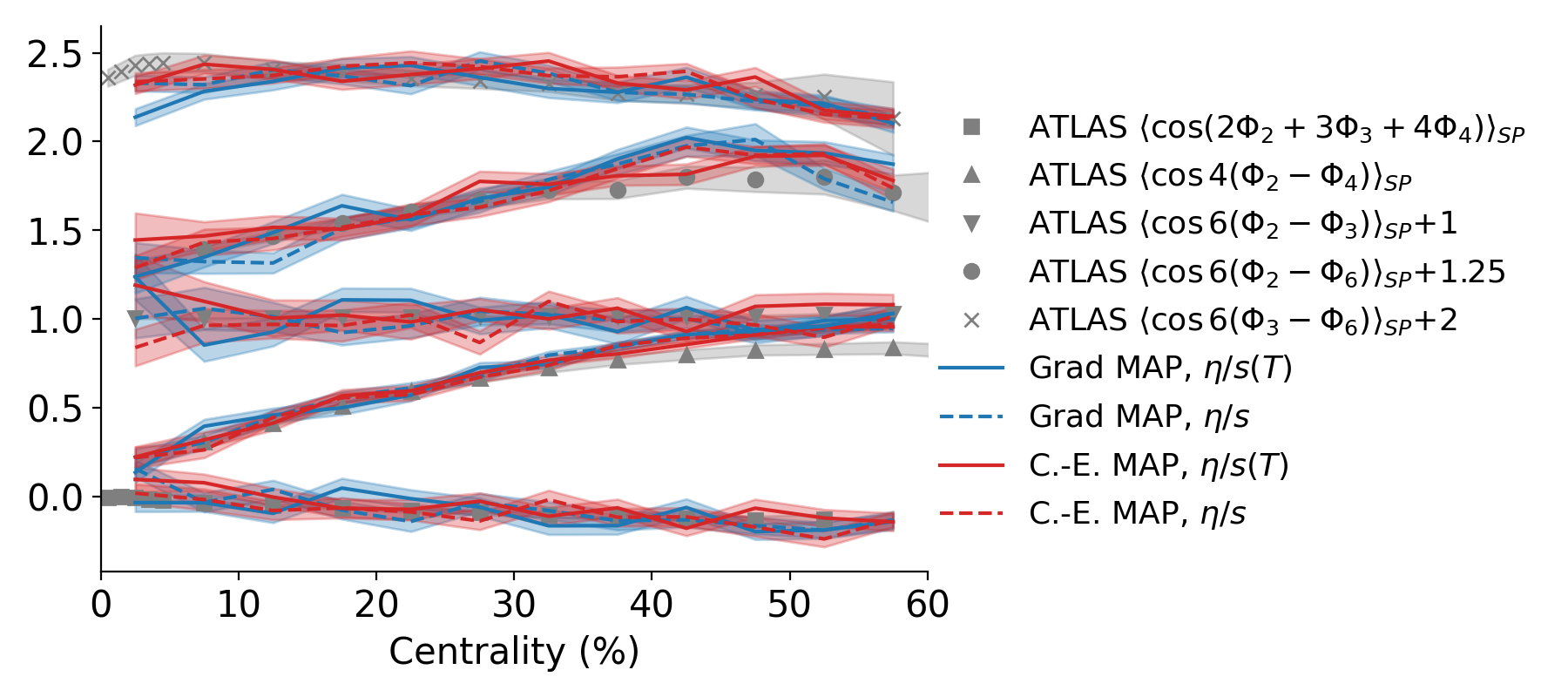}
        		\caption{Prediction of ATLAS event plane correlations at Maximum a Posteriori \cite{ATLAS:2014ndd}. Data and calculations are shifted for clarity.}
        		\label{fig:full-study-MAP-predictions-correlators-ATLAS}
        	\end{centering}
        \end{figure}

        The comparisons begin with measures of event plane correlation from ALICE in Fig.~\ref{fig:full-study-MAP-predictions-correlators-ALICE} and ATLAS in Fig.~\ref{fig:full-study-MAP-predictions-correlators-ATLAS}. In both cases, the model predictions are very well-aligned with experimental results. Both $\rho_{532}$ and $\rho_{633}$ are accurately predicted within experimental uncertainty, while $\rho_{6222}$ is accurately predicted below 30\% centrality. With respect to the ALICE measurements, the MAP calculations are broadly indistinguishable. A similarly indistinguishable picture is painted by comparison to ATLAS measurements, where $\langle \cos (2\Phi_2 + 3\Phi_3 + 4\Phi_4)\rangle_{SP}$, $\langle \cos 6(\Phi_2 - \Phi_3 )\rangle_{SP}$, and $\langle \cos 6(\Phi_3 - \Phi_3 )\rangle_{SP}$ are very well predicted by the model. Two predictions that perform somewhat less successfully above 30\% centrality are $\langle \cos 4(\Phi_2 - \Phi_4 )\rangle_{SP}$ and $\langle \cos 6(\Phi_2 - \Phi_6 )\rangle_{SP}$. The consistent picture drawn from comparison to both experiments suggests that the event planes produced by an IP-Glasma initial state are better suited to collisions below 30\% despite successful comparison to observables across the whole centrality range. These predictions outperform previous predictions made by a hybrid model with IP-Glasma \cite{McDonald:2016vlt}.

\begin{figure}[!htb]
        	\begin{centering}
        		\includegraphics[width=\columnwidth]{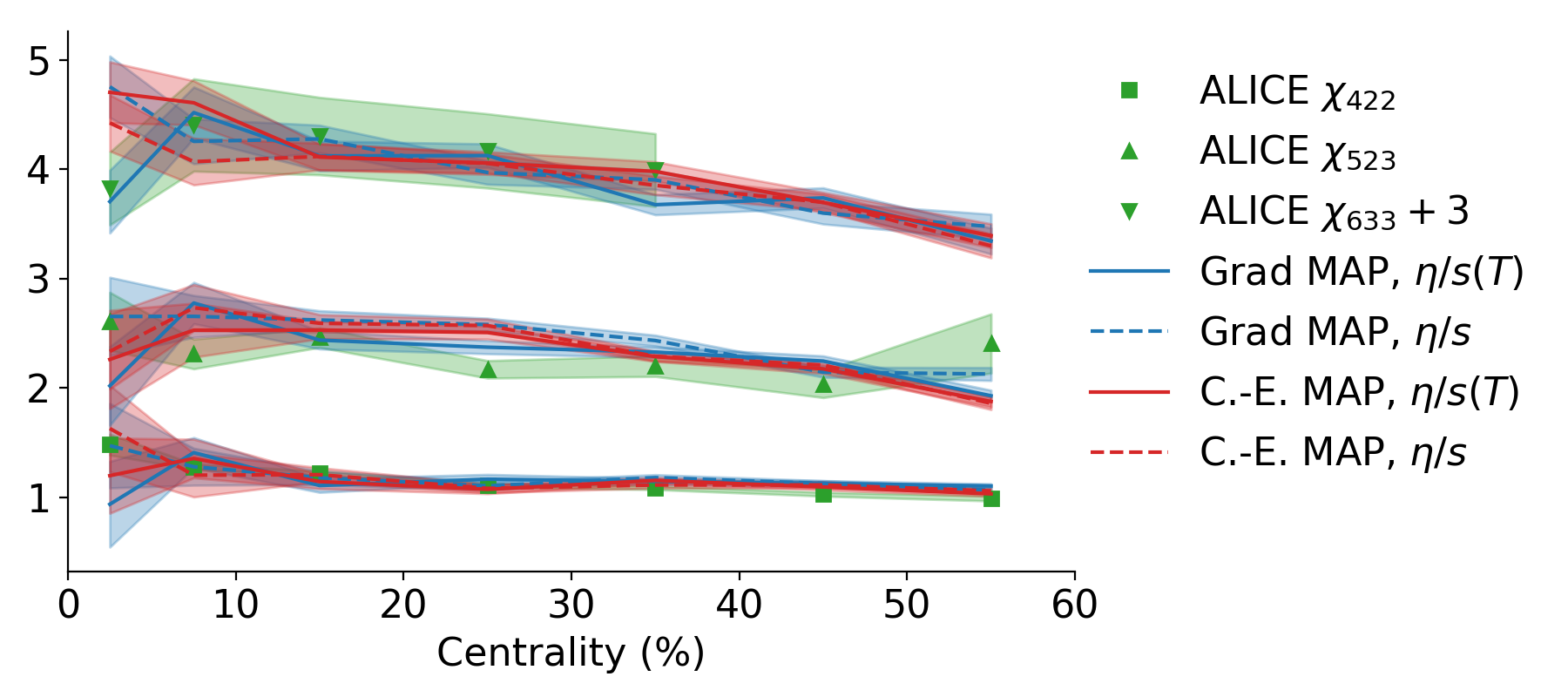}
        		\caption{Prediction of ALICE nonlinear response coefficients at Maximum a Posteriori \cite{Acharya:2017zfg}. Data and calculations are shifted for clarity.}
        		\label{fig:full-study-MAP-predictions-nonlinear-ALICE}
        	\end{centering}
        \end{figure}

\begin{figure}[!htb]
        	\begin{centering}
        		\includegraphics[width=\columnwidth]{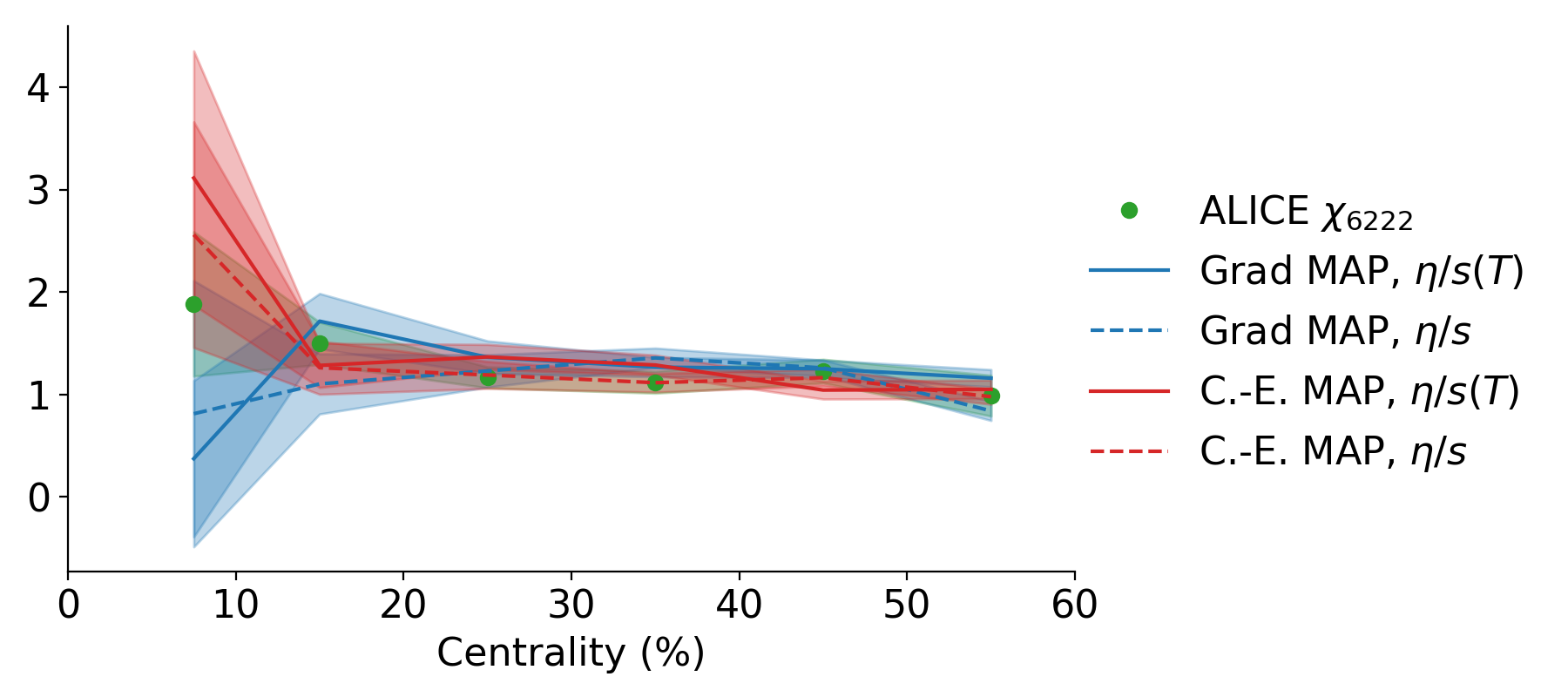}
        		\caption{Prediction of the ALICE $\chi_{6222}$ nonlinear response coefficients at Maximum a Posteriori \cite{Acharya:2017zfg}. Data and calculations are shifted for clarity.}
        		\label{fig:full-study-MAP-predictions-nonlinear-ALICE-6222}
        	\end{centering}
        \end{figure}

        A motivation for the use of IP-Glasma as a pre-equilibrium model was its success in simultaneous description of next generation observables, particularly both the event plane correlations and nonlinear response coefficients. With demonstrated success in prediction of event plane correlations not used in model-to-data comparison, predictions for nonlinear response coefficients are shown in Figs.~\ref{fig:full-study-MAP-predictions-nonlinear-ALICE} and \ref{fig:full-study-MAP-predictions-nonlinear-ALICE-6222}. These broadly describe the experimental results within experimental uncertainty, with slight overestimation in $\chi_{5,23}$ between 20 and 40\% centrality and peripheral $\chi_{4,22}$. In this case, the model with $\eta/s(T)$ slightly outperforms predictions with constant $\eta/s$, although they are often consistent within standard error. This demonstrates that a multistage model with an IP-Glasma pre-equilibrium stage is able to produce simultaneous, accurate predictions of the event plane correlations and hydrodynamic response with a initial geometry broadly fixed by low-energy nuclear correlations. This strongly suggests that the hydrodynamic phase is accurately described and the hydrodynamic response to geometry matches that seen in experiment. The centrality dependence is also often accurately captured, such as in $\chi_{5,23}$, which was not the case in previous calculations.
        
\begin{figure}[!htb]
        	\begin{centering}
        		\includegraphics[width=\columnwidth]{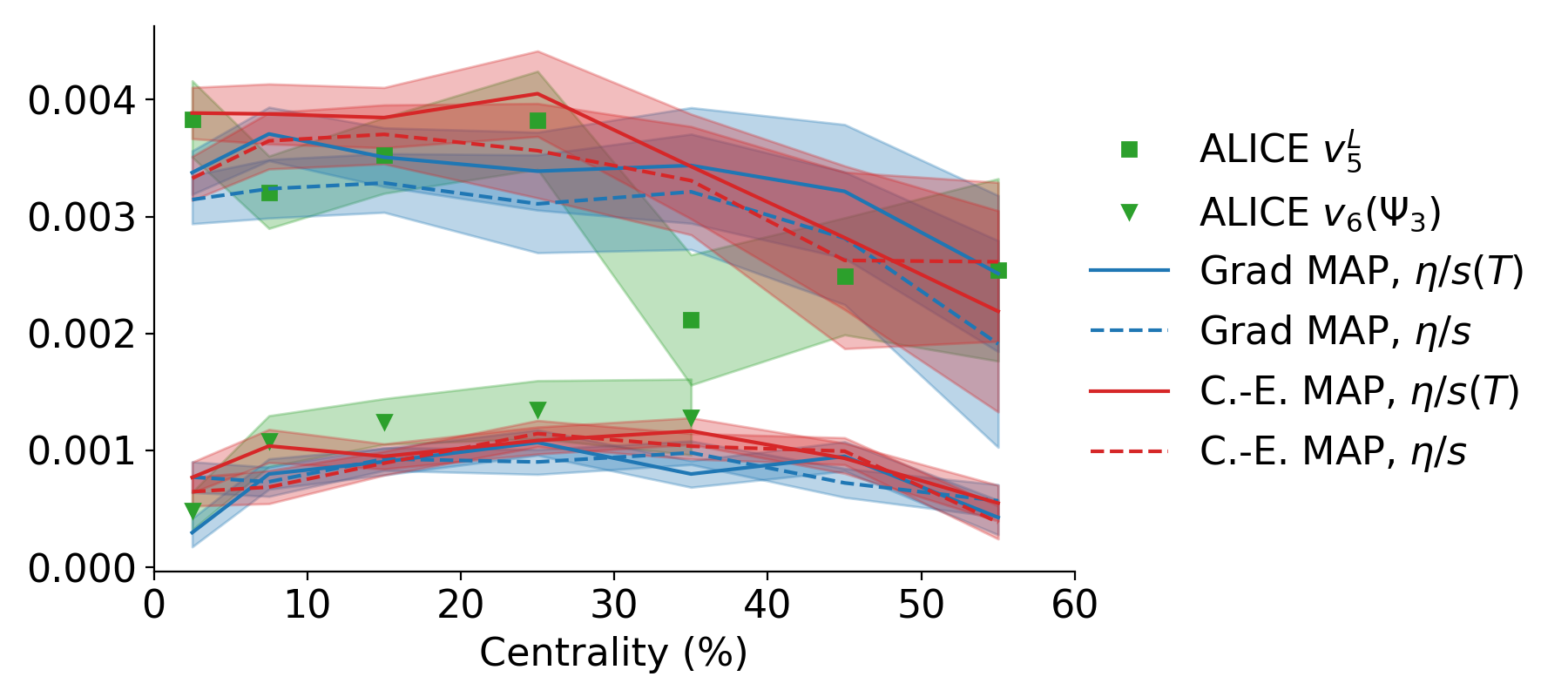}
        		\caption{Prediction of ALICE linear and nonlinear flow at Maximum a Posteriori \cite{Acharya:2017zfg}.}
        		\label{fig:full-study-MAP-predictions-vn-decomposition}
        	\end{centering}
        \end{figure}

        Predictions for the final category of observables used in the analysis are shown in Fig.~\ref{fig:full-study-MAP-predictions-vn-decomposition} for the linear  and nonlinear flow decomposition. These predictions are accurate and are clearly consistent with experimental results within uncertainties, save for $30-40\%$ $v_5^L$. This demonstrates the continuing success of the hybrid model with IP-Glasma as it is able to both describe and predict a wide range of observables. In the $v_5^L$ predictions, the constant $\eta/s$ prediction is more consistent with the experimental measurement, further supporting an inconclusive preference for one model over the other as the quality of predictions depends on which observable is considered.
        
\begin{figure}[!htb]
        	\begin{centering}
        		\includegraphics[width=\columnwidth]{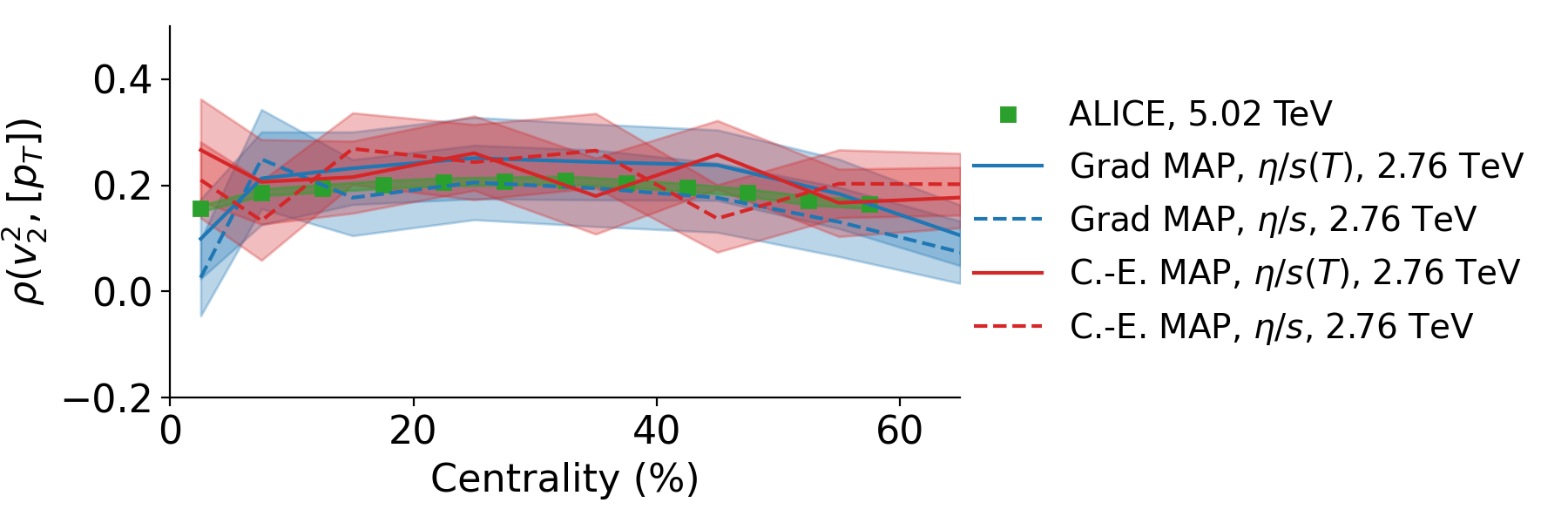}
        		\caption{Prediction of correlation between $v_2^2$ and $p_T$ at Maximum a Posteriori, compared to data from a higher-energy collision~\cite{ALICE:2021gxt} where uncertainties are shown by a shaded region. Note that data are at $\sqrts=5.02$ TeV while the MAP predictions are at $\sqrts=2.76$ TeV. }
        		\label{fig:full-study-MAP-predictions-rho2}
        	\end{centering}
        \end{figure}

        The final $p_T$-integrated prediction is made for the modified Pearson correlation between $v_2^2$ and $p_T$, shown in Fig.~\ref{fig:full-study-MAP-predictions-rho2}. As no experimental results at this energy are available, preliminary results for a higher Pb-Pb collision energy system ($\sqrts = 5.02$ TeV) are used ~\cite{ALICE:2021gxt} for comparison. The predictions made at $\sqrts = 2.76$ TeV describe the higher-energy data and its centrality dependence well. The current study has not utilized sub-nucleonic degrees of freedom.
        There is no significant difference seen between predictions with different viscous corrections or between $\eta/s$ and $\eta/s(T)$. Even with variation of the nucleon width in previous calculations of this quantity with IP-Glasma and T$_R$ENTo-based hybrid models, successful prediction of the value and centrality dependence has proved elusive. Hybrid models with T$_R$ENTo + freestreaming initial states, as well as previous calculations with IP-Glasma, have sign changes as they become increasingly peripheral. This feature is not seen in the data, nor in this prediction.  Based on the prediction in Fig.~\ref{fig:full-study-MAP-predictions-rho2}, there is no anticipated collision-energy dependence of this correlation and the IP-Glasma initial state at maximum a posteriori is able to successfully describe this observable. The lack of collision-energy dependence is supported by the comparison of Pb-Pb at $\sqrts = 5.02$ TeV data compared to Xe-Xe collisions at $\sqrts = 5.44$ TeV data from ALICE \cite{ALICE:2021gxt}. Of note is that it appears to not yield further constraint on the temperature dependence of $\eta/s$. Nonetheless, comparing it directly to the previous state-of-the-art Bayesian study using a T$_R$ENTo + freestreaming initial state, it appears  that the microscopic physics of the IP-Glasma pre-equilibrium stage plays an important role. This represents a true prediction as data at $\sqrts=2.76$ TeV has yet to be published.
        
        Up to this point, only $p_T$-integrated observables have been considered. Differential observables also exist and provide interesting and discriminating probes of the soft sector. However, the boundary between the soft sector and the hard sector (such as jets and jet-medium interactions) is unclear. By considering the integrated quantities up to now, the sensitivity of the inference to the precise location of this boundary is reduced and predictions can be made. This sensitivity is reduced because integrated observables are weighted by the multiplicity, which drops exponentially. By considering each differential $p_T$ bin, this exponentially-decreasing weighting would be removed and each bin would be treated on an equal footing, in turn giving the bins on the boundary of the soft and hard sectors a higher proportional weighting.
        
\begin{figure}[!htb]
        	\begin{centering}
        		\includegraphics[width=\columnwidth]{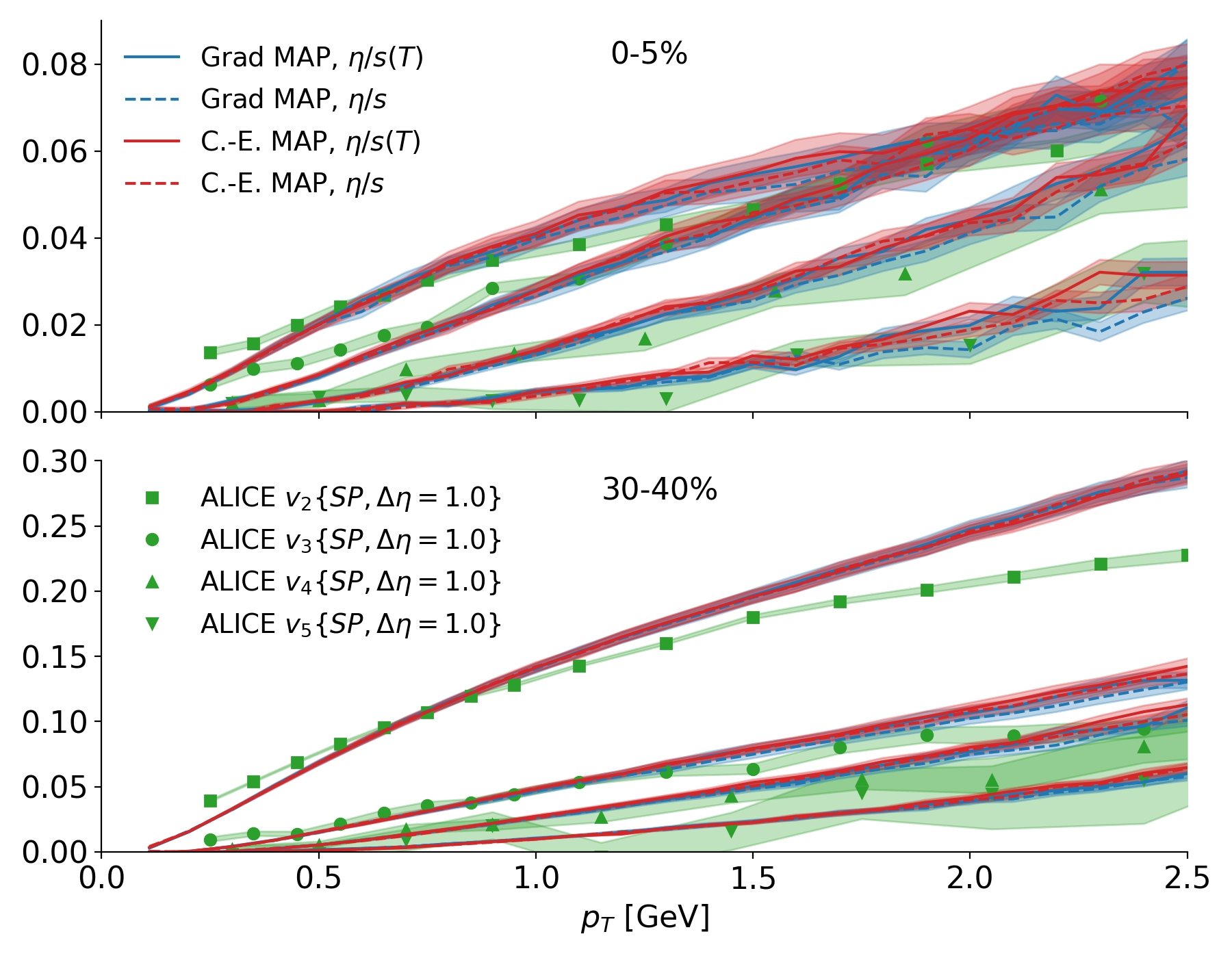}
        		\caption{Prediction of differential $v_n\{2\}$ at Maximum a Posteriori for the $0-5$\% centrality bin (upper panel) and the $30-40\%$ centrality bin (lower panel).}
        		\label{fig:full-study-MAP-predictions-vndiff}
        	\end{centering}
        \end{figure}

        The first differential observable investigated is the differential charged hadron $v_n\{2\}$, with predictions shown in Fig.~\ref{fig:full-study-MAP-predictions-vndiff} compared to experimental measurements from ALICE \cite{ALICE:2011ab}. Tension is clearly present in reproducing the spectra, with predictions from integrated observables often undershooting at lower transverse momentum and overshooting at higher momenta. Nonetheless, the majority of predictions are consistent with experimental measurements for the first time or the distance from the prediction to measurement has been greatly reduced from the previous IP-Glasma state-of-the-art \cite{McDonald:2016vlt}. The greatest tension is observed in the differential $v_2\{2\}$ in the $0-5\%$ and $30-40\%$ centrality bins and low-$p_T$ $v_3\{2\}$ in more peripheral collisions. This low-momentum region is expected to be the region best described by hydrodynamics, suggesting that relevant physics remains missing from the hybrid model. As $v_3\{2\}$ is primarily fluctuation driven, this suggests that fluctuation structure is missing. The underestimate of $v_2\{2\}$ in contrast suggests that a geometric aspect is not included or an aspect of the conversion between position-space and momentum-space geometry remains incomplete. This is not necessarily a concern for the validity of the hydrodynamic description, as the higher-order differential $v_n$ are well-described, but instead suggests that additional physics may be at play. Recent works including the differential momentum spectra suggest that their inclusion in systematic model-to-data comparison can yield insight, but various analysis errors and inclusion of momentum bins in regions where unincluded physics is relevant hinders the interpretation of results \cite{Nijs:2020ors, Nijs:2020roc}. The posterior predictive distribution, rather than single MAP predictions, may provide more insight into the present apparent mismatch of the model predictions and data.
        
\begin{figure}[!htb]
        	\begin{centering}
        		\includegraphics[width=\columnwidth]{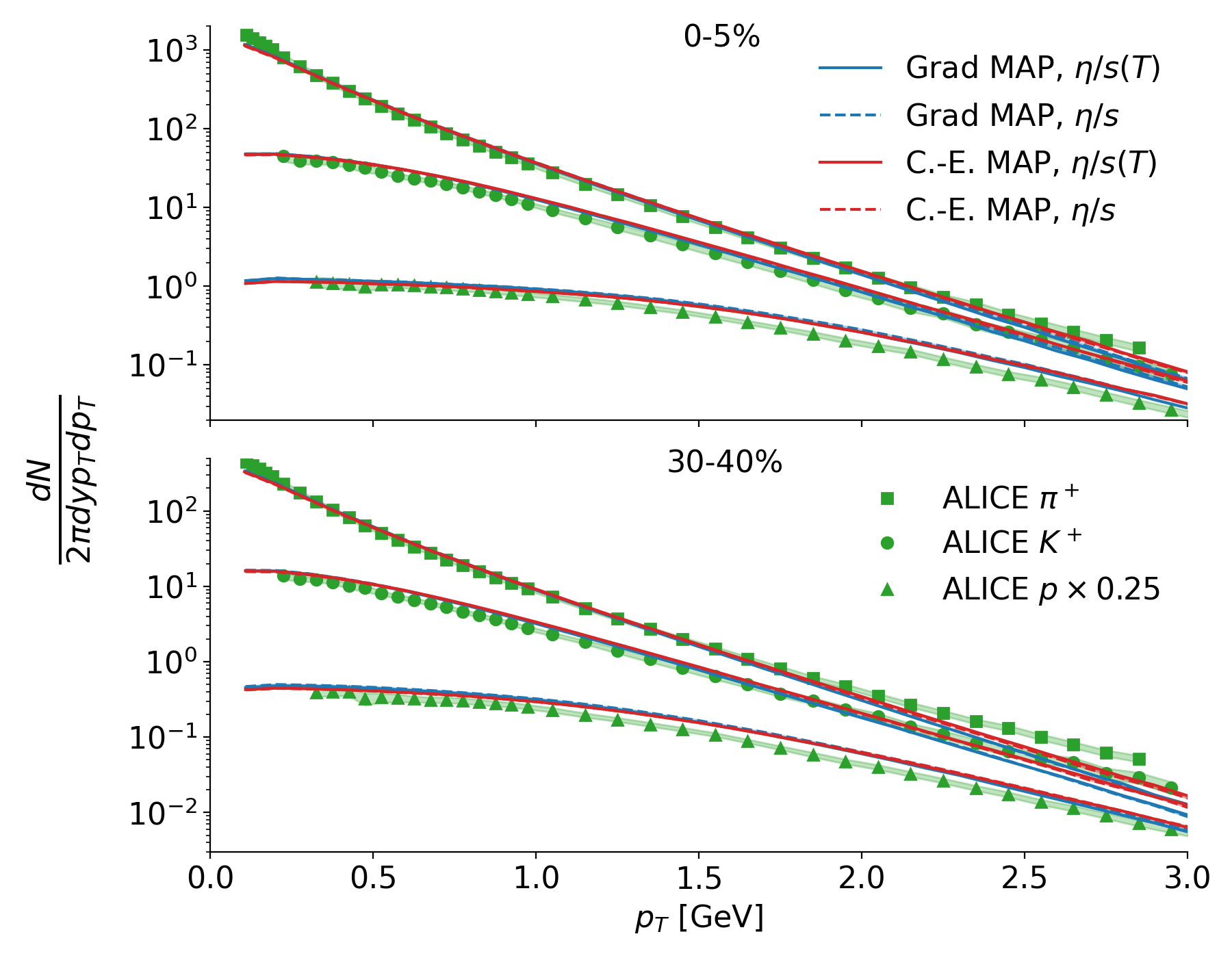}
        		\caption{Prediction of differential light hadron multiplicity spectra at Maximum a Posteriori for the $0-5$\% centrality bin (upper panel) and the $30-40\%$ centrality bin (lower panel).}
        		\label{fig:full-study-MAP-predictions-dNdiff}
        	\end{centering}
        \end{figure}

        The light hadron multiplicity spectra, shown in Fig.~\ref{fig:full-study-MAP-predictions-dNdiff} for selected central and mid-central centrality bins, paints a complementary picture to that of the integrated multiplicity in Fig.~\ref{fig:full-study-MAP-dNdy}. The integrated proton and kaon multiplicity were overestimated, while the pion multiplicity was slightly underestimated; the same is found here. The momentum dependence of the spectra, however, remains well-predicted until the higher momentum region ($p_T > 1.75$ GeV), where mini-jets, jet showers, and other hard-sector considerations begin to gain relevance. Beginning in this region, all the identified light hadrons are underpredicted. To include these additional effects is a matter of ongoing theoretical effort and is beyond the scope of this investigation.
        
        Postdictions and predictions using four MAP calculations have been shown, comparing Grad and Chapman-Enskog viscous corrections with and without temperature-dependent shear viscosity. The inconclusive preference between viscous correction models is consistent when comparing MAP parameter sets, as is the inconclusive preference for or against temperature-dependent $\eta/s$, in keeping with the Bayesian model comparison.  
        
\section{Bayesian Model Selection}
\label{BMS}
    
    Bayesian model comparison can be used to determine if data exhibit a preference for one model or another, if additional complexity is justified by the model, or even if the model can differentiate between pseudodata and experimental data. This is extremely valuable as it does not attempt to falsify a model, but rather puts it to a binary test to determine which model is the most useful in describing the data.
     
    To test, as always, with self-consistency, the first use of Bayesian Model Comparison is to determine if the model can differentiate between pseudodata used for the previous self-consistency testing and experimental data. This hypothesizes the following scenario: a ``true'' model underlies the experimental data just as a known model underlies the pseudodata generated to test self-consistency. A distinct model is never expected to systematically defeat the true underlying model and, if it did, would be a sign of systematic bias. As a result, the Bayes evidence for the pseudodata is expected to be greater than the Bayes evidence for the experimental data and strong preference is expected from the Bayes factor. This is found when comparing the model estimate of the Bayes evidence for pseudodata and true data using Grad viscous corrections: the natural logarithm of the Bayes Factor ($\ln B$) determining which data the model is best suited to ranges\footnote{There is a spread because pseudodata from different points in parameter space have different Bayes factors.} from $118.7 \pm 3.1$ to $147.9 \pm 2.3$ in favor of the pseudodata, corresponding to odds of around $2 \times 10^{56}:1$ to $10^{62}:1$ differentiating the two data sources. Comparable, albeit slightly reduced preference is found using Chapman-Enskog viscous corrections ($\ln B \approx 90 \pm 5)$. This is an overwhelming validation of the model's ability to differentiate the data and demonstrates the self-consistency of the Bayesian model selection. It remains a sobering revelation of just how much information is not yet captured by the model. 
    
    The self-consistent Bayesian model comparison can now be used to determine if the model exhibits a preference for a variety of features. For example, by fixing the high and low temperature slopes to 0 (and fixing the kink temperature to any value in the prior range, since it is meaningless with no change in slope), Bayesian model comparison can be used to test if the model demands a temperature-dependent shear viscosity in the two viscous correction models considered. With Grad viscous corrections, performing this comparison yields  $\ln B = 0.3 \pm 0.2$ in favor of temperature-dependent shear viscosity; the Bayes factor is $\ln B = 0.9 \pm 0.4$ when Chapman-Enskog viscous corrections are used instead of Grad. On the Jeffreys' scale (Table ~\ref{tab:jeffreys}), which provides an odds-based scale for Bayesian model comparison, these are both consistent with no preference. The evidence is thus inconclusive \emph{in favor or against} temperature-dependent $\eta/s$ given the data considered in this study.  The same conclusion had been reached in Ref.~\cite{SIMSPRC}. This suggests that further studies are required to conclusively demonstrate the temperature dependence (or lack thereof) of $\eta/s$ in heavy ion collisions. This inconclusiveness arises from the balance between better explanation of the data and a penalty for increased complexity in the form of additional model parameters.

    \begin{table}[htb!]
	\centering
	\begin{tabular}{rrll}
		\hline$\hspace*{0.3cm}\left|\ln B_{01}\right|$ & Odds & Probability & Strength of evidence \\
		\hline$<1.0$\hspace*{0.25cm} & $\leq 3: 1$ & $<0.750$ & Inconclusive \\
		1.0\hspace*{0.25cm} & $\approx 3: 1$ & \hspace*{0.35cm}0.750 & Weak evidence \\
		2.5\hspace*{0.25cm} & $\approx 12: 1$ & \hspace*{0.35cm}0.923 & Moderate evidence \\
		5.0\hspace*{0.25cm} & $\approx 150: 1$ & \hspace*{0.35cm}0.993 & Strong evidence \\
		\hline
	\end{tabular}
	\caption{The Jeffreys' Scale, reproduced from \cite{BayesintheSky}.}
	\label{tab:jeffreys}
\end{table}
    
    The lack of such preference for or against $\eta/s(T)$ is not surprising. Hybrid models with IP-Glasma have demonstrated considerable success in describing experimental results using a constant specific shear viscosity and the viscous posteriors in this study are themselves consistent with a constant value. In the study requiring a constant $\eta/s$, the result is well-constrained  -- $\eta/s = 0.137^{+0.025}_{-0.028}$ for Grad viscous corrections and $\eta/s = 0.125^{+0.021}_{-0.022}$ for Chapman-Enskog viscous corrections, where the uncertainty denotes the 95\% C.I. -- and with minimal covariance. By inspection, it is apparent that this is entirely consistent with the $\eta/s(T)$ posteriors in Fig.~\ref{fig:full-study-grad-and-CE-data-posterior} and nearly spans the full width at the narrowest point.
    
    As many Bayesian works require $\eta/s(T)$ to strictly increase or be constant below a fixed kink temperature, this is also a useful comparison and is performed with only Grad viscous corrections as both models are consistently in agreement. To do this, $a_{\eta,low}$ is fixed to zero as it is in those studies and $T_{\eta,kink}$ is fixed to 0.154 GeV. Finally, $a_{\eta,high}$'s prior range is reduced to require it to be positive definite. Comparing the evidence for this configuration to the full study produces $\ln B = 3.8 \pm 2.6$ in favor of the full study allowing for a negatively-sloped $\eta/s(T)$. This corresponds to moderate-to-strong evidence on the Jeffreys' Scale in Table~\ref{tab:jeffreys}. Comparing the requirement of a positive-definite slope to $\eta/s(T)$ to a constant $\eta/s$, the Bayes factor is $\ln B = 3.6 \pm 2.6$ in favor of the constant specific shear viscosity. Because the Bayes factor penalizes complexity, the additional complexity is not justified by the data.
    
    Next generation observables are employed in this study in the hope of determining the features of $\eta/s$ and $\zeta/s$ with greater accuracy and precision. Some studies use next generation correlations that require much greater computational expenditure to attempt to find this constraint, but use parametric initial conditions \cite{Parkkila:2021yha}. A conclusion from these Bayesian model comparisons is that success in learning the physical specific viscosity of strongly-interacting matter will only come from combining realistic initial conditions and well-chosen observables. A promising candidate for increased constraint are $v_n-p_T$ correlations, which are not readily calculable at the precision of this study, but further couple pre-equilibrium geometry to the hydrodynamic evolution \cite{Giacalone:2020dln, Giacalone:2021clp}. This is investigated later in this work as a prediction made at Maximum a Posteriori.
     
    Recent Bayesian works with a T$_R$ENTo + freestreaming initial state have been finding success with small specific bulk viscosity \cite{Nijs:2020ors, Nijs:2020roc, Parkkila:2021yha}, contrasting with prior non-Bayesian studies using IP-Glasma initial conditions that indicated a need for a significant $\zeta/s$ to reproduce hadronic observables. 
    By fixing $(\zeta/s)_{max}$ to zero and holding the other parameters fixed to arbitrary values as they no longer have any impact, it is straightforward to assess the demand for nonzero $\zeta/s$. This comparison results in $\ln B = 34.4 \pm 2.4$ in favor of non-zero $\zeta/s$ when using Grad viscous corrections, corresponding to odds of $\approx 8 \times 10^{14}:1$. With Chapman-Enskog viscous corrections, this preference for the inclusion of bulk viscosity increases to $\ln B = 61 \pm 5$, conclusively demonstrating that bulk viscosity is strongly justified when using IP-Glasma initial conditions, no matter the viscous corrections at particlization. The physical impacts of the lack of bulk viscosity are an enhancement of the identified particle  $\langle p_T \rangle$ and the momentum fluctuations $\delta p_T / \langle p_T \rangle$ with simultaneous suppression of $v_3\{2\}$ and the three-plane correlators. The particlization temperature is also forced to the highest possible temperature allowed in the prior while the hydrodynamic initialization time is required to be as short as possible. This arises from a need to preserve as many initial-state fluctuations as possible as they must reproduce fluctuation-driven final-state observables. The high particlization temperature additionally preserves fluctuations by allowing for less viscous dissipation in the hydrodynamic phase.
    
    Comparing the relative likelihood of the viscous correction models is a useful way to assess model applicability and begin to quantify the uncertainty introduced by the choice of viscous correction. Comparing Grad and Chapman-Enskog viscous corrections to data with none of the parameters held fixed, the relative preference for the Grad over the Chapman-Enskog RTA viscous corrections is $\ln B = -0.2 \pm 0.3$ in imperceptibly-slight favor of Grad viscous correction, although this should be interpreted as the models being indistinguishable in this analysis as it does not rise to the level of even weak evidence. When we hold $\eta/s$ to be constant, the relative preference for Grad over Chapman-Enskog RTA viscous corrections is $\ln B = 0.5 \pm 0.3$: again, no preference between the two models even rises to weak evidence. 
    
    This indistinguishable nature of the viscous correction models deserves further study. The posteriors, as shown previously, are quite similar but not identical, and are equally well-suited to experimental measurements. As a result, the viscous corrections chosen in a study are an important source of theoretical uncertainty to quantify and not doing so results in an artificially precise posterior. Progress in adding additional constraining observables must not neglect quantification of uncertainty as a parallel goal lest analyses fall into the trap of the bias-variance tradeoff. The goal is not to constrain these quantities the most precisely, but to do so both accurately \emph{and} precisely. By not including sources of theoretical uncertainties, an analysis focuses on the latter and sacrifices the former.
    
    Model preference between the Grad and Chapman-Enskog viscous corrections was seen to be strong in previous work \cite{SIMSPRL, SIMSPRC}, but are indeterminate in this study. In the earlier studies, the bulk viscosity was larger at particlization and as a result, enhanced the effect of the corrections. A role may also be played by the more realistic initial state model. 
    
     As the Bayesian model comparison exhibits no preference for or against temperature-dependent specific shear viscosity, estimates of the MAP are provided for both temperature dependence and a lack thereof in Table~\ref{tab:MAP-Grad-and-CE}.

    \subsection{Bayesian Model Averaging}
   
        In Bayesian model comparison, the question under investigation is ``which model is best suited to the data?'' This informs which model to use and how best to use it. A related question is ``given two models, how does one best estimate the truth?'' For this, Bayesian model averaging (BMA) is employed. In Bayesian model averaging, two posteriors are combined using a weighted average in which the weights are the Bayes evidence \cite{fragoso2018bayesian}. In a simplified example, if two models are equally likely, then the truth is most likely to be in the region where the model posteriors overlap. This is formalized as
        \begin{equation}
        	p_{BMA}(x | y) \propto \sum_i p_i(y) p_i(x|y) \label{eq:BMA}
        \end{equation}
        for models indexed $i$.
        
\begin{figure*}[!htb]
        	\begin{centering}
        		\includegraphics[width=1.5\columnwidth]{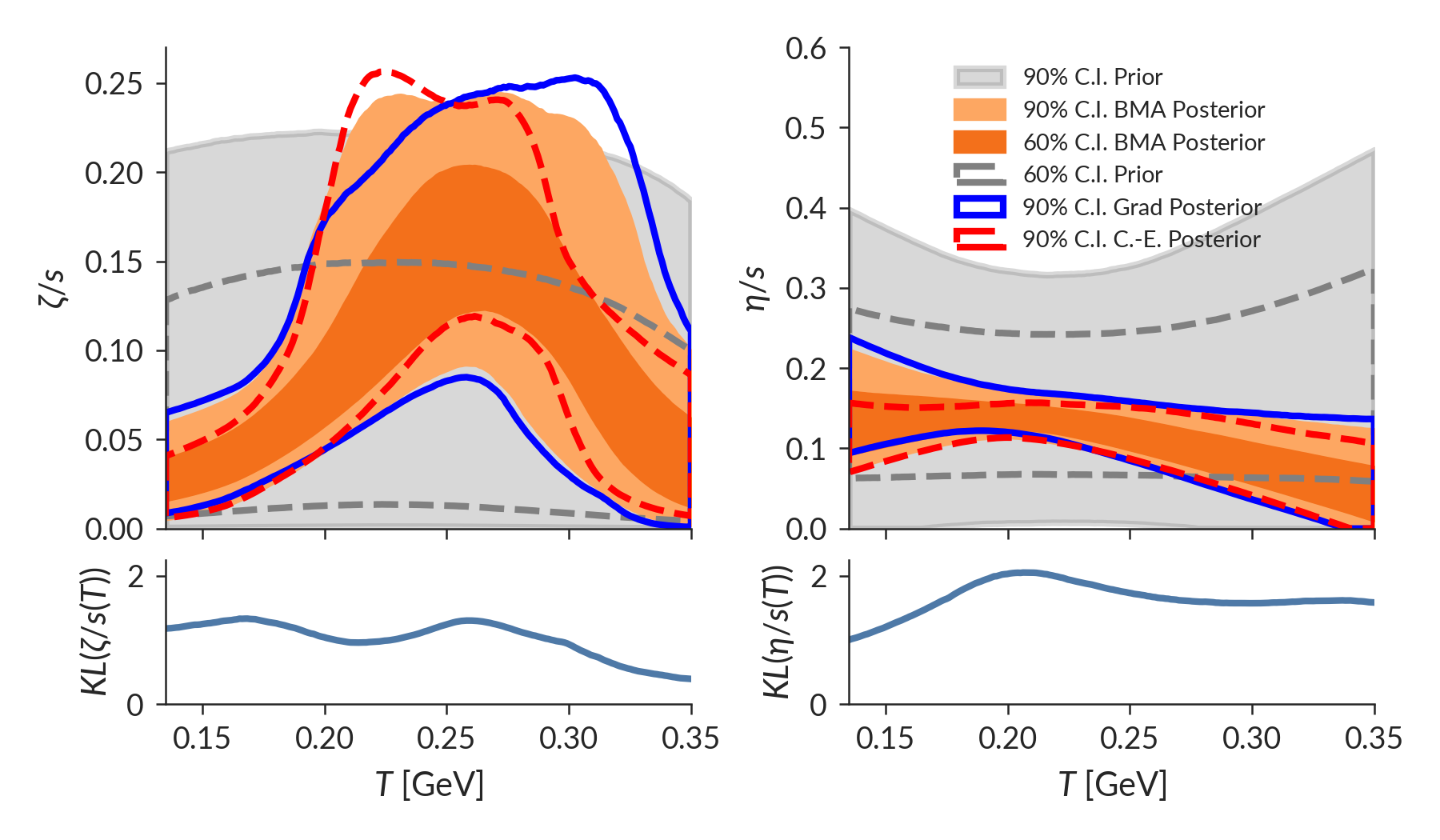}
        		\caption{Bayes Model Averaged viscous posterior shown with Grad 90\% credible interval (blue) and Chapman-Enskog 90\% credible interval (red) and the Kullback-Leibler Divergence quantifying information gain from the priors to the BMA posterior in bits (bottom panels).}
        		\label{fig:full-study-BMA-viscous-posterior}
        	\end{centering}
        \end{figure*}

\begin{figure}[!htb]
        	\begin{centering}
        		\includegraphics[width=\columnwidth]{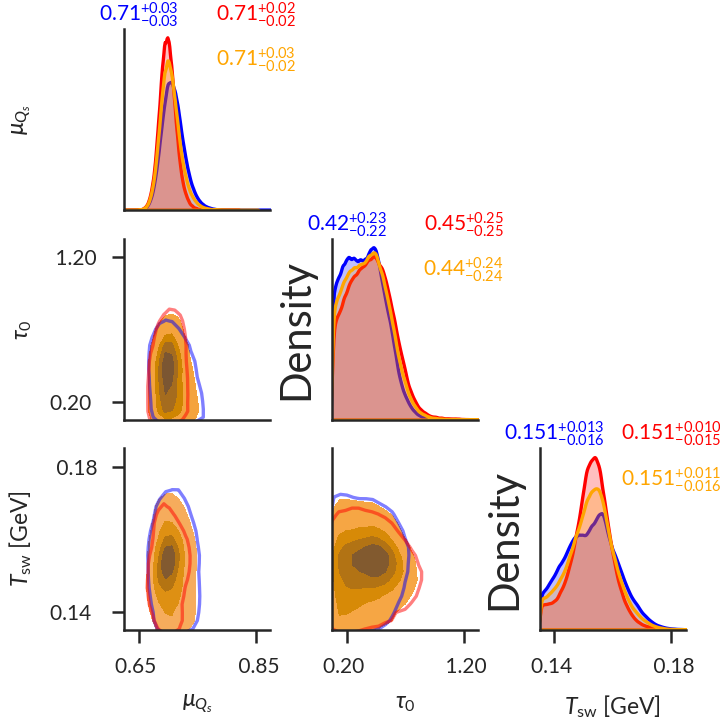}
        		\caption{Bayes Model Averaged posterior for non-viscous parameters (orange) shown with Grad (blue) and Chapman-Enskog (red). The lowest contour shown is the $5^{th}$ percentile.}
        		\label{fig:full-study-BMA-posterior}
        	\end{centering}
        \end{figure}

        Bayesian model averaging was first used in heavy ion collisions to perform model averaging of the transport coefficients and later for model averaging of non-viscous parameters \cite{SIMSPRL, Everett:2021ruv}. The BMA viscous posteriors are shown in Fig.~\ref{fig:full-study-BMA-viscous-posterior} along with the Kullback-Leibler divergence, which quantifies the distance between two distributions and is used here to calculate the information gained from the prior to the BMA posterior \cite{10.1214/aoms/1177729694}. The BMA posterior for non-viscous parameters is shown in Fig.~\ref{fig:full-study-BMA-posterior}.
        
        The BMA viscous posterior clearly demonstrates the value of accounting for the uncertainty due to viscous corrections at particlization by showing the state of knowledge by considering both simultaneously. 
        The two models contribute their constraint throughout the temperature evolution of both $\zeta/s$ and $\eta/s$, although the impact is clearer in the specific bulk viscosity due to the differences in constraint between the two models. Particularly of interest is that BMA leverages the information content of both models to produce a more-constrained 60\% C.I. than either model independently, demonstrating how to address the bias-variance tradeoff with multiple models in a rigorous way.
        
        The Kullback-Leibler divergence in Fig.~\ref{fig:full-study-BMA-viscous-posterior} is non-zero for the entire temperature range shown on the figure.
        This contrasts with the Kullback-Leibler divergence from Ref.~\cite{SIMSPRL}, which was small already at $T\approx 200$--$250$~MeV, consistent with the very limited constraints on the viscosities at these higher temperatures.
        Since the hydrodynamic, particlization, and hadronic cascade stages were intentionally chosen to be identical to those of Ref.~\cite{SIMSPRL}, this difference in constraints can be ascribed to the difference between the pre-hydrodynamic models, IP-Glasma, and \trento{} with free-steaming. 
        This difference in dynamics is most pronounced at early times in the evolution, roughly corresponding to higher temperatures, where the increased constraints on the viscosities are found.  
        
        The non-viscous posteriors for the Grad and Chapman-Enskog viscous correction models are quite similar, demonstrating that these are robust to a modeling choice intended to be a small correction. The largest difference between the two models is in the particlization temperature,  which is robustly accounted for in the BMA posterior, and the median value remains consistent between the models.

\section{Summary and Conclusion}

       This study has implemented rigorous Bayesian model-to-data comparison with IP-Glasma for the first time and incorporated transfer learning for the first time in a full Bayesian model-to-data comparison. 
       It found inconclusive evidence of temperature-dependent compared to temperature-independent shear viscosity, as in Ref.~\cite{SIMSPRC}. However, unlike this previous work, the current study did not find conclusive preference for a model of viscous correction at particlization. A large number of postdictions and predictions are shown at maximum a posteriori and should be considered the new state-of-the-art theoretical result to which future measurements and calculations should be compared. The posterior distributions are the main results of this study and are our current best estimate of the properties of strongly-interacting matter in ultra-relativistic heavy ion collisions.
        From a performance point of view, the improved sampling procedure (the ordered Maximum Projection Latin Hypercube, compared with the maximin Latin Hypercube) resulted in a more rapid sampling that covered the design space coupled with an increased fidelity of the surrogate modeling. This large scale simulation involved varying parameters of IP-Glasma, the transport coefficients of the relativistic fluid dynamics phase, and the particlization temperature. 

        A hybrid model with an IP-Glasma initial state was constructed, and closure tests were able to recover input parameters of IP-Glasma, demonstrating self-consistency. This important step has shown that the sensitivity of the chosen final state hadronic observables to the pre-equilibrium phase was sufficient to establish to reliably extract accurate information. The self-consistency of the subsequent phases and elements of the hybrid model had been established in earlier studies, but this work establishes the utility of IP-Glasma in a large scale statistical study. 

        We have used Bayesian model comparison and Bayesian model averaging to establish the most likely values of the physical quantities included in this work. Special emphasis was put on the specific shear and bulk viscosity coefficients. The temperature dependence of the specific shear viscosity remains statistically indeterminate -- i.e. statistically consistent with being flat. Note that previously, several calculations with viscous hydrodynamics following IP-Glasma have produced successful phenomenology with a constant specific shear viscosity~\cite{Schenke:2012wb,Gale:2013da,Ryu:2015vwa,McDonald:2016vlt,Ryu:2017qzn,Schenke:2020mbo,Gale:2021emg}.        
    On the other hand, the specific bulk viscosity, $\zeta/s$, was found to be larger than in similar previous studies \cite{Bernhard:2019bmu,Nijs:2020ors,Nijs:2020roc,Parkkila:2021tqq,Parkkila:2021yha} and peaked during the hydrodynamic phase. Importantly, this study concludes that the bulk viscosity of strongly-interacting matter is inconsistent with zero.

        It is clear that the theoretical effort in the field is moving closer to true {\it ab initio} modelling of relativistic heavy-ion collisions. What this works also makes clear is that the physical quantities deduced from the analysis of the final states are influenced by the physics of the very early stages of the hadronic reaction. This is true in the case of hadrons, as emphasized in this study, as it is for electromagnetic variables \cite{Vujanovic:2016anq,Churchill:2020uvk,Coquet:2021lca,Gale:2021emg}. 

        As is often the case in fields with a plenitude of data, Bayesian model averaging remains the current state-of-the-art in heavy-ion collisions for leveraging the information in multiple models to best constrain the physical understanding of strongly-interacting matter without over-fitting (see also Bayesian model mixing~\cite{Everett:2021ruv,Phillips:2020dmw}). This is only the second study in this field, following \cite{SIMSPRL} and elaborated in \cite{Everett:2021ruv}, to utilize BMA for improving uncertainty quantification and has further demonstrated its importance. Further sources of unquantified uncertainty still exist in heavy ion collisions, usually at the interface between models at each stage in the evolution of the fireball, but how to incorporate such interface effects in BMA is not yet clear. A strong focus in studying the strongly-interacting matter produced in heavy ion collisions has been to improve the precision of the models; it should be emphasized that the pursuit of arbitrary precision without accounting for sources of uncertainty using techniques such as BMA is a perilous path: it does not fully leverage the information available, and could lead to bias. Simultaneous consideration of observables and uncertainty quantification are required for reliable inference of the physical properties of strongly-interacting matter.

        \acknowledgements
        We are happy to acknowledge useful exchanges with members of the JETSCAPE Collaboration. In addition, we are grateful for useful discussions with S. Bass, D. Everett, D. Liyanage, S. McDonald, N. Miro-Fortier, and M. Singh. 
        This work was funded in part by the Natural Sciences and Engineering Research Council of Canada, in part by the U.S. Department of Energy Grant no. DE-FG-02-05ER41367 and DE-SC0024347, and in part by Vanderbilt University. Computations were made on the B\'eluga, Cedar, Graham, Niagara, and Narval supercomputers managed by Calcul Qu\'ebec, SciNet, WestGrid, and other members of the Digital Research Alliance of Canada.

\bibliography{Bibliography}

\appendix

\section{Sensitivity Analysis} 
\label{appendix:A}

    First physics results can be produced once a surrogate model has been trained and validated with a reliable set of observables. Using the surrogate model, the global sensitivity of the model to variation of input parameters is analyzed. This analysis is an example of analysis of variance (ANOVA) that decompose the total variance of a model into variance of model parameters (at first order), pairs of model parameters (second order), and so on. The first-order Sobol indices quantify the global variance in model observable due to variance in model parameters \cite{SOBOL2001271} and are readily available \cite{ Iwanaga2022, Herman2017}.
    
    For a given observable output $y$, suppose it can be represented as a function of model parameters $\mathbf{x}$, $y = f(\mathbf{x})$. A prior predictive distribution $p(y)$ is produced for each output by marginalization,
    \begin{equation}
        p(y) = \int d\mathbf{x} p(y|\mathbf{x})p(\mathbf{x}).
    \end{equation}
    The quantity of interest is, however, the variance associated with a single parameter. In this case, suppose one fixed a single parameter $x_i$ to take a particular value $a$. The variance of the resulting distribution of outputs can be readily computed,
    \begin{equation}
        p(y|x_i = a) = \int dx_1 \dots dx_{i-1} dx_{i+1}\dots dx_n p(y|\mathbf{x})p(\mathbf{x})
    \end{equation}
    where $\mathbf{x}$ consists of $n$ elements. The variance of this distribution is $\text{Var}(y)|_{a} \equiv \text{Var}(p(y|x_i = a))$ and is the variance of the observable $y$ due to varying all parameters except $x_i$, \emph{i.e.} conditional on  $x_i = a$. By marginalizing over possible values of $x_i$, determined in turn by the prior, the variance due to variation of $x_i$ is found,
    \begin{equation}
        \text{Var}(y)|_{x_i} = \int da \text{Var}(y)|_a p(a).
    \end{equation}
    The first-order Sobol sensitivity index $S_1$ for a parameter $x_j$ and observable $y$ is then
    \begin{equation}
        S_1[x_j] \equiv \frac{\text{Var}(y) - \text{Var}(y)|_{x_j}}{\text{Var}(y)},
    \end{equation}
    the \emph{fractional} variance in the observable from variation of parameter $x_j$ alone. Therefore, if $S_1[x_j] = 0.7$, this is interpreted as 70\% of the global variation in this observable being ascribed to variation of $x_j$ alone.
    
\begin{figure*}[!htb]
        \begin{center}
            \includegraphics[width=0.85\textwidth]{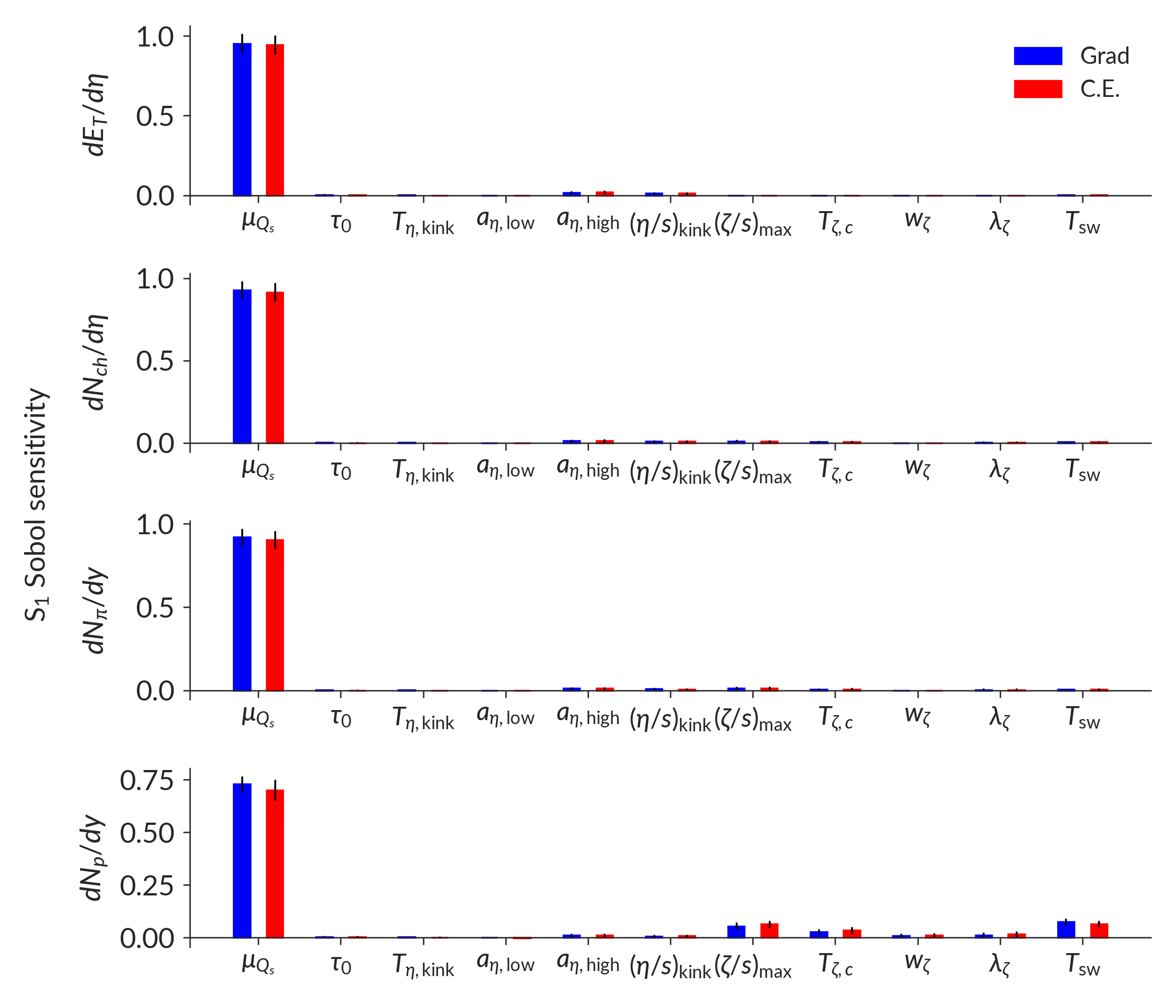}
            \caption{First-order Sobol sensitivity of charged hadron multiplicity, identified particle multiplicity, and transverse energy to input parameters.}
            \label{fig:full-study-sensitivity-multiplicity}
        \end{center}
    \end{figure*}

\begin{figure*}[!htb]
        \begin{center}
            \includegraphics[width=0.85\textwidth]{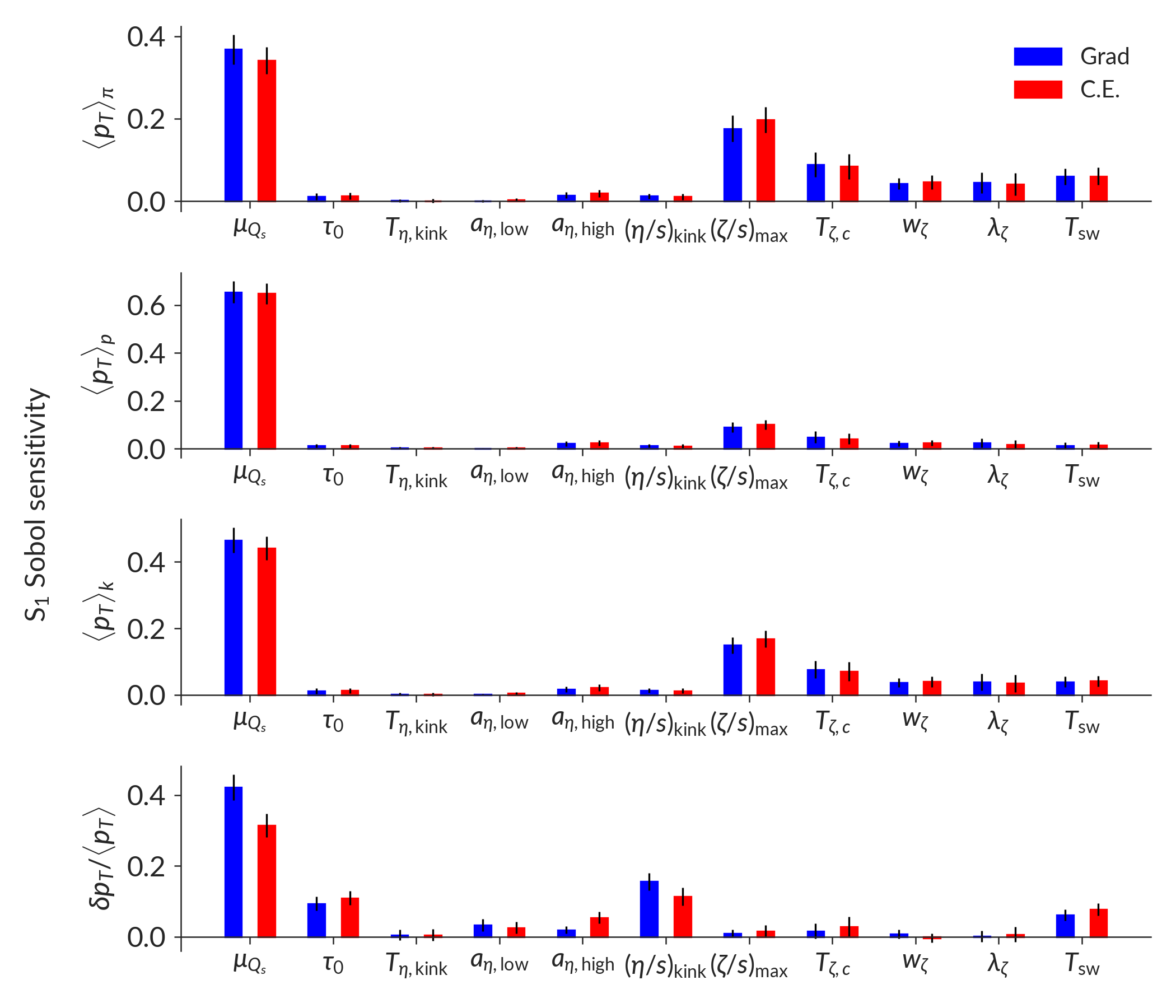}
            \caption{First-order Sobol sensitivity of identified particle mean transverse energy and correlated momentum fluctuations to input parameters.}
            \label{fig:full-study-sensitivity-pT}
        \end{center}
    \end{figure*}

\begin{figure*}[!htb]
        \begin{center}
            \includegraphics[width=0.85\textwidth]{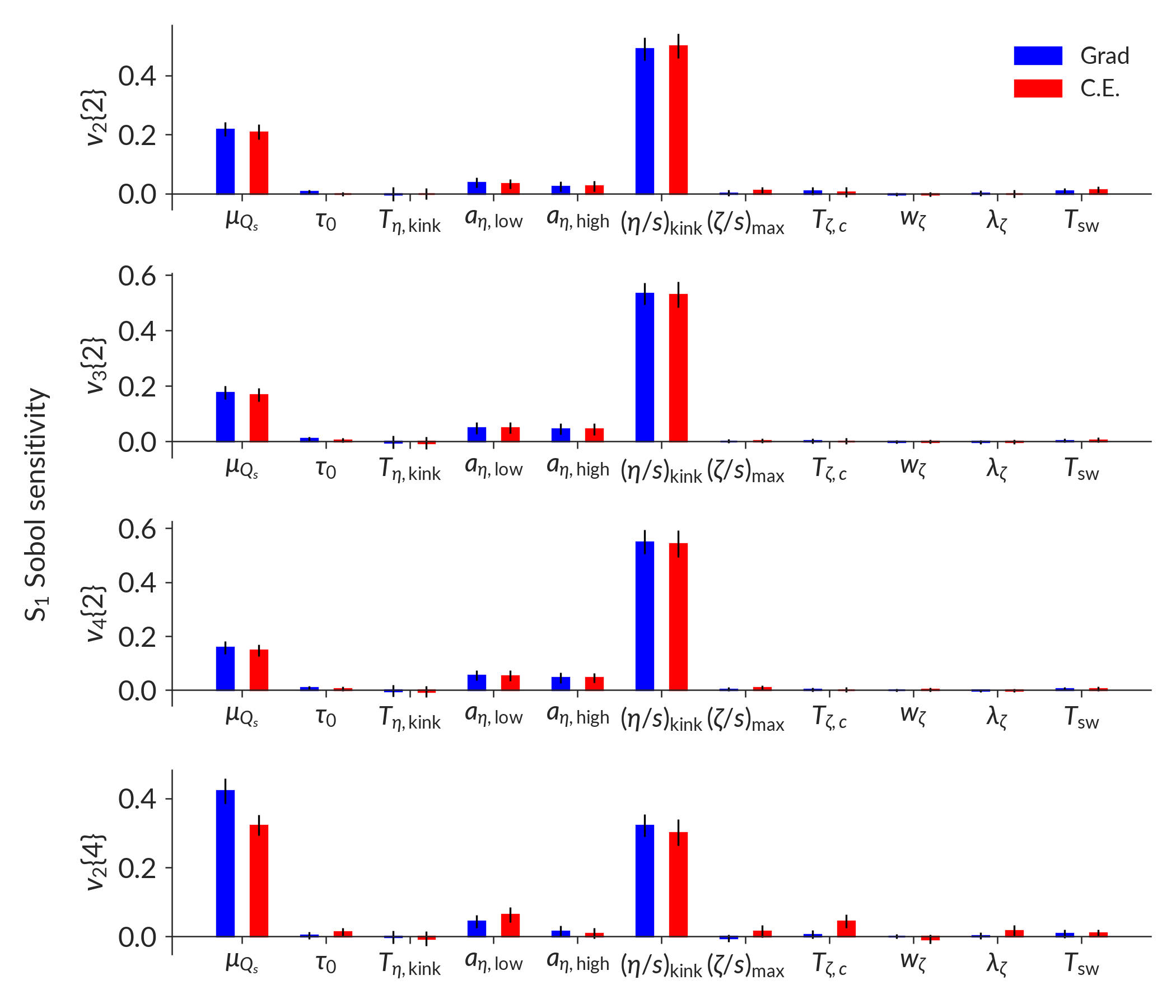}
            \caption{First-order Sobol sensitivity of charged hadron anisotropic flow coefficients to input parameters.}
            \label{fig:full-study-sensitivity-vn}
        \end{center}
    \end{figure*}

\begin{figure*}[!htb]
        \begin{center}
            \includegraphics[width=0.85\textwidth]{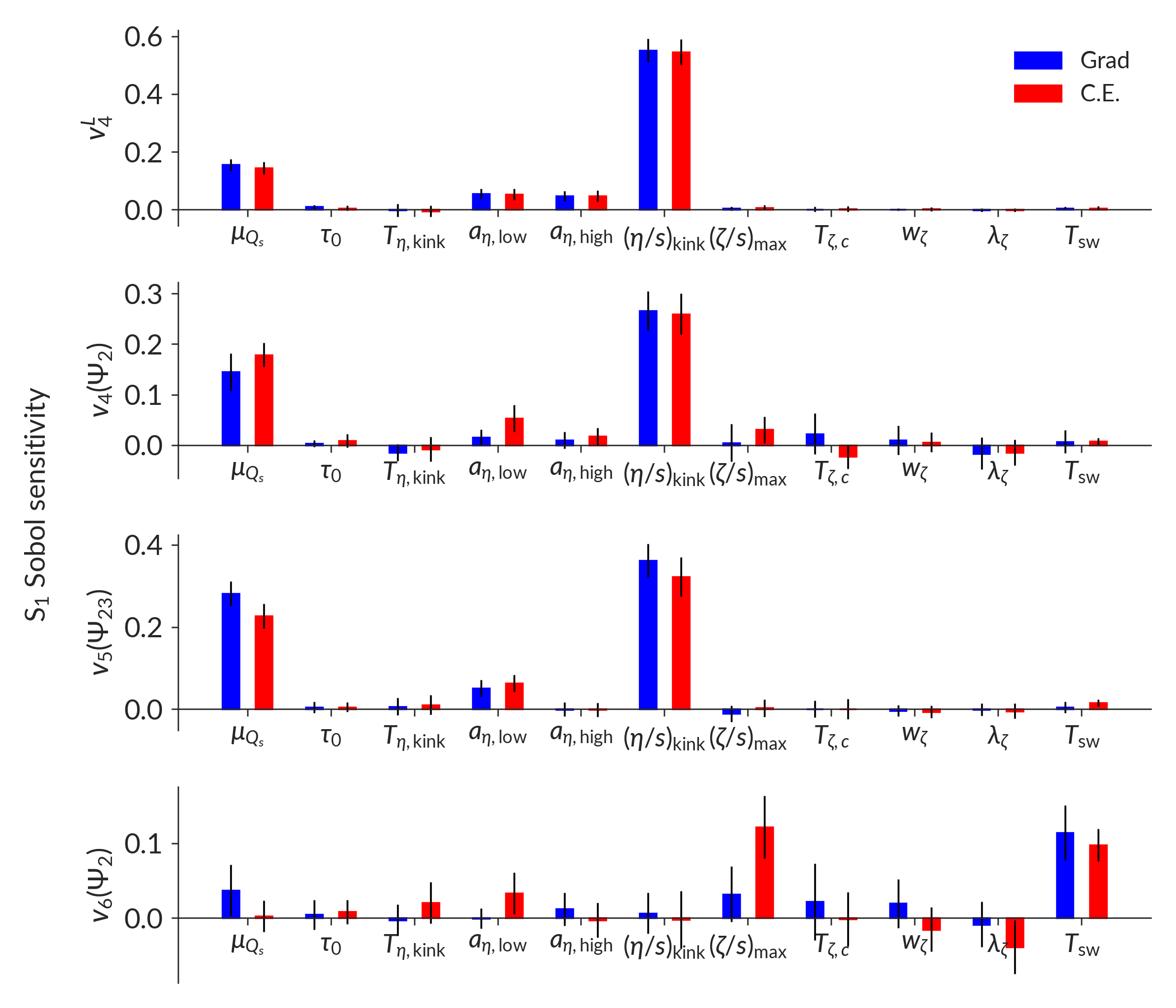}
            \caption{First-order Sobol sensitivity of charged hadron linear and nonlinear flow modes to input parameters.}
            \label{fig:full-study-sensitivity-linear-and-nonlinear-flow}
        \end{center}
    \end{figure*}

\begin{figure*}[!htb]
        \begin{center}
            \includegraphics[width=0.85\textwidth]{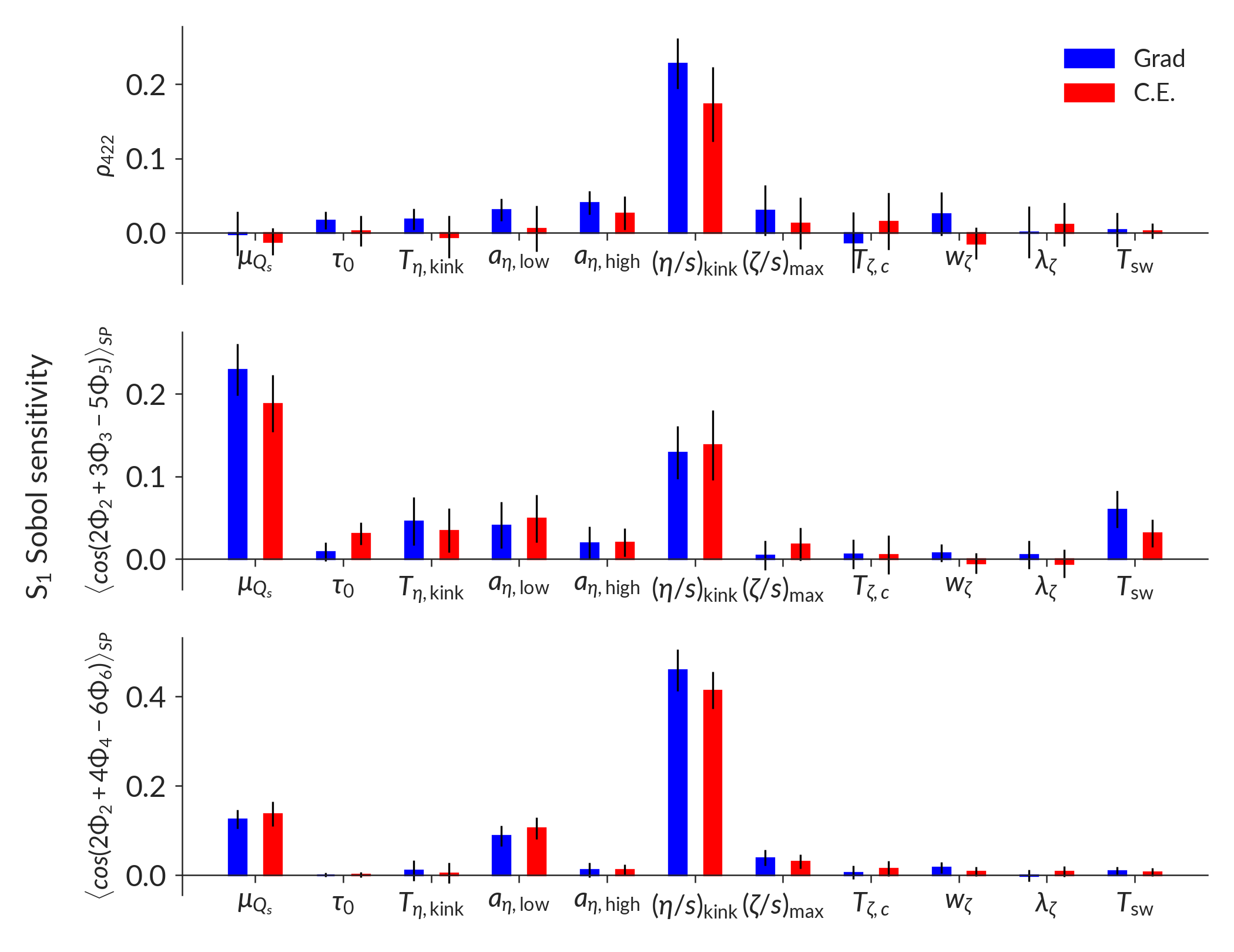}
            \caption{First-order Sobol sensitivity of charged hadron event plane correlators to input parameters.}
            \label{fig:full-study-sensitivity-correlators}
        \end{center}
    \end{figure*}

    The first-order Sobol sensitivities of the observables in the most central centrality bin are shown in Figs.~\ref{fig:full-study-sensitivity-multiplicity}-\ref{fig:full-study-sensitivity-correlators}. Due to length, they are divided into the following groups of observables: multiplicities and transverse energy in Fig.~\ref{fig:full-study-sensitivity-multiplicity}; identified hadron mean transverse momentum and correlated transverse momentum fluctuations in Fig.~\ref{fig:full-study-sensitivity-pT}; anisotropic flow in Fig.~\ref{fig:full-study-sensitivity-vn}; flow modes $v_4^L$, $v_4(\Psi_2)$, $v_5(\Psi_{23})$, $v_6(\Psi_2)$ in Fig.~\ref{fig:full-study-sensitivity-linear-and-nonlinear-flow}; and the plane correlations $\rho_{422}$, $\langle \cos(2\Phi_2+3\Phi_3-5\Phi_5) \rangle$, and $\langle \cos(2\Phi_2+4\Phi_4-6\Phi_6) \rangle$ in Fig.~\ref{fig:full-study-sensitivity-correlators}. The similar sensitivity of the two viscous correction models, almost always overlapping, demonstrates that the models are similar, but not identical, and that the viscous corrections are small compared to the overall effect of parameter variation.
    
    In these Sobol sensitivities, physical intuition is confirmed - an important step in further verifying that the model behaves as expected. The multiplicities and transverse energy (Fig.~\ref{fig:full-study-sensitivity-multiplicity}) are dominantly sensitive to the overall normalization ($\mu_{Q_s}$) and the viscosity presents a small correction. The proton multiplicity, one of the most sensitive observables to the chemistry of the system, is the more sensitive than the rest to the bulk viscosity and the switching temperature.
    
    The mean transverse momentum (Fig.~\ref{fig:full-study-sensitivity-pT}) again confirm the prior expectation that the dominant sensitivity is to the overall normalization and the bulk viscosity. To date, however, no study using an IP-Glasma initial state has been able to reproduce experimental results for $\delta p_T/\langle p_T \rangle$. 
    This sensitivity analysis reveals that while the dominant sensitivity is to the overall normalization, the fluctuations are primarily sensitive to the shear viscosity, onset of hydrodynamics, and the switching temperature. This study is the first with IP-Glasma to consider simultaneous variation of these parameters -- in concert with all observables -- to learn what can be learned about and from this observable.
    
    The anisotropic flow coefficients (Fig.~\ref{fig:full-study-sensitivity-vn}) reveal, as expected, that the dominant sensitivity is to shear viscosity and overall normalization. The overall normalization is related to the lifetime of the hydrodynamic phase, in turn allowing for more time for the shear viscosity to act. Interestingly, the difference in sensitivity between the two-particle and four-particle anisotropic flow is isolated to an increased sensitivity to overall normalization and a decreased sensitivity to the high-temperature slow of the shear viscosity. However, the $v_n\{m\}$ are broadly insensitive to the bulk viscosity and particlization temperature, as anticipated. 
    
    In summary of the first generation observables, the expectations set by both hand-tuning and previous studies is confirmed: the normalization, shear viscosity, and bulk viscosity are broadly sensitive to separate parameter families, but yield constraints across the input parameter space. Disappointingly perhaps, the switching time between the pre-equilibrium and hydrodynamic stage does not appear to be a dominant factor in the variance of any of these observables.
    
    Consideration of the global sensitivity of next generation observables begins with linear and nonlinear flow modes in Fig.~\ref{fig:full-study-sensitivity-linear-and-nonlinear-flow}. These observables have less defined and smaller sensitivity to the input parameters, alternately suggesting that they are insensitive to the parameters or that the IP-Glasma initial state by construction contains the information needed to reproduce these quantities. The exception in this case is to the shear viscosity, whose kink value dominates the constraint of these quantities. Nonetheless, physical expectations suggest that these quantities couple the pre-equilibrium and hydrodynamic stages in a way that similar first generation observables do not -- a feature that can be readily seen by their different relations to the principal components in Fig.~\ref{fig:full-study-pca-vectors-variance}. 
    
    The correlators (Fig.~\ref{fig:full-study-sensitivity-correlators}) are also less sensitive to variation of the parameters than the first generation observables. Additionally, the large uncertainties on the Sobol indices suggests that the sensitivity is less uniform across the space than in the first generation observables. Nonetheless, these observables further couple the initial state geometry to the hydrodynamic phase in a way poorly quantified by the first generation observables alone. 
    
    The sensitivity analysis provides the first glimpse into the response of an IP-Glasma + MUSIC + iS3D + SMASH hybrid model to parameter variation. This result, with a validated surrogate model, yields the first physics insights by calculating the global sensitivity of these observables with the leading physics model of the pre-equilibrium stage and how a realistic hydrodynamic medium responds. The PCA and sensitivity analysis confirm, before any inference has taken place, that the model outputs will yield constraints and that there is information in the next generation observables not contained in the first generation observables that have been the focus of most previous studies. Combined with a physical understanding of the observables themselves, this suggests strongly that exciting opportunities lie ahead for learning through model-to-data comparison.

\section{MCMC}

    This work uses \texttt{ptemcee} \cite{ptemcee, ptemcee-code} to implement parallel tempering Markov Chain Monte Carlo. This has been found to give excellent convergence behavior to the target distribution in heavy ion collisions \cite{SIMSPRC} with low autocorrelation. Low autocorrelation ensures that the model is sampling from the target, which is further supported by visualizing the trace of the MCMC chain to look for coordinated walks. As demonstrated in Fig. \ref{fig:full-study-grad-data-mcmc}, the chain is well-behaved for these applications.

\end{document}